\documentstyle[12pt]{article}
\newlength{\abstractwidth}
\renewcommand{\thanks}[1]{\footnote{#1}} 

\newcommand{\be}{\begin{equation}}
\newcommand{\bea}{\begin{eqnarray}}
\newcommand{\eea}{\end{eqnarray}}
\newcommand{\ee}{\end{equation}}
\newcommand{\N}{{\cal N}}
\newcommand{\<}{\langle}
\renewcommand{\>}{\rangle}
\def\ba{\begin{eqnarray}}
\def\ea{\end{eqnarray}}

\def\A{{\cal A}}

\def\M{{\cal M}}
\def\N{{\cal N}}
\def\R{{\cal R}}
\def\O{{\cal O}}

\def\X{{\cal X}}
\def\Z{{\cal Z}}

\def\Im{{\rm Im}}

\def\det{{\rm det}}

\def\half{ {1\over 2}}

\def\14{{1 \over 4}}
\def\p{\partial}

\def\tet{\vartheta}

\def\chiz{{\chi _{\bar z} ^+}}
\def\chiw{{\chi _{\bar w} ^+}}

\begin{document}
\baselineskip=16pt

\begin{flushright}
UCLA/01/TEP/27 \\
Columbia/Math/01
\end{flushright}

\bigskip

\begin{center}
{\Large \bf TWO-LOOP SUPERSTRINGS \ IV }
 \\
\bigskip
{\large \bf The Cosmological
Constant and Modular Forms
\footnote{Research supported in
part by National Science Foundation grants PHY-98-19686 and DMS-98-00783,
and by the Institute for Pure and Applied Mathematics under NSF grant
DMS-9810282.}}

\bigskip\bigskip

{\large Eric D'Hoker$^a$ and D.H. Phong$^b$} \\ 

\bigskip

$^a$ \sl Department of Physics and \\
\sl Institute for Pure and Applied Mathematics (IPAM) \\
\sl University of California, Los Angeles, CA 90095 \\
$^b$ \sl Department of Mathematics \\ 
\sl Columbia University, New York, NY 10027

\end{center}

\bigskip\bigskip

\begin{abstract}

The slice-independent gauge-fixed superstring chiral measure in genus 2
derived in the earlier papers of this series for each spin structure
is evaluated explicitly in terms of theta-constants.
The slice-independence allows an arbitrary choice of superghost insertion
points $q_1, q_2$ in the explicit evaluation, and the most effective one
turns out to be the {\sl split gauge} defined by $S_{\delta}(q_1,q_2)=0$.
This results in expressions involving bilinear theta-constants $\M$. The
final formula in terms of only theta-constants follows from new
identities between $\M$ and theta-constants which may be interesting in
their own right. The action of the modular group Sp(4,{\bf Z}) is worked
out explicitly for the contribution of each spin structure to the
superstring chiral measure.  It is found that there is a unique choice of
relative phases which insures the modular invariance of the full chiral
superstring measure, and hence a unique way of implementing the GSO
projection for even spin structure. The resulting cosmological constant
vanishes, not by a Riemann identity, but rather by the genus 2 identity
expressing any modular form of weight 8 as the square of a modular form
of weight 4. The degeneration limits for the contribution of each spin
structure are determined, and the divergences, before the GSO projection,
are found to be the ones expected on physical grounds.

\end{abstract}

\vfill\eject

\baselineskip=15pt
\setcounter{equation}{0}
\setcounter{footnote}{0}

\vfill\eject

\section{Introduction}
\setcounter{equation}{0}

This paper is the fourth of a series whose goal is to establish
unambiguous, explicit formulas for superstring multiloop amplitudes. 
In \cite{dpI}, a summary of the main results was presented.
In \cite{dpII}, a gauge-fixed expression for the superstring partition
function in genus 2 was obtained, which is invariant under
variations of the gauge slice $\chi_{\alpha}$
by local reparametrizations and local supersymmetry.
Specializing to Dirac-like gauge slices of the form 
$\chi_{\alpha}(z)=\delta(z,q_{\alpha})$, the gauge-fixed
expression was shown directly in \cite{dpIII} to be independent
of the points $q_{\alpha}$. For the superstring partition function,
the measure on moduli space is of the form
\be
\label{meas}
{\bf A}
=
\int \ (\det\,\Im\Omega)^{-5}d\mu(\Omega) 
\wedge \overline{d\mu(\Omega)}
\ee
Here the chiral superstring measure $d\mu(\Omega)$ is given by
\be
d\mu(\Omega)=\sum_{\delta}\eta_{\delta}\,d\mu[\delta](\Omega)
\ee
with $\eta_{\delta}$ the
relative phases between different spin structures,
yet to be determined, 
and $d\mu[\delta](\Omega)$ the contribution
to the superstring measure of each even spin
structure $\delta$. Explicit expressions
for $d\mu[\delta](\Omega)$
have been obtained in \cite{dpIII}
in terms of $\tet$-{\sl}functions, but not yet
in terms of $\tet$-{\sl constants}.

\medskip

The purpose of the present paper is to show that, exploiting the
invariance of the gauge-fixed expression from the points $q_{\alpha}$, 
the chiral measure $d\mu[\delta](\Omega)$
can actually be rewritten remarkably simply in terms of $\tet$-constants.
In fact, we show that
\bea
\label{dmu}
d\mu[\delta](\Omega)
=
{\tet^4[\delta](0, \Omega ) \ \Xi _6 [\delta ] (\Omega ) 
\over 
16\pi^6 \Psi _{10} (\Omega)} \ \prod_{I\leq J}d\Omega_{IJ}
\eea
where $\Psi _{10}$ is the 
well-known genus 2 modular form of modular weight 10
\be
\label{Psi10}
\Psi _{10} (\Omega )  \equiv  \prod _\epsilon  \tet ^2 [\epsilon ]
(0,\Omega )
\ee
The product in $\Psi _{10}$ is over all 10 even spin structure
$\epsilon$. 
The object $\Xi _6 [\delta ](\Omega)$ is the following
spin structure dependent combination of
$\tet$ constants, and is of modular weight 6
\be
\Xi _6 [\delta ] (\Omega ) \equiv \sum _{1\leq i<j\leq 3} 
\< \nu _i | \nu _j\> \prod _{k=4,5,6} \tet ^4 [\nu _i + \nu _j + \nu _k]
(0,\Omega )
\ee
In the definition of $\Xi _6[\delta]$, the even spin
structure $\delta$ is written as a sum of three distinct odd spin
structure $\delta = \nu _1 + \nu _2 + \nu _3$, and $\nu_4$, $\nu_5$,
$\nu _6$ denote the remaining 3 distinct odd spin structures.
Finally, the signature of a pair $\kappa$, $\lambda$ of spin structures
(even or odd) is defined by
\be
\< \kappa | \lambda \> \equiv \exp \{ 4 \pi i (\kappa ' \lambda '' -
\kappa '' \lambda ') \}
\ee
We would like to stress that $\Xi_6[\delta](\Omega)$ 
{\sl is not a modular form}, since the spin structure $\delta$
transforms under modular transformations. 
More precisely, it turns out that
modular transformations map the spin structure $\delta \to \tilde
\delta$, the period matrix $\Omega \to \tilde \Omega$, and the chiral
superstring measure $d \mu [\delta]$ as follows,
\bea
\tilde \Omega  & = & (A\Omega + B) (C \Omega + D)^{-1}
\hskip 1in 
\pmatrix{A&B\cr C&D\cr} \in Sp(4,{\bf Z})
\nonumber \\
d\mu [\tilde \delta ] (\tilde \Omega ) & = & \det (C\Omega +D)^{-5} d \mu
[\delta ] (\Omega )
\eea
A crucial feature of superstring theory in the RNS formulation 
\cite{rns} is the
imposition of the Gliozzi-Scherk-Olive (GSO) projection
\cite{gso}, which is to be
carried out independently in the two chiral sectors. This is done by
summing over all spin structures the chiral amplitudes for given spin
structure with an assignment of relative phases. This phase assignment
must be consistent with modular invariance.  From the above modular
transformation laws of chiral measure, it follows that there is a {\sl
unique modular invariant choice of relative phases}, namely $\eta_{\delta}=1$ 
for each
even spin structure $\delta$. Since 
$d\mu(\Omega)=\sum_{\delta}d\mu[\delta](\Omega)$
is now by construction a modular form,
we can identify by examining its degeneration
behavior, using the classification of modular forms of
Igusa \cite{igusa1,igusa2,igusa3}.
We find that the
chiral superstring measure $d\mu(\Omega)$
itself vanishes,
\be
\label{sumdeltadmu}
\sum _\delta d\mu [\delta ](\Omega ) =0
\ee
and this implies in turn that the {\sl two-loop cosmological constants
for both Type II and heterotic strings vanish pointwise on moduli space}.

\subsection{The Starting Point of the Derivation}

The starting point for this paper is the gauge-fixed expression
for the superstring chiral measure $d \mu [\delta ] (\Omega )$ on moduli
space obtained in \cite{dpIII} for each spin structure $\delta$,
\bea
\label{finamppq}
d\mu [\delta ] (\Omega ) & = & \prod _{I\leq J} d \Omega _{IJ} \int \prod
_\alpha d\zeta ^\alpha \A [\delta] (\Omega , \zeta)
\nonumber \\
{\bf A}[\delta] (\Omega, \zeta)
& = &
\Z \biggl \{ 1+\X_1+\X_2+\X_3+\X_4+\X_5+\X_6 \bigg \}
\eea
Here $\Z$ is the chiral matter and superghost correlation
function
\be
\label{firstzee}
\Z= {\< \prod_{a=1}^3b(p_a)\prod_{\alpha=1}^2\delta
\big(\beta(q_{\alpha})\big)\>
\over
\det\,\omega_I\omega_J(p_a) }
\ee
where $p_a$ are $3$ arbitrary points on the worldsheet $M$, and $\omega_I$ 
are holomorphic 1-forms with the usual normalization conditions on a
canonical basis
$(A_I,B_I)_{I=1}^h$ of homology cycles
\be
\oint_{A_I}\omega_J=\delta_{IJ},
\quad\quad
\oint_{B_I}\omega_J=\Omega_{IJ}.
\ee
The terms $\X_i$ are given by
\bea 
\label{peeque}
\X_1 + \X _6 &=&
{\zeta ^1 \zeta ^2 \over 16 \pi ^2}
\biggl [
-10 S_\delta (q_1,q_2) \p _{q_1} \p _{q_2} \ln E(q_1,q_2)
 \\
&&
\qquad \
 - \p _{q_1} G_2 (q_1,q_2) \p \psi ^* _1 (q_2)
 + \p _{q_2} G_2 (q_2,q_1) \p \psi ^* _2 (q_1)
\nonumber \\
&&
\qquad \
+ 2   G_2 (q_1,q_2) \p \psi ^* _1 (q_2)  f_{3/2} ^{(1)} (q_2)
- 2   G_2 (q_2,q_1) \p \psi ^* _2 (q_1)  f_{3/2} ^{(2)} (q_1)
\biggr ]
\nonumber \\
\X _2 &=&
{\zeta ^1 \zeta ^2 \over 16 \pi ^2}
\omega _I(q_1) \omega _J(q_2) S_\delta (q_1,q_2) 
\biggl [  \p _I \p _J \ln {\tet [\delta ](0)^5 \over \tet
[\delta ](D_\beta )} +   \p _I \p _J \ln \tet (D_b ) \biggr ]
\nonumber \\
\X _3 &=&
{\zeta ^1 \zeta ^2 \over 8 \pi ^2}
S_\delta (q_1,q_2) \sum _a  \varpi _a  (q_1, q_2) \biggl [ B_2(p_a) +
B_{3/2}(p_a) \biggr ]
\nonumber \\
\X_4 &=& {\zeta ^1 \zeta ^2 \over 8 \pi ^2}
S_\delta (q_1,q_2) \sum _a \biggl [
\p _{p_a} \p _{q_1} \ln E(p_a,q_1) \varpi ^* _a(q_2)
+ \p _{p_a} \p _{q_2} \ln E(p_a,q_2) \varpi ^* _a(q_1) \biggr ]
\nonumber \\
\X _5 &=&
{\zeta ^1 \zeta ^2 \over 16 \pi ^2}
\sum _a \biggl [ 
S_\delta (p_a, q_1) \p _{p_a} S_\delta (p_a,q_2) 
- S_\delta (p_a, q_2) \p _{p_a} S_\delta (p_a,q_1) \biggr ] 
\varpi _a  (q_1,q_2) \, .
\nonumber
\eea
Here, the quantity $\p \psi ^* _1 (q_2)$ is a tensor, given by a
multiplicative formula
\be
\p \psi ^* _1 (q_2) =
{\tet [\delta ] (2 q_2 -2 \Delta ) \over \tet [\delta ] (q_1 + q_2 - 2
\Delta) E(q_1,q_2)}{ \sigma (q_2) ^2 \over \sigma (q_1)^2}\, ,
\ee 
Finally, $\varpi_a$ are finite-dimensional determinants of
holomorphic forms defined by
\bea
\varpi  _a (u,v) &=& {\det \omega _I \omega _J(p_b \{ u,v;a \} )
\over \det \omega _I\omega _J(p_b)},
\nonumber \\
\omega _I \omega _J (p_b\{ u,v; a\} ) 
&=& \left \{ \matrix{
\omega _I \omega _J(p_b) & \qquad b\not= a\cr
\half (\omega _I(u)\omega _J(v) + \omega _I(v) \omega _J(u) )  &
\qquad b=a
\cr}
\right .
\eea
All other quantities, such as the holomorphic 2-forms $B_2(w)$,
$B_{3/2}(w)$, the holomorphic 1-forms $\varpi ^* _a (w)$ and the
meromorphic 1-forms $f_{3/2}^{(\alpha)} (w)$ in (\ref{finamppq}) and
(\ref{peeque}) were defined in III, and will not be repeated here since
their precise form will not be needed here.

\subsection{Key Steps of the Procedure}

All terms in the gauge-fixed expression of (\ref{finamppq}) and
(\ref{peeque}) can be written explicitly in terms of $\tet$-{\sl functions}.
However, although the expression (\ref{finamppq}) has been shown to
be independent of the points $p_a,q_{\alpha}$ \cite{dpIII},
its apparent dependence on $p_a,q_{\alpha}$ is a major impediment to understand 
its modular transformations, and ultimately, to 
its effective evaluation.   
Our main task is then to recast 
(\ref{finamppq}) in terms of $\tet$-{\sl constants}.
The key steps in this procedure are the following. 

\medskip

$\bullet$
{\sl The split gauge condition}. The first delicate but very useful choice
is that of the points $q_{\alpha}$. In view of the appearance of the
Szeg\"o kernel $S_{\delta}(q_1,q_2)$ as an overall factor in the terms
$\X_2$, $\X_3$, and $\X_4$, it is very convenient to require the points
$q_1,q_2$ to satisfy the following
condition
\be
\label{split}
S_{\delta}(q_1,q_2)=0.
\ee
which we shall refer to as the {\sl split gauge condition}
\footnote{The split
gauge condition is more properly a one complex parameter family of gauge
conditions, as for example the point $q_1$ may be chosen freely, leaving
a two-fold solution for $q_2$.}.
This condition is also natural from another viewpoint. Recall that
the genus 2 period matrix $\Omega_{IJ}$ and the super period matrix
$\hat\Omega_{IJ}$ are related by (see \cite{dp88,dp89}) 
\be
\hat\Omega_{IJ}
=
\Omega_{IJ}
-  { i  \over 8 \pi}  \int \! d^2z \int \! d^2w\
\omega _I (z) \chiz S_{\delta}(z,w) \chiw \omega _J (w).
\ee
For $\chiz=\sum_{\alpha=1}^2\zeta^{\alpha}\delta(z,q_{\alpha})$, the
condition $S_{\delta}(q_1,q_2)=0$ implies that
$\Omega_{IJ} = \hat \Omega_{IJ}$. With (\ref{split}), 
we have $\X_2=\X_3=\X_4=0$, and it remains to
evaluate only $\X_1+\X_6$ and $\X_5$.

\medskip

$\bullet$ 
{\sl The hyperelliptic representation}. The difficulty with the gauge
choice (\ref{split}) is that the two points $q_1$ and $q_2$ are now
related in a complicated moduli and spin structure dependent way.
Fortunately, in genus 2, it turns out that the relation (\ref{split})
between $q_1$ and $q_2$ can be solved explicitly, by making use of the
hyperelliptic representation. In this representation, the surface is
described as a double cover of the complex plane, with three branch
cuts supported at six branch points $u_i$, $i=1,\cdots ,6$. There is a
one-to-one map between the six odd spin structures $\nu _i$ and the branch 
points $u_i$; there is also a one-to-one map between the 10 even spin
structures $\delta$ and partitions of the set of six branch points into
two disjoint sets of 3 branch points each.

\medskip

$\bullet$
{\sl The points $p_a$, $a=1,2,3$, at branch points }. The correlation
function $\Z$ is by itself independent of the ghost insertion points
$p_a$, so there is a great flexibility in setting $p_a$ at various
special points of the Riemann surface $\Sigma$. It is very useful to make 
a further gauge choice and put the three points $p_a$ at the three branch
points (or the complementary three branch points) of the partition of
branch points associated with spin structure $\delta$ in the
hyperelliptic representation of $\Sigma$. With this choice and the
explicit hyperelliptic solution to the split gauge condition $S_\delta
(q_1,q_2)=0$, we find
\be
\X_1+\X_6=0 \, .
\ee 
This choice also leads to an explicit formula for $\Z \X_5$, and hence for
the chiral measure $d\mu[\delta]$, where the residual dependence on
the points $q_1,q_2$ explicitly cancels out. The resulting
formula is expressed
entirely  in terms of the following {\sl bilinear $\tet$-constants} 
\be
\M_{\nu_i\nu_j} \equiv \M_{\nu_i\nu_j} (\Omega) \equiv 
  \p_1 \tet [\nu _i] (0,\Omega) \p_2 \tet [\nu _j] (0,\Omega)
- \p_2 \tet [\nu _i] (0,\Omega) \p_1 \tet [\nu _j] (0,\Omega)
\ee
and it is
\be
\label{dmuM}
d\mu[\delta](\Omega) 
= \prod_{I\leq J}d\Omega_{IJ}
\,\tet [\delta ](0) ^4 \cdot 
{
\<\nu_1 |\nu _2\> \M _{\nu _1 \nu _2}^4 +
\<\nu_2 |\nu _3\> \M _{\nu _2 \nu _3}^4 + 
\<\nu_3 |\nu _1\> \M _{\nu _3 \nu _1}^4 
\over 
16\pi^2\,\M _{\nu _1 \nu _2}^2 \M _{\nu _2 \nu _3}^2
\M _{\nu _3 \nu _1}^2} 
\ee

\medskip

$\bullet$  
{\sl Relation between bilinear $\tet$-constants and $\tet$-constants}.
The bilinear $\tet$-constants $\M_{\nu_i\nu_j}$ still involve derivatives
of $\tet$-functions. We can however establish a very powerful identity
which expresses $\M_{\nu_i\nu_j}$ directly in terms of $\tet$-constants
\be
\M_{\nu_i\nu_j}
= \pm \pi ^2 \prod _{k\not= i,j} \tet [\nu _i + \nu _j + \nu _k]
(0,\Omega)
\ee
This identity is proven by noticing that both sides have modular
weight 2, that the right hand side vanishes only on the boundary
of moduli space and that the asymptotic behaviors of both sides
at the boundary of moduli space match. With this identity,
we obtain the desired expression (\ref{dmu}) for $d\mu [\delta]$.

\medskip

$\bullet$ {\sl Relative phase factors associated with the GSO projection}.
Under modular transformations, the expressions $\M_{\nu_i\nu_j}$
transform covariantly into each other. By working out the effect of each
generator of $SL(4,{\bf Z})$, we can show that the unique choice leading
to a modular form for the sum over spin structures is when the relative
phase factors for $d\mu[\delta]$ are all $\eta_{\delta}=1$.

\medskip

$\bullet$ {\sl The Cosmological Constant}.
The expression 
\be
\label{sumdeltaxi}
\sum _\delta \tet ^4 [\delta ](0,\Omega) \Xi _6 [\delta ](\Omega)
\ee
entering into (\ref{sumdeltadmu}) is a modular form of weight $8$. 
As we noted before, the
ring of modular forms in genus 2 has been identified completely by
Igusa \cite{igusa1,igusa2,igusa3}. Modular forms of weight 8 must be 
proportional to the
square of the unique form of weight 4, $\Psi_4(\Omega) ^2$. Now
$\Psi_4 (\Omega)$ does not vanish along the degeneration divisor, where
the Riemann surface degenerates into two disjoint tori. One shows that
(\ref{sumdeltaxi}) vanishes, since it tends to $0$ along this divisor.
More precisely, upon using the Riemann relations for genus 2,
(\ref{sumdeltaxi}) may be rearranged in the following way,
\be
\label{sumdeltaxibis}
\sum _\delta \tet ^4 [\delta ](0,\Omega) \Xi _6 [\delta ](\Omega)
=
2 \sum _\delta \tet ^{16} [\delta ] - \half \biggl ( \sum _\delta \tet ^8
[\delta ] \biggr ) ^2
\ee
an expression which is known to vanish, but not just by the use of the
Riemann relations.

\bigskip 

This paper is organized as follows. 
In Section \S 2, we recall some basic facts about hyperelliptic Riemann
surfaces and their spin structures. We also establish some important
formulas for the sequel, including Szeg\"o kernel identities, the
relations between the Szeg\"o kernel and the Green's function on
$2$-forms, and Thomae-type identities for the constants
$\M_{\nu_i\nu_j}$. 
Section \S 3 is devoted to the reduction of the full gauge-fixed
expression (\ref{finamppq}) to the expression (\ref{dmuM}) purely in
terms of $\M_{\nu_i\nu_j}$. The steps include the evaluation of $\Z$ by
setting the $p_a$'s at different branch points, the proof of the
vanishing of $\X_1+\X_6$, the explicit evaluation of $\Z\X_5$, using
Thomae-formulae for $\M_{\nu_i\nu_j}$, and the concrete realization of the
gauge condition $S_{\delta}(q_1,q_2)=0$ in the
hyperelliptic representation. 
In Section \S 4, we show how our considerations lead heuristically to the
key identity relating $\M_{\nu_i\nu_j}$ to $\tet$-constants.
Assuming this relation for the moment, we derive the formula (\ref{dmu}).
Section \S 5 is devoted to the proof of the required $\tet$ identities,
namely the identity between ${\cal M}_{\nu_i\nu_j}$ and $\tet$-constants
mentioned above, and another identity, the ${\cal M}$ product formula. 
For this, we need to examine in detail the degeneration behavior
of ${\cal M}_{\nu_i\nu_j}$, both when the degenerating cycle
is or is not a separating cycle.
In Section \S 6, we begin by determining explicitly
the effect of modular transformations on the $\Xi_6 [\delta] (\Omega)$,
with particular care about the phases. The net outcome is a proof that
the ``zero relative phases" is the only possible choice leading to
modular invariance for the superstring chiral measure, so that there is
one and exactly one way of implementing the GSO projection.
The vanishing of the cosmological constant is established by
examining the degeneration limits of $d\mu(\Omega)$.
In Section \S 7, the chiral measure for the heterotic string is shown to
follow at once, and the heterotic cosmological constant is shown to
vanish. In this section, we take the opportunity to show how our
methods also give readily the well-known chiral measure for the bosonic
string, by direct evaluation and without appealing this time to Igusa's
classification of genus 2 modular forms.
In Section \S 8, we verify the consistency of the chiral measure
$d\mu[\delta](\Omega)$ with the degeneration behavior expected on
physical grounds. We do find the expected tachyon and massless
intermediate state divergences.

\vfill\eject

\section{Genus 2 Riemann Surfaces}
\setcounter{equation}{0}

We begin by collecting fundamental facts about genus 2 Riemann surfaces,
their spin structures, their holomorphic and meromorphic differentials
and modular properties, both in the $\tet$-function formulation and in the
hyperelliptic representation.

\subsection{$\tet$-Characteristics and Spin Structures}

On a Riemann surface $\Sigma$ of genus 2, there are 16 spin structures, of
which 6 are odd (usually denoted by the letter $\nu$) and 10 are even
(usually denoted by the letter $\delta$). Each spin structure $\kappa$
(even or odd) can be identified with a $\tet$-characteristic $\kappa =
(\kappa '|\kappa'')$, where
$\kappa ', \kappa '' \in \{0,\half\}^2$, and we shall represent those here
by {\it column matrices}. The parity of the spin structure $\kappa$ is the 
same as the parity of the integer $4\kappa' \cdot \kappa''$. The
corresponding
$\tet$-function is an entire function, defined by
\be
\tet [\kappa ] (\zeta, \Omega) 
\equiv 
\sum _{n \in {\bf Z}^2} \exp \{\pi i (n+\kappa ')\Omega (n+\kappa')
+ 2\pi i (n+\kappa ') (\zeta + \kappa '') \}
\ee
which is even or odd depending on the parity of the spin structure. It is
convenient to list here the following useful periodicity relations for
$\tet[\kappa ](\zeta,\Omega)$, in which $M,N \in {\bf Z}^2$
\bea
\tet [\kappa ] (\zeta + M + \Omega N, \Omega ) 
&=&
\tet [\kappa ](\zeta , \Omega) \ \exp \{ -i \pi N \Omega N
- 2 \pi i N (\zeta +\kappa '') + 2 \pi i \kappa ' M \}
\nonumber \\
\tet [\kappa ' +N, \kappa '' +M ] (\zeta , \Omega) 
&=&
\tet [\kappa ',\kappa ''] (\zeta, \Omega) \ \exp \{2 \pi i \kappa ' M\}
\eea
The standard $\tet$-function is defined by $\tet (\zeta, \Omega)= \tet[0]
(\zeta, \Omega)$. We have the following relation between $\tet$-functions
with different characteristics
\be 
\tet [\kappa ] (\zeta , \Omega) 
=
\tet (\zeta + \kappa '' + \Omega \kappa ', \Omega)
\ \exp \{\pi i \kappa ' \Omega \kappa ' +  2 \pi i \kappa '(\zeta + \kappa
'') \} \nonumber
\ee
For each odd spin structure $\nu$ we have $\tet [\nu ](0,\Omega)=0$. For
each even spin structure $\delta$ one defines the particularly important 
$\tet$-constants, 
\be
\tet[\delta] 
\equiv  \tet [\delta] (0 , \Omega ) \, .
\ee

Every genus 2 surface admits a hyperelliptic representation, given by a
double cover of the complex plane with three quadratic branch cuts
supported by 6 branch points, which we shall denote $u_i$, $i=1,\cdots,6$.
The full surface $\Sigma$ is obtained by gluing together two copies of
${\bf C}$ along, for example, the cuts from $u_{2j-1}$ to $u_{2j}$,
$1\leq j\leq 3$.   The surface is then parametrized by\footnote{It is
customary to introduce a local coordinate system $z(x)=(x,s(x))$, which is
well-defined also at the branch points. Throughout, the formulas in the
hyperelliptic representation will be understood in this way. However, to
simplify notation, the local coordinate $z(x)$ will not be exhibited
explicitly.}
\be
\label{hyperelliptic}
s^2 = \prod _{i=1} ^6 (x-u_i)
\ee
In the hyperelliptic representation, there is another
convenient way of identifying spin structures. Each spin structure can be
viewed then as a partition of the set of branch points $u_i$, $i=1,\cdots
,6$ into two disjoint subsets, in the following way.
\bea
\nu \ {\rm odd } & \Leftrightarrow & {\rm branch \
point } \ u_i
\\ 
\delta \ {\rm even } & \Leftrightarrow & {\rm
partition} \ A \cup B, \qquad
A = \bigg \{ u_{i_1} , u_{i_2}, u_{i_3} \bigg \}, \ 
B= \bigg \{ u_{i_4} , u_{i_5}, u_{i_6} \bigg \}
\nonumber 
\eea
where $(i_1, i_2, i_3, i_4, i_5, i_6 )$ is a permutation of
$(1,2,3,4,5,6)$.

\medskip

A typical explicit relation between even spin structures as identified
with $\tet$ characteristics and spin structures as identified with
partitions $\{u_i\} = A \cup B$ of the set of branch points in the
hyperelliptic representation is a Thomae formula for $\tet$-constants
\be
\label{thomae}
\tet[\delta]^8
=
c  \prod_{i<j \in A}(u_i - u_j)^2 \prod_{k<l \in B}(u_k - u_l)^2
\hskip .7in 
c=\det^4 \biggl (\oint_{A_I}{x^{J-1}dx\over s} \biggr )
\ee
Here $c$ is a spin structure $\delta $ independent quantity
\cite{fay,lm}.

\medskip  

The signature assignment between (even or odd)
spin structures $\kappa$ and $\lambda$ is defined by
\be
\label{signature}
\< \kappa | \lambda \> \equiv \exp \{4 \pi i (\kappa ' \lambda '' -
\kappa '' \lambda ') \}
\ee
and has the following properties,
\begin{itemize}
\item If $\nu _1$ and $\nu _2$ are odd then 
\bea
\<\nu _1|\nu_2\> = +1 &\Leftrightarrow & \nu _1 - \nu _2 \quad {\rm even}
\nonumber \\
\<\nu _1|\nu_2\> = -1 &\Leftrightarrow & \nu _1 - \nu _2 \quad {\rm odd}
\eea

\item If $\nu_1$, $\nu_2$ and $\nu_3$ are odd and all distinct, then
\be
\<\nu _1|\nu_2\> \<\nu _2|\nu_3\> \<\nu _3|\nu_1\> = -1
\ee
\end{itemize}

\subsection{Relations between even and odd spin structures}

There exists in genus 2 a simple relation between even and odd spin
structures, i.e. between even and odd $\tet$-characteristics. We shall
need this relation extensively and thus discuss it in detail. To see
this type of relation as explicitly as possible, it is very convenient
to choose a homology basis. In the next subsection, we shall exhibit
the behavior of spin structure and $\tet$-characteristics under modular
transformations.

\medskip

The odd spin structures may be labeled as follows,\footnote{The pairs for
which $\<\nu_i|\nu_j\>=-1$ are 14, 16, 23, 25, 35, 46; all others give
$\<\nu_i|\nu_j\>=+1$.} 
\bea
\label{listodd}
2\nu_1 =\left (\matrix{0 \cr  1\cr} \bigg | \matrix{0 \cr 1\cr} \right )
\qquad
2\nu_3 =\left (\matrix{0 \cr  1\cr} \bigg | \matrix{1 \cr 1\cr} \right )
\qquad
2\nu_5 =\left (\matrix{1 \cr  1\cr} \bigg | \matrix{0 \cr 1\cr} \right )
\nonumber \\
2\nu_2 =\left (\matrix{1 \cr  0\cr} \bigg | \matrix{1 \cr 0\cr} \right )
\qquad
2\nu_4 =\left (\matrix{1 \cr  0\cr} \bigg | \matrix{1 \cr 1\cr} \right )
\qquad
2\nu_6 =\left (\matrix{1 \cr  1\cr} \bigg | \matrix{1 \cr 0\cr} \right )
\eea
while the even ones may be labeled by,
\bea
2\delta_1 =\left (\matrix{0 \cr 0\cr} \bigg | \matrix{0 \cr 0\cr} \right )
\qquad
2\delta_2 =\left (\matrix{0 \cr 0\cr} \bigg | \matrix{0 \cr 1\cr} \right )
\qquad
2\delta_3 =\left (\matrix{0 \cr 0\cr} \bigg | \matrix{1 \cr 0\cr} \right )
\qquad
2\delta_4 =\left (\matrix{0 \cr 0\cr} \bigg | \matrix{1 \cr 1\cr} \right )
\nonumber \\
2\delta_5 =\left (\matrix{0 \cr 1\cr} \bigg | \matrix{0 \cr 0\cr} \right )
\qquad
2\delta_6 =\left (\matrix{0 \cr 1\cr} \bigg | \matrix{1 \cr 0\cr} \right )
\qquad
2\delta_7 =\left (\matrix{1 \cr 0\cr} \bigg | \matrix{0 \cr 0\cr} \right )
\qquad
2\delta_8 =\left (\matrix{1 \cr 0\cr} \bigg | \matrix{0 \cr 1\cr} \right )
\nonumber \\
2\delta_9 =\left (\matrix{1 \cr 1\cr} \bigg | \matrix{0 \cr 0\cr} \right )
\qquad
2\delta_0 =\left (\matrix{1\cr 1\cr} \bigg | \matrix{1\cr 1\cr} \right)
\eea
The first relation between even and odd $\tet$-characteristics states that
the sum of all odd spin structures is a specific double period, 
\be
\label{doubleperiod}
\nu _1 + \nu _2 + \nu _3  + \nu _4 + \nu _5 + \nu _6  = 4 \delta _0\, .
\ee 
The second relation makes it clear that every even spin sructure
$\delta$ may be viewed as a partition of the set of 6 branch points into
two disjoint sets of 3 branch points,
\bea
\label{evenodd}
\nu _1 + \nu _2 + \nu _3  & = & \delta _7  + 2 \nu _3 
\qquad \qquad
\nu _1 + \nu _2 + \nu _4   =  \delta _5  + 2 \nu _4 
\nonumber \\
\nu _1 + \nu _2 + \nu _5  & = & \delta _3  + 2 \nu _5 
\qquad \qquad
\nu _1 + \nu _2 + \nu _6   =  \delta _2  + 2 \nu _6 
\nonumber \\
\nu _1 + \nu _3 + \nu _4  & = & \delta _8  + 2 \nu _3 
\qquad \qquad
\nu _1 + \nu _3 + \nu _5   =  \delta _0  + 2 \nu _1 
\nonumber \\
\nu _1 + \nu _3 + \nu _6  & = & \delta _9  + 2 \nu _3 
\qquad \qquad
\nu _1 + \nu _4 + \nu _5   =  \delta _4  + 2 \nu _5 
\nonumber \\
\nu _1 + \nu _4 + \nu _6  & = & \delta _1  + 2 \delta _0 
\qquad \qquad 
\nu _1 + \nu _5 + \nu _6   =  \delta _6  + 2 \nu _5 
\eea
Thus, each even spin structure $\delta$ can be written as $\delta =
\nu_{i_1}+\nu_{i_2}+\nu_{i_3}$, where the $\nu_{i_a}$, $a=1,2,3$ are odd
and pairwise distinct. The mapping $\{\nu_{i_1},\nu_{i_2},\nu_{i_3}\} \to
\delta$ is 2 to 1, with $\nu_{i_1}+\nu_{i_2}+\nu_{i_3}$ and its complement
$\nu_1 + \cdots +\nu_6 - ( \nu_{i_1}+\nu_{i_2}+\nu_{i_3})$
corresponding to the same even spin structure, in view of
(\ref{doubleperiod}).

\subsection{The Action of Modular Transformations}

Modular transformations $M$ form the infinite discrete group $Sp(4,{\bf
Z})$, defined by
\be
M=\left ( \matrix{A & B \cr C & D \cr} \right )
\qquad \qquad
\left ( \matrix{A & B \cr C & D \cr} \right ) 
\left ( \matrix{0 & I \cr -I & 0 \cr} \right )
\left ( \matrix{A & B \cr C & D \cr} \right ) ^t
=
\left ( \matrix{0 & I \cr -I & 0 \cr} \right )
\ee
where $A,B,C,D$ are integer valued $2 \times 2$ matrices and the
superscript ${}^t$ denotes transposition. The group is generated by the
following elements
\bea
M_i &=& \left ( \matrix{I & B_i \cr 0 & I \cr} \right ) 
\qquad \ \ \ 
B_1 = \left ( \matrix{1 & 0 \cr 0 & 0 \cr} \right )
\quad 
B_2 = \left ( \matrix{0 & 0 \cr 0 & 1 \cr} \right )
\quad 
B_3 = \left ( \matrix{0 & 1 \cr 1 & 0 \cr} \right )
\nonumber \\
S &=& \left ( \matrix{0 & I \cr -I & 0 \cr} \right ) 
\\
\Sigma &=& \left ( \matrix{\sigma & 0 \cr 0 & -\sigma \cr} \right ) 
\qquad \ \ \ 
\sigma = \left ( \matrix{0 & 1 \cr -1 & 0 \cr} \right )
\nonumber \\
T &=& \left ( \matrix{\tau _+ & 0 \cr 0 & \tau _- \cr} \right )
\qquad \quad 
\tau _+ = \left ( \matrix{1 & 1 \cr 0 & 1 \cr} \right )
\quad 
\tau _- = \left ( \matrix{1 & 0 \cr -1 & 1 \cr} \right )
\nonumber 
\eea
To exhibit the action of the modular group on 1/2 characteristics
$\kappa$ (even or odd), it is convenient to assemble the 1/2
characteristics into a single column of 4 entries and the action of the
modular group is then given by \cite{igusa3}
\be
\left (\matrix{ \tilde \kappa' \cr \tilde \kappa ''\cr}  \right )
=
\left ( \matrix{D & -C \cr -B & A \cr} \right )
\left ( \matrix{ \kappa ' \cr \kappa '' \cr} \right )
+ \half \ {\rm diag} 
\left ( \matrix{CD^T  \cr AB^T \cr} \right )
\ee
Here and below, ${\rm diag} (M)$ of a $n \times n$ matrix $M$ is an
$1\times n$ column vector whose entries are the diagonal entries on $M$. 
With the above generators, this action of the modular group on
characteristics reduces to the following expressions, (mod 1), (returning
to our previous notation for the characteristics as $2 \times 2$ matrices)
\bea
M_i \left (\kappa ' |\ \kappa '' \right )
&=& 
\left ( \kappa ' |\ \kappa '' + B_i \kappa ' + \half {\rm diag} (B_i)
\right )
\nonumber \\
S \left ( \kappa ' |\ \kappa '' \right )
&=& 
\left ( \kappa '' |\ \kappa ' \right ) 
\nonumber \\
\Sigma \left (\kappa ' |\ \kappa ''  \right )
&=& 
\left (\sigma \kappa ' |\ \sigma \kappa ''  \right ) 
\nonumber \\
T \left (\kappa ' |\ \kappa '' \right )
&=& 
\left ( \tau _- \kappa ' |\ \tau _+ \kappa ''  \right ) 
\eea
On the period matrix, the transformation acts by
\be
\tilde \Omega = (A\Omega + B ) (C\Omega + D)^{-1}
\ee
while on the Jacobi $\tet$-functions, the action is given by
\be
\tet [\tilde \kappa ] \biggl ( \{(C\Omega +D)^{-1} \}^t \zeta , \tilde
\Omega \biggr ) =
\epsilon (\kappa, M) \det (C\Omega + D) ^\half 
e^{ i \pi \zeta ^t (C\Omega +D)^{-1} C \zeta }
\tet [ \kappa ] (\zeta, \Omega)
\ee
where $\kappa = (\kappa ' |\kappa '')$ and $\tilde \kappa = (\tilde
\kappa ' | \tilde \kappa '')$. The phase factor $\epsilon (\kappa, M)$
depends upon both $\kappa $ and the modular transformation $M$ and obeys 
$\epsilon (\kappa , M )^8=1$.
We shall be most interested in the modular transformations of
$\tet$-constants $\tet ^4 [\delta]$ and thus in even spin structures
$\delta$ and the fourth powers of $\epsilon$, which are given by 
\bea
&&
\epsilon ^4 (\delta, M_i)  = \exp \{ 4 \pi i \delta ' B_i \delta '\}
\qquad i=1,2
\nonumber \\
&&
\epsilon ^4 (\delta, M_3) = \epsilon ^4 (\delta , S) = \epsilon ^4
(\delta, \Sigma) = \epsilon ^4 (\delta, T)=1  
\eea
A convenient way of establishing these values is by first analyzing the
case of the shifts $M_i$, whose action may be read off from the
definition of the  $\tet$-function,
\be
\epsilon (\delta, M_i) 
= \exp \{ -i \pi \delta ' B_i \delta ' - i \pi \delta ' {\rm diag} (B_i)
\}
\ee
and then of the transformations $S$, $\Sigma$ and $T$ by letting the
surface undergo a separating degeneration $\Omega _{12} \to 0$, and
using the sign assignments of genus 1 $\tet$-functions. 
 The non-trivial entries for $\epsilon ^4$ are listed in Table 2.

\begin{table}[t]
\begin{center}
\begin{tabular}{|c||c|c|c|c|c|c|} \hline 
 $\nu$ & $M_1$  & $M_2$ & $M_3$ & $S$ & $\Sigma$ & $T$
                \\ \hline \hline
              $\nu _1$  
            & $\nu _3$ 
            & $\nu _1$        
            & $\nu _3$ 
            & $\nu _1$       
            & $\nu _2$
            & $\nu _3$ 
 \\ \hline
              $\nu _2$  
            & $\nu _2$ 
            & $\nu _4$        
            & $\nu _4$ 
            & $\nu _2$       
            & $\nu _1$
            & $\nu _6$ 
 \\ \hline
              $\nu _3$  
            & $\nu _1$ 
            & $\nu _3$        
            & $\nu _1$ 
            & $\nu _5$       
            & $\nu _4$
            & $\nu _1$ 
 \\ \hline
              $\nu _4$  
            & $\nu _4$ 
            & $\nu _2$        
            & $\nu _2$ 
            & $\nu _6$       
            & $\nu _3$
            & $\nu _5$ 
 \\ \hline
              $\nu _5$  
            & $\nu _5$ 
            & $\nu _5$        
            & $\nu _6$ 
            & $\nu _3$       
            & $\nu _6$
            & $\nu _4$ 
 \\ \hline
              $\nu _6$  
            & $\nu _6$ 
            & $\nu _6$        
            & $\nu _5$ 
            & $\nu _4$       
            & $\nu _5$
            & $\nu _2$ 
 \\ \hline
\end{tabular}
\end{center}
\caption{Modular transformations of odd spin structures }
\label{table:1}
\end{table}

\begin{table}
\begin{center}
\begin{tabular}{|c||c||c|c|c|c|c|c||c|c|} \hline 
$\sum _i \nu_i$   & $\delta$ & $M_1$  & $M_2$ & $M_3$ & $S$ & $\Sigma$ &
$T$ & $\epsilon ^4 (\delta, M_1)$ & $\epsilon ^4(\delta, M_2)$
                \\ \hline \hline
$\nu_1 + \nu_4 + \nu_6$  
            & $\delta _1$  
            & $\delta _3$ 
            & $\delta _2$        
            & $\delta _1$ 
            & $\delta _1$       
            & $\delta _1$
            & $\delta _1$ & + & +
 \\ \hline
$\nu_1 + \nu_2 + \nu_6$  
            & $\delta _2$  
            & $\delta _4$ 
            & $\delta _1$        
            & $\delta _2$ 
            & $\delta _5$       
            & $\delta _3$
            & $\delta _4$ & + & +
 \\ \hline
$\nu_1 + \nu_2 + \nu_5$  
            & $\delta _3$  
            & $\delta _1$ 
            & $\delta _4$        
            & $\delta _3$ 
            & $\delta _7$       
            & $\delta _2$
            & $\delta _3$ & + & +
 \\ \hline
$\nu_1 + \nu_4 + \nu_5$  
            & $\delta _4$  
            & $\delta _2$ 
            & $\delta _3$        
            & $\delta _4$ 
            & $\delta _9$       
            & $\delta _4$
            & $\delta _2$ & + & +
 \\ \hline
$\nu_1 + \nu_2 + \nu_4$  
            & $\delta _5$  
            & $\delta _6$ 
            & $\delta _6$        
            & $\delta _6$ 
            & $\delta _2$       
            & $\delta _7$
            & $\delta _5$ & + & $-$
 \\ \hline
$\nu_1 + \nu_5 + \nu_6$  
            & $\delta _6$  
            & $\delta _5$ 
            & $\delta _6$        
            & $\delta _5$ 
            & $\delta _8$       
            & $\delta _8$
            & $\delta _6$ & + & $-$
 \\ \hline
$\nu_1 + \nu_2 + \nu_3$  
            & $\delta _7$  
            & $\delta _7$ 
            & $\delta _8$        
            & $\delta _8$ 
            & $\delta _3$       
            & $\delta _5$
            & $\delta _9$ & $-$ & +
 \\ \hline
$\nu_1 + \nu_3 + \nu_4$  
            & $\delta _8$  
            & $\delta _8$ 
            & $\delta _7$        
            & $\delta _7$ 
            & $\delta _6$       
            & $\delta _6$
            & $\delta _0$ & $-$ & +
 \\ \hline
$\nu_1 + \nu_3 + \nu_6$  
            & $\delta _9$  
            & $\delta _9$ 
            & $\delta _9$        
            & $\delta _0$ 
            & $\delta _4$       
            & $\delta _9$
            & $\delta _7$ & $-$ & $-$
 \\ \hline
$\nu_1 + \nu_3 + \nu_5$  
            & $\delta _0$  
            & $\delta _9$ 
            & $\delta _0$        
            & $\delta _9$ 
            & $\delta _0$       
            & $\delta _0$
            & $\delta _8$ & $-$ & $-$ 
 \\ \hline
\end{tabular}
\end{center}
\caption{Modular transformations of even spin structures }
\label{table:2}
\end{table}

\subsection{The Riemann relations}

At various times, we shall make use of the Riemann relations. They may be
expressed as the following quadrilinear sum over all spin structures
\be
\sum _\kappa \<\kappa | \lambda \>
\tet [\kappa ](\zeta _1 ) \tet [\kappa ](\zeta _2)
\tet [\kappa ](\zeta _3 ) \tet [\kappa ](\zeta _4)
= 4\, 
\tet [\lambda ] (\zeta _1 ') \tet [\lambda ] (\zeta _2 ')
\tet [\lambda ] (\zeta _3 ') \tet [\lambda ] (\zeta _4 ')
\ee
where the signature symbol $\< \kappa | \lambda \>$ was introduced
in (\ref{signature}). There is one Riemann relation for each spin
structure $\lambda$. We have the following relations between the
vectors $\zeta$ and $\zeta '$, expressed in terms of a matrix $\Lambda$,
which satisfies $\Lambda ^2 = I$ and $2 \Lambda$ has only integer entries,
\be
\left ( \matrix{
\zeta _1 ' \cr  \zeta _2 ' \cr \zeta _3 ' \cr \zeta _4 ' \cr} \right )
= \Lambda 
\left ( \matrix{
\zeta _1  \cr  \zeta _2  \cr \zeta _3  \cr \zeta _4  \cr} \right )
\qquad \qquad
\Lambda =
\half \left (\matrix{
 1 &  1 &  1 &  1 \cr 
 1 &  1 & -1 & -1 \cr 
 1 & -1 &  1 & -1 \cr 
 1 & -1 & -1 &  1 \cr} \right )
\ee 
In the special case where $\zeta = \zeta '=0$, only even spin
structures  $\kappa =\delta $ contribute to the sum and we have one
Riemann identity for each odd spin structure $\lambda = \nu$ on the
Riemann constants 
\be
\sum _\delta \<\nu |\delta \> \tet [\delta ]^4 (0,\Omega )  =0
\ee
For later use, we shall list these 6 equations in the basis of
characteristics introduced previously. We make use of the
standard abbreviation \cite{igusa3}
\be
\label{abbre}
(i) = \tet [\delta _i] ^4 ,
\quad\quad i=0,1,\cdots ,9
\ee 
and find 
\bea
+ \quad \nu _1 \ : \quad && 
 (1) - (2) + (3) - (4) - (5) - (6) + (7) - (8) - (9) + (0) =0
\nonumber \\
- \quad \nu _2 \ : \quad &&  
 (1) + (2) - (3) - (4) + (5) - (6) - (7) - (8) - (9) + (0) =0
\nonumber \\
- \quad \nu _3 \ : \quad && 
 (1) - (2) + (3) - (4) - (5) - (6) - (7) + (8) + (9) - (0) =0
\nonumber \\
+ \quad \nu _4 \ : \quad && 
 (1) + (2) - (3) - (4) - (5) + (6) - (7) - (8) + (9) - (0) =0
\nonumber \\
- \quad \nu _5 \ : \quad && 
 (1) - (2) - (3) + (4) - (5) + (6) + (7) - (8) - (9) - (0) =0
\nonumber \\
+ \quad \nu _6 \ : \quad && 
 (1) - (2) - (3) + (4) + (5) - (6) - (7) + (8) - (9) - (0) =0
\nonumber 
\eea
There exists one linear relation between these 6 equations, obtained by
summing all six after multiplication by the sign factor appearing to the
left of each $\nu _i$.

\subsection{Holomorphic and Meromorphic Forms for Genus 2}

As in (\ref{hyperelliptic}), we represent the genus 2 Riemann surface
$\Sigma$ by a double sheeted cover of the plane, given by the equation
\bea
s^2 = (x-u_1) (x-u_2) (x-u_3) (x-u_4) (x-u_5) (x-u_6)
\nonumber 
\eea
It is convenient to label each branch point $u_{\nu _i}$ by the unique
corresponding odd spin structure $\nu _i$, using the Abel map, and the
Riemann constants $\Delta _I$ with base point $z_0$ (see for example
Appendix A of \cite{dpII}),
\bea
(\nu _i )_I = \int ^{u_{\nu _i}} _{z_0} \omega _I - \Delta _I 
\eea
A separation of the branch points into a partition $A=\{u_1,u_2,u_3\}$ 
and $B=\{u_4,u_5,u_6\}$ represents the choice of an even spin structure
$\delta \equiv \nu _1 + \nu _2 + \nu _3$. Which spin structure this is may be 
inferred from the assignment
of an odd spin structure $\nu _i$ to each of the branch points and then
using the above relation expressing uniquely any even spin structure
$\delta$ in terms of a partition of the six odd spin structures into
two groups of 3. It is convenient to define the following functions for
the partition associated with a spin structure $\delta$,
\bea
r_A(x) &=& (x-u_1) (x-u_2) (x-u_3) \nonumber \\
r_B(x) &=& (x-u_4) (x-u_5) (x-u_6)
\eea
The hyperelliptic representation of $\Sigma$ may be recast in the form
$s^2=r_A(x)r_B(x) $.

\medskip

In terms of these quantities, we have the following representation of
holomorphic differentials. For even spin structure, there are no
holomorphic 1/2 differentials. We denote by $\omega _{\nu _i} (x)$, the
unique (up to normalization) holomorphic 1-forms with double zero at the
branch point $u_i$ associated with an odd spin structure $\nu_i$. Given
its uniqueness properties, $\omega _{\nu _i} (x)$ may be readily
identified both in its $\tet$-function and hyperelliptic forms,
\bea
\label{omeganu}
\omega _{\nu _i} (z) \equiv \omega _I (z) \p _I \tet [\nu ](0) \equiv 
\N _{\nu _i}\ (x - u_i) {dx \over s(x)}
\eea
where $\N _{\nu _i}$ is a moduli and $\nu_i$ dependent normalization
factor. Finally, we denote by $\psi _A(x)$ (respectively $\psi _B (x)$)
the unique (up to normalization) holomorphic 3/2 form all of whose three
zeros are at the three points of the partition $A$ (respectively B). We
have the following explicit formulas,
\bea
\label{psiAB}
\psi _A(x)  \equiv  r_A (x) ^\half \left ({dx \over s(x)} \right )^{3/2}
\hskip .5in
\psi _B(x)  \equiv  r_B (x) ^\half \left ({dx \over s(x)} \right )^{3/2}
\eea
The $\tet$-function form of these quantities may be deduced from their
definition (\ref{psiAB}) and from the expressions for the square roots of
the 1-forms $\omega_{\nu _i}$ in (\ref{omeganu}). While the square root of
each $\omega_{\nu _i}$ is double valued on a surface with even spin
structure $\delta$, the square roots of the products of three 
$\omega_{\nu _i}$ with all three $\nu _i$ spanning either the $A$
partition or the $B$ partition associated with $\delta$ is single valued,
and proprtional to either
$\psi _A$ or $\psi _B$. Their precise normalizations involve $\N_\nu$ and
will not be needed here.

\medskip

Finally, the only meromorphic form we shall need explicitly is the Szeg\"
o kernel $S_\delta (z,w)$ for even spin structure $\delta$. Its
$\tet$-function form is standard,
\bea
S_\delta (z,w) = {\tet [\delta ](z-w) \over \tet [\delta ](0) E(z,w)}
\eea
and its hyperelliptic form may be found in \cite{fay},
\be
S_\delta (z,w) 
= \half 
{[r_A(x) r_B (y)]^\half + [r_A(y) r_B (x)]^\half \over
x-y} 
\left ({dx \over s(x)} {dy \over s(y)} \right )^\half
\ee
Recall that the notation subsumes local coordinates $z= (x,s(x))$ and
$w=(y,s(y))$ which distinguish between the two sheets of the surface
$\Sigma$. As expected, the Szeg\"o kernel has singularities only when
$z=w$, i.e., when $x=y$ and the points $z$ and $w$ are on the same sheet.
This is because the numerator above can be rewritten as
\be
[r_A(x) r_B (y)]^\half + [r_A(y) r_B (x)]^\half
=
\biggl ( {r_B(y)\over r_B(x)} \biggr )^{1/2}
\bigg ( s(x)+s(y){r_B(x)\over r_B(y)} \bigg )
\ee
so that it vanishes when $x=y$, and $z$ and $w$ are on
different sheets.

\subsection{The split gauge condition $S_{\delta}(q_1,q_2)=0$}

An important advantage of the hyperelliptic representation is that the
{\sl split gauge condition} $S_\delta (q_1,q_2)=0$ of (\ref{split}) can be
solved for, essentially explicitly. In fact, it is equivalent to the
cancellation of the numerator factor, 
\bea
\label{splitgauge1}
 [r_A(q_1) r_B (q_2)]^\half + [r_A(q_2) r_B (q_1)]^\half =0
& \Leftrightarrow &
\psi _A (q_1) \psi _B (q_2) + \psi _A (q_2) \psi _B (q_1) =0
\nonumber \\
& \Rightarrow &
r_A(q_1) r_B (q_2) = r_A (q_2) r_B (q_1)\, .
\eea
Given $q_1$, the last line is a degree 3 polynomial in $q_2$, with 3
roots. However, one root $q_2 = q_1$ does not actually correspond to a
zero of $S_\delta$ since it is neutralized by the denominator factor.
Thus, two solutions remain, as is expected.

\subsection{Ghost insertion points $p_a$ at branch points}

The hyperelliptic representation also allows us to make a special choice
for the ghost insertion points $p_a$. We consider now a fixed even spin 
structure $\delta$ on a surface of genus~2. Let $\nu_1$, $\nu_2$ and
$\nu_3$ be the three odd spin structures such that $\delta \equiv \nu_1+
\nu_2 + \nu_3$. In view of our previous discussion on spin structures,
each odd spin structure is uniquely associated with one of the branch
points, denoted $u_{\nu _i}$, and the even spin structure $\delta$ 
corresponds to a partition $A\cup B$ of the set of branch points into two
disjoint subsets $A=\{u_{\nu _i} ; i=1,2, 3\}$ and $B=\{u_{\nu_i} ; i=4,5,
6\}$ of 3 points each. We place the points $p_i$, $i=1,2,3$, at the three
branch points $u_{\nu _i}$ in one of the subsets of the partition, say
the subset $A$. To keep the notation symmetric, we shall denote the
points $u_{\nu _i}$ in the $B$-set by $p_i$, $i=4,5,6$, so that
\be
\label{choicep}
p_i=u_{\nu_i},\quad\quad i=1,2,3,4,5,6.
\ee
We shall continue to use the subscript $a$ in $p_a$ to denote only the
three $p$'s in the A-set.

\medskip

This very special choice produces a remarkable simplification in the form
of the $b,c$ ghost Green's function $G_2(z,w) = G_2(z,w;p_a)$, whose
definition  
\bea
\nabla_{\bar z}^{(2)}G_2(z,w;p_a)
&=& 2\pi\delta(z,w)
\nonumber\\
\nabla_{\bar w}^{(-1)}G_2(z,w;p_a)   
&=&
-2\pi\delta(z,w)
+
2\pi\sum_a \phi_a^{(2)*}(z) \delta(w,p_a)
\eea
involves the points $p_a$. The holomorphic 2-forms $\phi ^{(2)*} _a (z)$
are normalized by $\phi_a^{(2)*}(p_b)=\delta_{ab}$.
Viewed as a 2-form in $z$, $G_2(z,w;p_a)$ is meromorphic with a simple
pole at $z=w$ and three zeros at $p_a$. Having chosen the points $p_a$ at
branch points of the $A$-partition of the even spin structure
$\delta$, the holomorphic 3/2 form $\psi _A (z)$ now has its 3 zeros
precisely at the points $p_a$ and has no other zeros. Thus,
$G_2(z,w;p_a)/\psi _A(z)$ is a meromorphic 1/2 form with a single pole
at $z=w$, and must therefore be proportional to the Szeg\" o kernel
$S_\delta (z,w)$. Using these arguments, we readily find,
\be
\label{g2s}
G_2(z,w;p_a) = S_\delta (z,w) {\psi _A (z) \over \psi _A (w)}
\ee
The formula may also be proven directly from the well-known (see
Appendix A of \cite{dpII}) $\tet$-function expressions of both sides.

\subsection{The bilinear $\tet$-Constants $\M_{\nu_i\nu_j}$}

As an even more important application, we derive additional
Thomae-type formulas which will play a central role in the sequel.
These are formulas for the key {\it bilinear $\tet$-constants
$\M_{\nu_i\nu_j}$} defined as follows
\be
\M _{\nu _i \nu _j} 
\equiv 
\p _1 \tet [\nu _i] (0, \Omega) \p _2 \tet [\nu _j](0, \Omega ) - 
\p _2 \tet [\nu _i] (0, \Omega) \p _1 \tet [\nu _j](0, \Omega)
\ee
We shall often abbreviate $\p_I \tet [\nu]\equiv \p_I \tet [\nu]
(0,\Omega)$. We continue to use the correspondence between odd spin
structures $\nu_i$ and branch points $p_i$, $i=1,\cdots, 6$, introduced in
the preceding subsection. We consider the holomorphic 1-form $\omega
_{\nu_i} (z)$ with a double zero at the branch point $p_i$, given
(\ref{omeganu}). Since $\omega _{\nu_i} (p_i)=0$, we  have 
\bea
\omega _I (p_i) \p _I \tet [\nu_i ] = 0 \, .
\eea

First, evaluate the following scalar ratio, (for $i\not=j,k$)
\be
{\omega _{\nu _j} (p_i) \over \omega _{\nu_k} (p_i)}
=
{\omega _I (p_i) \p _I \tet [\nu _j] \over
\omega _I (p_i) \p _I \tet [\nu _k]}
\ee
For genus 2, the $\omega _I (p_i)$ may be eliminated because the
formula is homogeneous in them and their ratio is given by the
vanishing of $ \omega _I (p_i) \p _I \tet [\nu_i ]$. 
The result can be expressed in terms of $\M_{\nu_i\nu_j}$ as
\be
\label{Mratio}
{\N _{\nu _j} (p_i - p_j) \over \N _{\nu _k}
(p_i - p_k)}
={\omega _{\nu _j} (p_i) \over \omega _{\nu_k} (p_i)}
=
{\M _{\nu_i \nu _j} \over \M _{\nu _i \nu _k}}
\ee
Taking the cross ratio of four branch points (with $i,l \not=j,k$), the
normalization factors $\N_{\nu_i}$ cancel out and we get the desired 
identity
\be
\label{thomaeM}
{p_i - p_j \over p_i - p_k}
\cdot 
{p_k - p_l \over p_j - p_l}
=
{\M _{\nu _i \nu _j} \M _{\nu _k \nu _l} \over 
 \M _{\nu _i \nu _k} \M _{\nu _j \nu _l}}
\ee
This is clearly a Thomae-type formula, relating $\tet$-constants
to rational expressions of branch points. The existence of two Thomae-type
formulas, (\ref{thomae}) and (\ref{thomaeM}) suggests that there should
be a direct relation between the bilinear $\tet$-constants 
$\M_{\nu_i\nu_j}$ and standard $\tet [\delta]$ constants. Such a relation
indeed exists and will be discussed in detail in \S 5.

\vfill\eject

\section{The Chiral Measure via  Bilinear $\tet$-Constants}
\setcounter{equation}{0}

In this section, we choose the split gauge $S_\delta(q_1,q_2)=0$ for the
points $q_1$ and $q_2$, and place all three ghost insertion points $p_a$'s
at the three branch points of the $A$-partition associated with the even
spin structure $\delta$, as in (\ref{choicep}). With these choices, all
$\X_i$, except $\X_5$ will vanish, and the product of the overall factor
$\Z$ with $\X_5$ may be recast into a simple final expression involving
only the bilinear $\tet$-constants $\M_{\nu _i \nu _j}$ and the
standard $\tet[\delta]^4$ constants.

\subsection{The Chiral Partition Function Overall Factor}

We begin by evaluating the overall factor $\Z$ of (\ref{firstzee}), which
is the matter and superghost chiral partition function. Using the
expressions for the ghost and superghost correlators established in
\cite{vv87}, we obtain 
\bea
\label{zee}
{\cal Z} &=&
{ \< \prod _a b(p_a) \prod _\alpha \delta (\beta (q_\alpha)) \>
\over 
\det \omega _I \omega _J (p_a) } 
 \nonumber \\
&=& 
{\tet [\delta ](0)^5 \tet (D_b) \prod _{a<b} E(p_a, p_b) \prod _a \sigma
(p_a)^3 \over Z^{15}  \tet [\delta ](D_\beta) E(q_1,q_2) \prod _\alpha
\sigma (q_\alpha )^2  \det \omega _I \omega _J (p_a)}
\eea
where the chiral scalar partition function $Z$ is defined by
\be
Z^3 = {\tet (\sum _I z_I -w_0 -\Delta) \prod _{I<J} E(z_I,z_J) \prod _I
\sigma (z_I) \over \sigma (w_0) \prod _I E(z_I,w_0) \ \det \omega _I(z_J)}
\ee
for any triplet of distinct points $z_1,z_2,w_0$. Recall that 
\be
D_b = p_1+p_2+p_3-3\Delta
\qquad \qquad
D_\beta = q_1 + q_2 - 2 \Delta\, .
\ee
Notice that ${\cal Z}$ is independent of the three points $p_a$, but does
depend upon the points $q_1,q_2$. Upon evaluating this quantity, we may 
thus choose $p_a$ any way we want, while the $q$'s have to be the same as
the ones used throughout. In particular, just in this calculation, it is
convenient not to choose the $p_a$ as we did in (\ref{choicep}).

\medskip

In the expression (\ref{zee}), and in one factor $Z^3$ of its
denominator (leaving another factor $Z^{12}$ untouched), we set $z_1=p_1$,
$z_2=p_2$. Also, we place $p_3$ at a branch point, labeled by odd spin
structure $\nu_3$, $p_3 = \Delta + \nu_3$, but leave the points
$p_1$ and $p_2$ arbitrary. As a result, this $Z^3$ factor in the
denominator will cancel the factor $\tet (D_b)$ in the numerator, up to
an exponential factors arising from a shift by an integral period in the
$\tet$-function, 
\bea
\tet (p_1+p_2+p_3-3\Delta ) & = &
C \cdot \tet (p_1+p_2- p_3 -\Delta) 
\nonumber \\
C &=&  - \exp \{- 4 \pi i \nu _3 ' (p_1 +p_2 - 2 \Delta) \}
\eea
so that we have 
\bea
{\cal Z} =
C \cdot {\tet [\delta ](0)^5  E(p_1,p_3)^2 E(p_2,p_3)^2 \sigma (p_1)^2
\sigma (p_2)^2 \sigma (p_3)^4 \det\, \omega _I (p_1,p_2) 
 \over 
Z^{12}  \tet [\delta ](D_\beta) E(q_1,q_2) \sigma (q_1)^2 \sigma (q_2)^2
\det\, \omega _I \omega _J (p_a)}
\eea
Next, we let $p_2 \to p_1$, keeping the point $p_1$ still arbitrary, so
that the determinants simplify as follows
\be
{\det\, \omega _I (p_1,p_2) 
 \over 
\det\, \omega _I \omega _J (p_a)}
\to - \biggl (\omega _1 (p_3) \omega _2 (p_1) - \omega _1(p_1) \omega
_2(p_3) \biggr )^{-2}
\ee
and we are left with
\bea
{\cal Z} =
-C \cdot {\tet [\delta ](0)^5  E(p_1,p_3)^4 \sigma (p_1)^4 \sigma (p_3)^4 
 \over 
Z^{12}  \tet [\delta ](D_\beta) E(q_1,q_2) \sigma (q_1)^2 \sigma (q_2)^2
\bigl [ \omega _1 (p_3) \omega _2 (p_1) - \omega _1(p_1) \omega _2(p_3)
\bigr ] ^2}
\eea
Next, we evaluate the remaining factors of $Z$ in the following
way. We let $z_1 = p_1$ and $z_2 = p_3$, so that
\be
Z^3 = {\tet (p_1 + p_3 -w_0 -\Delta)  E(p_1,p_3) \sigma (p_1) \sigma (p_3)
\over \sigma (w_0) E(p_1,w_0) E(p_3,w_0) \ \det \omega _I(p_1,p_3)}
\ee
and now set $p_1 = \Delta +\nu _1$ as we already had $p_3=\Delta +\nu _3$.
We obtain two different but equivalent formulas by letting 
$w_0 \to p_3$ or $w_0 \to p_1$  respectively
\bea
Z^3 &=& - C_1 {\omega _{\nu _1} (p_3) \sigma (p_1) \over 
\omega _1 (p_3) \omega _2 (p_1) - \omega _1(p_1) \omega _2(p_3)}
\nonumber \\
Z^3 &=& + C_3 {\omega _{\nu _3} (p_1) \sigma (p_3) \over 
\omega _1 (p_3) \omega _2 (p_1) - \omega _1(p_1) \omega _2(p_3)}
\eea
with 
\be
\label{seei}
C_i = \exp \{ -i \pi \nu _i ' \Omega \nu _i ' - 2 \pi i \nu _i' \nu _i '' \}
\ee
In evaluating ${\cal Z}$, we use the first formula for one $Z^6$ factor,
while the second formula for the second $Z^6$ factor and we get
\bea
{\cal Z} =
-{C \over C_1 ^2 C_3 ^2} \cdot {\tet [\delta ](0)^5  E(p_1,p_3)^4
\sigma (p_1)^2 \sigma (p_3)^2 (\omega _1 (p_3) \omega _2 (p_1) - \omega
_1(p_1)
\omega _2(p_3))^2
 \over 
\tet [\delta ](D_\beta) E(q_1,q_2) \sigma (q_1)^2 \sigma (q_2)^2
\omega _{\nu _1}(p_3)^2 \omega _{\nu _3}(p_1)^2}
\eea
The ratio
\be
{\omega _1 (p_3) \omega _2 (p_1) - \omega _1(p_1) \omega _2(p_3)
 \over 
\omega _{\nu _1}(p_3) \omega _{\nu _3}(p_1)}
\ee
may be easily computed because it involves only the ratios $\omega
_2/\omega _1(p_1)$ and $\omega _2/\omega _1(p_3)$, and they are known
from the fact that $\omega _{\nu _1}(p_1) = \omega _{\nu _3} (p_3)=0$,
\be
{\omega _2 (p_1) \over \omega _1(p_1)}
 = - {\p _1 \tet [\nu _1] \over \p_2 \tet [\nu_1]}
\qquad \qquad
{\omega _2 (p_3) \over \omega _1(p_3)} 
= - {\p _1 \tet [\nu _3] \over \p_2 \tet [\nu_3]}
\ee
and we find
\be
\label{theMformula}
{\omega _1 (p_3) \omega _2 (p_1) - \omega _1(p_1) \omega _2(p_3)
 \over 
\omega _{\nu _1}(p_3) \omega _{\nu _3}(p_1)}
= {1 \over \M _{\nu _1 \nu _3}}
\ee
Putting all together, we have
\bea
{\cal Z} =
-{C \over C_1 ^2 C_3 ^2} \cdot {\tet [\delta ](0)^5  E(p_1,p_3)^4
\sigma (p_1)^2 \sigma (p_3)^2 
 \over 
\tet [\delta ](D_\beta) E(q_1,q_2) \sigma (q_1)^2 \sigma (q_2)^2}
{1 \over \M _{\nu _1 \nu _3} ^2}
\eea
We stress that at this moment, no choice for the $q_{\alpha}$
has been made as yet.
Instead of trying to simplify this factor further, we shall rather start
combining it with the $\X$ terms.

\subsection{Vanishing of $\X_1\! +\! \X_6$ in Split Gauge and $p_a$ at
Branch Points}

Combining the split gauge condition $S_\delta (q_1,q_2)=0$ on the
points $q_1$ and $q_2$ while placing the ghost insertion points $p_a$ at
the branch points belonging to the $A$-partition of the even spin
structure $\delta$ produces drastic simplifications. From the expression
for $\X_1+\X_6$ in (\ref{finamppq}) and (\ref{peeque}), it is clear that
the basic ingredients are $G_2(q_1,q_2)$, $\p _{q_1} G_2 (q_1,q_2)$
and  $\p _{q_2} G_2 (q_2,q_1)$, since the first line already vanishes in
split gauge. Actually, with this choice of points $p_a$, the ghost Green's
function $G_2(z,w;p_a)$ is given by (\ref{g2s}) and thus we readily have
in split gauge that 
\be
G_2(q_1,q_2)=0
\ee 
As a result, $\p _{q_1} G_2 (q_1,q_2)$
transforms as a tensor, which we now evaluate,
\bea
\label{derg2}
\p _{q_1} G_2 (q_1,q_2) = \p_{q_1} S_\delta (q_1,q_2) {\psi _A (q_1)
\over \psi _A (q_2)}
\eea
To evaluate $\p _{q_1} S_\delta (q_1,q_2)$, subject to the condition
$S_\delta (q_1,q_2)=0$, is usually not so easy, but the calculation is
feasible in the hyperelliptic representation.

\medskip

We begin by proving the following useful formula, valid when
$S_\delta (q_1,q_2)=0$,
\be
\label{derszego}
\p _{q_1} S_\delta (q_1,q_2) = \p \omega _1 ^* (q_2) \p \psi ^* _2(q_1)
\ee
where $\omega_\alpha ^*$ and $\psi_\alpha^*$ are the holomorphic 1-forms
and 3/2-forms with normalizations at the points $q_\beta$, given by
$\omega_\alpha^* (q_\beta) = \psi_\alpha^* (q_\beta)=\delta _{\alpha
\beta}$. To show (\ref{derszego}), one begins by calculating $\omega ^*$
and
$\psi ^*$ in the hyperelliptic representation
\bea
\omega ^* _1 (q) &=& {q - q_2 \over q_1-q_2} \cdot {dq \ s(q_1) \over 
dq_1 \ s(q)}
\nonumber \\
\psi ^* _2(q) &=& 
{r_A(q) ^\half r_B(q_2) ^\half + r_A(q_2) ^\half r_B(q) ^\half
 \over 2\, s(q_2)} \left ( {dq \ s(q_2) \over
dq_2 \ s(q)} \right ) ^{3/2}
\eea
Using the condition $S_{\delta}(q_1,q_2)=0$, the derivatives needed in the
formula may be  readily evaluated as well
\bea
\label{derdifferentials}
\p \omega ^* _1 (q_2) &=& {1 \over q_1-q_2 } {(dq_2) ^2 \cdot
s(q_1) \over dq_1 \cdot s(q_2) }
\nonumber \\
\p \psi ^* _2 (q_1) &=& \half \p \ln  s(q_1) 
{r_A (q_1) ^\half r_B (q_2) ^\half \over  s(q_2) }
\biggl ( {dq_1 \cdot s(q_2) \over dq_2 \cdot s(q_1) }  \biggr )^{3/2}
dq_1\nonumber \\
\p _{q_1} S_ \delta (q_1,q_2) &=&
{1 \over 2} \p \ln s(q_1)  {r_A (q_1) ^\half r_B (q_2) ^\half \over
q_1 -q_2}
  (dq_1) ^{3/2} (dq_2) ^\half
\eea
The above formula (\ref{derszego}) now follows immediately.

\medskip

Combining (\ref{derg2}) with (\ref{derszego}), and substituting the
result into $\X_1 + \X_6$ of (\ref{peeque}), the term $\X_1+\X_6$ is
found to  reduce to
\be
\X _1 + \X_6
= {\zeta ^1 \zeta ^2 \over 16 \pi ^2}
\p \psi ^* _1 (q_2)  \p \psi ^* _2 (q_1) \biggl [
\p \omega ^* _2 (q_1) {\psi _A (q_2) \over \psi _A (q_1)} -
\p \omega ^* _1 (q_2) {\psi _A (q_1) \over \psi _A (q_2)} \biggr ]
\ee
It remains to evaluate the terms within the brace. We shall do so
by using the hyperelliptic representation. 
We begin by considering (for any points $q_1$ and $q_2$, not necessarily
in split gauge) the expressions 
\bea
\p \omega ^* _1(q_2) & = & {1 \over q_1 - q_2} {(dq_2)^2 \cdot s(q_1)
\over dq_1 \cdot s(q_2) }
\nonumber \\
\p \omega ^* _2(q_1) & = & {1 \over q_2 - q_1} {(dq_1)^2 \cdot s(q_2)
\over dq_2 \cdot s(q_1) }
\eea
Upon taking their ratio, we obtain
\be
{\p \omega ^* _1 (q_2) \over \p \omega ^* _2 (q_1)}
=
-{(dq_2) ^3 /s(q_2)^2 \over (dq_1)^3 /s(q_1)^2}\, .
\ee
Clearly, this object is the ratio of a holomorphic 3-form evaluated at
$q_2$ and evaluated at $q_1$. As a function of $q_2$, this 3-form has
its 6 simple zeros at all 6 branch points, and by inspection, it may be
rewritten as
\be
{\p \omega ^* _1 (q_2) \over \p \omega ^* _2 (q_1)}
=
- {\psi _A (q_2) \psi _B (q_2) \over \psi _A (q_1) \psi _B (q_1)}\, .
\ee
Note that this formula was derived {\it without} assuming any relation
between  $q_1$ and $q_2$.

\medskip

Next, recall the relation  $\psi _A (q_1) \psi _B (q_2) + \psi
_A(q_2) \psi _B (q_1) =0$, which was derived in (\ref{splitgauge1}) and
holds whenever $S_\delta (q_1,q_2)=0$. We use this relation to  rewrite
the above ratio as
\be
{\p \omega ^* _1 (q_2) \over \p \omega ^* _2 (q_1)}
=
 {\psi _A (q_2)^2 \over \psi _A (q_1)^2 }\, .
\ee
It is trivially seen that this relation makes $\X_1+\X_6=0$ when $S_\delta 
(q_1,q_2)=0$.

\subsection{First Evaluation of $\X_5$}

The overall factor ${\cal Z}$ did not depend upon the points $p_a$ and so
they may be taken to be anything. In particular, in combining ${\cal Z}$
with $\X_5$, we shall let the points $p_1$ and $p_3$ in the expression
for $\Z$ depend upon the term labeled by $a$ in the following way: $p_1
\to p_b$ and $p_3 \to p_a$ where $b\not= a$. Starting from 
\be
\X _5 = {\zeta ^1 \zeta ^2 \over 16 \pi ^2} \sum _a \varpi  _a(q_1,q_2)
{\tet [\delta ] (q_1+q_2-2 p_a) E(q_1,q_2) \over \tet [\delta ](0)
E(q_1,p_a)^2 E(q_2,p_a)^2}
\ee
using the fact that 
\bea
C' & \equiv &
{\tet [\delta ](q_1+q_2-2 p_a) \over \tet [\delta ](q_1+q_2-2 \Delta)} 
\\
& = & 
\exp \{-4 \pi i \nu _a ' \Omega \nu _a' + 4 \pi i \nu _a '(\delta ''+q_1
+q_2 - 2 \Delta ) - 4 \pi i \delta ' \nu _a '' \}
\nonumber 
\eea
and the following two relations (established by using the explicit
representations of a ratio of $\sigma$-functions, see
Appendix A of \cite{dpII} for more details),
\bea
{E(q_\alpha ,p_a)^2 \sigma (q_\alpha)^2 \over E(p_b,p_a)^2 \sigma (p_b)^2}
=
{\omega _{\nu _a} (q_\alpha) \over \omega _{\nu _a}(p_b)}
\exp \{ 4 \pi i \nu _a ' (q_\alpha -p_b) \}
\qquad \alpha =1,2
\eea
we find
\be
{\cal Z} \X_5 = {\zeta ^1 \zeta ^2 \over 16 \pi ^2} \tet [\delta ](0)^4
\sum _a {
\varpi _a  (q_1,q_2) \over \omega _{\nu _a}(q_1) \omega _{\nu _a}(q_2)}
\cdot {\sigma (p_a)^2 \omega _{\nu _a}(p_b)^2 \over \sigma (p_b)^2}
\cdot {{\cal C} \over \M _{\nu _a \nu _b}^2}
\ee
where ${\cal C}$ collects all the exponential factors and is given by
\bea
{\cal C} &=& - {C C' \over C_a ^2 C_b ^2} \exp \{- 4 \pi i (q_1+q_2 - 2
p_b) \}
\nonumber \\
&=& \exp \{ 2 \pi i \nu _b ' \Omega \nu '_b - 2 \pi i \nu _a' \Omega
\nu'_a + 4 \pi i (\nu _a ' \delta '' - \nu _a '' \delta ') \}
\eea
with $C_a$ given in (\ref{seei}). Furthermore, we have 
\bea
{\sigma (p_a) \over \sigma (p_b)}
=
{\tet (r+s-p_a-\Delta) E(p_b,r) E(p_b,s) \over \tet (r+s-p_b-\Delta)
E(p_a,r) E(p_a,s)}
=
- {\omega _{\nu _b} (p_a) \over \omega _{\nu _a} (p_b)}
{C_b \over C_a}
\eea
so that
\be
{\cal C} {\sigma (p_a)^2 \over \sigma (p_b)^2}
= 
\<\nu _a |\delta\> {\omega _{\nu _b} (p_a)^2 \over \omega _{\nu _a} 
(p_b)^2}
\ee
where we recall from (\ref{signature}) that
the signature $\< \nu |\delta\> $ of two spin structures is defined
by $ \< \nu |\delta \> \equiv \exp 4\pi i \{\nu ' \delta '' - \nu ''
\delta '\}$. Putting all together, we obtain a first formula for
${\cal Z}\X_5$ 
\be
\label{first}
{\cal Z} \X_5 = {\zeta ^1 \zeta ^2 \over 16 \pi ^2} \tet [\delta ](0)^4
\sum _a {
\varpi _a  (q_1,q_2) \over \omega _{\nu _a}(q_1) \omega _{\nu _a}(q_2)}
\cdot { \omega _{\nu _b}(p_a)^2 \over  \M _{\nu _a \nu _b}^2} 
\<\nu _a |\delta \>
\ee
As a result of (\ref{Mratio}), this expression is in fact independent of
the choice of the point $p_b$ and its associated spin structure $\nu_b$.

\subsection{Good Formulas for $\varpi_a$}

The expression for ${\cal Z}\X_5$ will now be rendered more explicit 
by deriving better formulas for $\varpi_a $.
We set $p_a= \Delta + \nu _a$, $a=1,2,3$ where $\delta \equiv \nu _1 + \nu
_2 +\nu_3$ and use the holomorphic 1-forms with double zeros at $p_a$,
$\omega _{\nu _a}(z)$, which obey one linear dependence relation
\be
A_1 \omega _{\nu _1} (z) + A_2 \omega _{\nu _2} (z)  + A_3 \omega _{\nu _3}
(z)  =0
\ee
with $A_a\not=0$. The coefficients $A_a$ will be identified
later.
We begin with the denominator of the determinants for
$\varpi  _a(q_1,q_2)$,
\be
D = \left |\matrix{
\omega _{\nu _1} \omega _{\nu _1} (p_1) &
\omega _{\nu _2} \omega _{\nu _2} (p_1) &
\omega _{\nu _1} \omega _{\nu _2} (p_1) \cr
\omega _{\nu _1} \omega _{\nu _1} (p_2) &
\omega _{\nu _2} \omega _{\nu _2} (p_2) &
\omega _{\nu _1} \omega _{\nu _2} (p_2) \cr
\omega _{\nu _1} \omega _{\nu _1} (p_3) &
\omega _{\nu _2} \omega _{\nu _2} (p_3) &
\omega _{\nu _1} \omega _{\nu _2} (p_3) \cr} \right |
\ee
To the first column, add the $A_2 /A_1$ times the third column; to the
second column, add $A_1/A_2$ times the third column and use the linear
dependence relation. Now using the fact that $\omega _{\nu _a} (p_a)=0$, we
readily find a nicely factorized expression
\be
D = - {A_3 ^2 \over A_1 A_2} \ 
\omega _{\nu _1} (p_2) \omega _{\nu _1} (p_3) 
\omega _{\nu _2} (p_1) \omega _{\nu _2} (p_3) 
\omega _{\nu _3} (p_1) \omega _{\nu _3} (p_2) 
\ee
Next, we look at the numerator for $\varpi  _1(q_1,q_2)$, which is
$$
D \varpi  _1(q_1,q_2) = \left |\matrix{
\omega _{\nu _1} (q_1) \omega _{\nu _1} (q_2) &
\omega _{\nu _2} (q_1) \omega _{\nu _2} (q_2) &
\half \omega _{\nu _1} (q_1) \omega _{\nu _2} (q_2) +
\half \omega _{\nu _1} (q_2) \omega _{\nu _2} (q_1)
 \cr
\omega _{\nu _1} \omega _{\nu _1} (p_2) &
\omega _{\nu _2} \omega _{\nu _2} (p_2) &
\omega _{\nu _1} \omega _{\nu _2} (p_2) \cr
\omega _{\nu _1} \omega _{\nu _1} (p_3) &
\omega _{\nu _2} \omega _{\nu _2} (p_3) &
\omega _{\nu _1} \omega _{\nu _2} (p_3) \cr} \right |
$$
We perform exactly the same linear combinations as we did on the
denominator and after some simplifications find the following results
\bea
\varpi  _1 (q_1, q_2) &=&  {
\omega _{\nu _2} (q_1) \omega _{\nu _3} (q_2) +
\omega _{\nu _2} (q_2) \omega _{\nu _3} (q_1)
\over 
2 \omega _{\nu _2} (p_1) \omega _{\nu _3} (p_1)}
\nonumber \\
\varpi  _2 (q_1, q_2) &=&  {
\omega _{\nu _3} (q_1) \omega _{\nu _1} (q_2) +
\omega _{\nu _3} (q_2) \omega _{\nu _1} (q_1)
\over 
2 \omega _{\nu _3} (p_2) \omega _{\nu _2} (p_2)}
\nonumber \\
\varpi  _3 (q_1, q_2) &=&  {
\omega _{\nu _1} (q_1) \omega _{\nu _2} (q_2) +
\omega _{\nu _1} (q_2) \omega _{\nu _2} (q_1)
\over 
2 \omega _{\nu _1} (p_3) \omega _{\nu _2} (p_3)}
\eea
For later use, we calculate the coefficients $A_i$ up to an overall
factor by letting $z=p_a$ for all three $a=1,2,3$, and we find
\be
\M _{\nu _2 \nu _3} \omega _{\nu _1} (z) +
\M _{\nu _3 \nu _1} \omega _{\nu _2} (z) +
\M _{\nu _1 \nu _2} \omega _{\nu _3} (z) =0\, .
\ee

\subsection{Elimination of $\varpi _a $}

We return to the expression (\ref{first}) obtained for ${\cal Z}\X_5$.
Recall that this expression was independent of the choice of $p_b$. For
definiteness, we concentrate on the $p$-dependence of, for example, the
term $a=1$. It is given by
\be
{\omega _{\nu _b}(p_1)^2 \over \omega _{\nu _2}(p_1) \omega _{\nu _3}(p_1) }
{1 \over \M _{\nu _1 \nu _b}^2} 
\< \nu _1 |\delta \>
\ee
Without loss of generality, let us take $b=2$, so that we need the ratio
\be
\label{Mratio1}
{\omega _{\nu _2} (p_1) \over \omega _{\nu _3} (p_1)}
={\M _{\nu _1 \nu_2} \over \M _{\nu _1 \nu _3}}
\ee
of (\ref{Mratio}), and we get
\be
{\omega _{\nu _b}(p_1)^2 \over \omega _{\nu _2}(p_1) \omega _{\nu _3}(p_1) }
{\<\nu _1|\delta \> \over \M _{\nu _1 \nu _b}^2} 
=
{\<\nu _1 |\delta\>
\over \M _{\nu _1 \nu _2} \M _{\nu _1 \nu _3}}
\ee
The expression for the spin structure $\delta = \nu _1 + \nu _2 + \nu _3$
may be used to simplify the exponential further 
\be
\<\nu _1 |\delta \>
=
\exp 4\pi i \{\nu _1 ' \nu _2 '' + \nu _1 \nu _3 '' - \nu _1 '' \nu _2 ' -
\nu _1 '' \nu _3 ' \} 
=\< \nu _1 |\nu _2\> \<\nu _1 |\nu _3\>
\ee
and we notice that this product of signatures factorizes along with the
$\M$ factors, so that
\be
{\omega _{\nu _b}(p_1)^2 \over \omega _{\nu _2}(p_1) \omega _{\nu _3}(p_1) }
{\<\nu _1|\delta\> \over \M _{\nu _1 \nu _b}^2} 
=
{\<\nu _1 |\nu _2\>
\over \M _{\nu _1 \nu _2} }
\cdot
{\<\nu _1 |\nu _3\>
\over \M _{\nu _1 \nu _3} }
\ee
Finally, putting all together, we obtain a second formula
for ${\cal Z}\X_5$, 
\bea
\label{exfive}
{\cal Z} \X_5 &=& {\zeta ^1 \zeta ^2 \over 32 \pi ^2} \tet [\delta ](0)^4
\biggl [
+{\omega _{\nu _2}(q_1) \omega _{\nu _3} (q_2) 
   \over \omega _{\nu _1}(q_1) \omega _{\nu _1}(q_2)} 
\cdot 
{\<\nu _1 |\nu _2\>\<\nu _1 |\nu _3\> 
   \over \M _{\nu _1 \nu _2}\M _{\nu _1 \nu _3} } 
\nonumber \\
&& \qquad \qquad \qquad 
+ {\omega _{\nu _3}(q_1) \omega _{\nu _1} (q_2) 
   \over \omega _{\nu _2}(q_1) \omega _{\nu _2}(q_2)} 
\cdot 
{\<\nu _2 |\nu _3\>\<\nu _2 |\nu _1\> 
   \over \M _{\nu _2 \nu _3}\M _{\nu _2 \nu _1} } 
\nonumber \\
&& \qquad \qquad \qquad
+ {\omega _{\nu _1}(q_1) \omega _{\nu _2} (q_2) 
   \over \omega _{\nu _3}(q_1) \omega _{\nu _3}(q_2)} 
\cdot 
{\<\nu _3 |\nu _1\>\<\nu _3 |\nu _2\> 
   \over \M _{\nu _3 \nu _1}\M _{\nu _3 \nu _2} } 
\biggr ] + (q_1 \leftrightarrow q_2)
\eea
Although it may appear longer, this expression is much more
explicit than the earlier one in (\ref{first}).

\subsection{The $\M$ Product Formula}

The split gauge condition $S_{\delta}(q_1,q_2)=0$ on the points
$q_{\alpha}$ still allows for a one-parameter family of choices. The full
amplitude ${\cal Z}\sum_{i=1}^6\X_i$ was shown to be independent of the
points $q_{\alpha}$. Therefore, it remains independent of any residual
choice of points $q_1,q_2$ satisfying $S_{\delta}(q_1,q_2)=0$.  Actually,
it is instructive to perform a further consistency check and verify this
residual independence on $q_1$ and $q_2$ in split gauge of the term 
${\cal Z}\X_5$, to which the full amplitude reduces  in this gauge. In
the process of doing so, we shall come across new $\tet$-function
identities which will play a key role in the sequel.

\subsubsection{The hyperelliptic representation}

Since the variables $q_1$ and $q_2$ are related by $S_\delta (q_1,q_2)=0$, 
we  must consider their simultaneous variation. It is convenient to do
this in the  hyperelliptic representation, where the dependence on all
points is expressed  via rational functions. The final expression for
${\cal Z} \X_5$ in  (\ref{exfive}) is not manifestly  a combination of
cross ratios and so does not  manifestly admit an expression in terms of
rational functions on the  hyperelliptic curve. This is easily remedied
by using the formula (\ref{Mratio1}).
The first term in (\ref{exfive}) may be recast in the form
\bea
&& 
{\omega _{\nu _2}(q_1) \omega _{\nu _3} (q_2) 
   \over \omega _{\nu _1}(q_1) \omega _{\nu _1}(q_2)} 
\cdot 
{\<\nu _1 |\nu _2\>\<\nu _1 |\nu _3\> 
   \over \M _{\nu _1 \nu _2}\M _{\nu _1 \nu _3} } 
   \nonumber \\
&& \qquad =
{\omega _{\nu _2}(q_1) \omega _{\nu _1} (p_3) 
   \over \omega _{\nu _1}(q_1) \omega _{\nu _2}(p_3)} 
\cdot 
{\omega _{\nu _3}(q_2) \omega _{\nu _1} (p_2) 
   \over \omega _{\nu _1}(q_2) \omega _{\nu _3}(p_2)} 
\cdot 
{(-) \<\nu _1 |\nu _2\>\<\nu _1 |\nu _3\> \M_{\nu _2 \nu _3}^2 
   \over \M _{\nu _1 \nu _2} ^2 \M _{\nu _1 \nu _3}^2 }
   \nonumber \\
&& \qquad =
{q_1-p_2 \over q_1-p_1} \cdot {p_3 - p_1 \over p_3 - p_2} \cdot
{q_2 - p_3 \over q_2 - p_1} \cdot {p_2 - p_1 \over p_2 - p_3}
    {(-) \<\nu _1 |\nu _2\>\<\nu _1 |\nu _3\> \M_{\nu _2 \nu _3}^2 
   \over \M _{\nu _1 \nu _2} ^2 \M _{\nu _1 \nu _3}^2 }
\eea      
Carrying out similar manipulations on the other two terms, and 
symmetrizing in  $q_1$ and $q_2$, we may recast the result in the
following form
\be
{\cal Z} \X_5 = {\zeta ^1 \zeta ^2 \over 32 \pi ^2}
{\tet [\delta ](0) ^4 \over \M _{\nu _1 \nu _2}^2 \M _{\nu _2 \nu _3}^2 \M 
_{\nu _3 \nu _1}^2} \cdot \R
\ee
where $\R$ is given by
\bea
\label{scriptR}
\R 
&=& + \<\nu_2 |\nu _3\> \M _{\nu _2 \nu _3}^4 
{p_3 - p_1 \over p_3 - p_2} \cdot {p_2 - p_1 \over p_2 - p_3} 
 \biggl \{ {q_1-p_2 \over q_1-p_1} \cdot {q_2 - p_3 \over q_2 - p_1} +
           {q_2-p_2 \over q_2-p_1} \cdot {q_1 - p_3 \over q_1 - p_1} 
\biggr \}
           \nonumber \\
           && \nonumber \\
&& + \<\nu_3 |\nu _1\> \M _{\nu _3 \nu _1}^4 
{p_1 - p_2 \over p_1 - p_3} \cdot {p_3 - p_2 \over p_3 - p_1} 
 \biggl \{ {q_1-p_3 \over q_1-p_2} \cdot {q_2 - p_1 \over q_2 - p_2} +
           {q_2-p_3 \over q_2-p_2} \cdot {q_1 - p_1 \over q_1 - p_2} 
\biggr \}
           \nonumber \\
           && \nonumber \\
&& + \<\nu_1 |\nu _2\> \M _{\nu _1 \nu _2}^4 
{p_2 - p_3 \over p_2 - p_1} \cdot {p_1 - p_3 \over p_1 - p_2} 
 \biggl \{ {q_1-p_1 \over q_1-p_3} \cdot {q_2 - p_2 \over q_2 - p_3} +
           {q_2-p_1 \over q_2-p_3} \cdot {q_1 - p_2 \over q_1 - p_3} 
\biggr \}
           \nonumber \\
 \eea 
This formula has all the symmetry properties manifest.

\subsubsection{Absence of singularities}

The combination $\R$, and thus the full amplitude $\Z \X_5$ in this gauge,
appears to exhibit poles when $q_\alpha \to p_1,p_2,p_3$. These poles
must of course cancel since the expression should be independent of the
$q_\alpha$, in split gauge. We begin by checking that such poles indeed
cancel. We may let $q_1 \to p_1$, without loss of generality. Then, it
follows from the split gauge condition that $q_2$ must tend either to
$p_2$ or to $p_3$, namely to one of the other two points in the same
$A$-partition of the spin structure $\delta$ as $p_1$ belongs to.
Choosing $q_2 \to p_2$, it is convenient to parametrize this  joint
solution for $q_1$ and $q_2$ in the following  way
\bea
q_1 &=& p_1 + F_1 t^2 \nonumber \\
q_2 &=& p_2 + F_2 t^2 
\eea
where $F_1$ and $F_2$ depend on $p_i$ and on $t^2$. For small $t^2$, as we 
are  using in the vicinity of the branch points, $F_{1,2}$ effectively
reduce to their value for $t=0$, which are given by solving the
equation (\ref{splitgauge1}) recalled below,
\be
r_A(p_1 + F_1 t^2)^\half r_B(p_2+F_2 t^2) ^\half
+
r_B(p_1 + F_1 t^2)^\half r_A(p_2+F_2 t^2) ^\half =0\, .
\ee
 Using the fact that $r_A(p_1)=r_A(p_2)=0$, we get
\be
F_1 r_A '(p_1) r_B(p_2) = F_2 r_A '(p_2) r_B (p_1)
\ee
The pole contributions in (\ref{scriptR}) arise only from the first term
in the brace of the  first  line and the first term in the brace of the
second line, and is given by
\be
\R \bigg |_{\rm pole} =
 - \<\nu_2 |\nu _3\> \M _{\nu _2 \nu _3}^4 
 {(p_3 - p_1)(p_2-p_1) \over (p_3-p_2) \ F_1 \ t^2}
- \<\nu_3 |\nu _1\> \M _{\nu _3 \nu _1}^4 
 {(p_1 - p_2)(p_3-p_2) \over (p_3-p_1) \ F_2 \ t^2}\, .
\ee
Proving the vanishing of this quantity is equivalent to showing that the
ratio  of the two terms on the right hand side equals $-1$. The ratio
equals
\bea
&&
- {\<\nu_2 |\nu _3\> \M _{\nu _2 \nu _3}^4 
\over \<\nu_3 |\nu _1\> \M _{\nu _3 \nu _1}^4 } \cdot
{(p_3-p_1) ^2 \over (p_3-p_2)^2} \cdot {F_2 \over F_1}
 \\
&& \hskip 1in  =
+\<\nu_1 |\nu _2\> {\M _{\nu _2 \nu _3}^4 \over \M _{\nu _3 \nu _1}^4 } 
\cdot {(p_3-p_1) ^2 \over (p_3-p_2)^2} \cdot {r_A'(p_1) r_B(p_2) \over
r_A'(p_2)  r_B(p_1)}
\nonumber \\
&& \hskip 1in =
- \<\nu_1 |\nu _2\> {\M _{\nu _2 \nu _3}^4 \over \M _{\nu _3 \nu _1}^4 } 
\cdot {(p_3-p_1) ^3 \over (p_3-p_2)^3} \cdot \prod _{i=4,5,6} {p_2-p_i
\over p_1-p_i}
\nonumber
\eea
This expression may be written purely in terms of the $\M$-functions by 
using  again the cross-ratio formula
\be
{p_3 - p_1 \over p_3 - p_2} \cdot {p_i - p_1 \over p_i - p_2}
=
{\M _{\nu _3 \nu _1} \M _{\nu _2 \nu _i} \over
\M _{\nu _3 \nu _2} \M _{\nu _1 \nu _i}}
\ee 
and we find that the condition for the cancellation of the pole is
\be
\label{Mproductformula}
\<\nu_1 |\nu _2\> \prod _{i=3,4,5,6} 
{\M _{\nu _2 \nu _i} \over \M _{\nu _1 \nu _i} } =1
\ee
This identity is written completely in terms of $\tet$-functions. 
We shall refer to it as the {\sl $\M$ product formula}. In view of our
discussion, the $\M$ product formula follows at once from the independence
of ${\cal Z}\X_5$ from any choice of $q_1,q_2$ satisfying
$S_{\delta}(q_1,q_2)=0$. We shall presently give a direct proof of it,
-- up to overall signs --  using the classical Thomae formula. 
Later, we shall obtain an explicit formula for
the bilinear $\tet$-constant ${\cal M}_{\nu_i\nu_j}$ itself,
which will imply the full product identity including signs.

\subsubsection{Proof of the $\M$ product formula -- up to overall signs}

We translate the desired identity -- up to overall signs -- into the
hyperelliptic representation, using the normalizations of holomorphic
Abelian differentials  with double zeros appearing in the Thomae formula.
The starting point is formula (\ref{Mratio}), adapted here to the spin
structures $\nu_1$, $\nu_2$ and $\nu_3$,
\bea
{\omega _{\nu _2} (p _i) \over \omega _{\nu _1} (p_i)}
= {\M _{\nu _2 \nu _i} \over \M _{\nu _1 \nu _i}}
={{\cal N}_{\nu _2} (p_i - p_2) \over 
{\cal N}_{\nu _1} (p_i - p_1)}
\nonumber 
\eea
with the normalization factors related by (see also \cite{lm})
\be
{\N _{\nu _1} ^4 \over \N _{\nu _2}^4 } =
\pm \prod _{i \not= 1,2} {p_1 - p_i \over p_2 - p_i}
\ee
Taking the product over $i=3,4,5,6$, we have
\be
\prod _{i=3,4,5,6} {\M _{\nu _2 \nu _i} \over \M _{\nu _1 \nu _i}}
= \pm {{\cal N}_{\nu _2} ^4 \over {\cal N}_{\nu _1}^4 } \ 
\prod _{i=3,4,5,6} {p_2 - p_i \over p_1 - p_i} =\pm 1\, .
\ee
This proves the identity up to a $\pm$ sign.

\subsection{The Chiral Measure in terms of ${\cal M}_{\nu_1\nu_2}$}

The expression (\ref{exfive}) which we have obtained so far for
${\cal Z}\X_5$ mixes both $\tet$-functions (as encoded in
${\cal M}_{\nu_i\nu_j}$) and the hyperelliptic representation
(as encoded in the branch points).
We proceed now to simplify it further.
The strategy is to eliminate directly any reference to the
points $q_{\alpha}$, using the relation (\ref{Mproductformula}). 
Multiplying numerator and denominator by $\M _{\nu _1 \nu _3}^4$, and
using the cross ratio formula in terms of $\M$'s, we deduce that
\bea
{\M _{\nu _3 \nu _1} ^4 \over \M _{\nu _2 \nu _3} ^4}
&=&
\<\nu_1 |\nu_2\> {(p_3 - p_1)^3 \over (p_3 - p_2)^3} 
\prod _{i=4,5,6} {(p_2 - p_i)^3 \over (p_1 - p_i)^3}
\nonumber \\
&& \nonumber \\ 
{\M _{\nu _1 \nu _2} ^4 \over \M _{\nu _2 \nu _3} ^4}
&=&
\<\nu _1 |\nu_3\> {(p_1 - p_2)^3 \over (p_3 - p_2)^3} 
\prod _{i=4,5,6} {(p_3 - p_i)^3 \over (p_1 - p_i)^3}\, .
\eea
The relation (\ref{splitgauge1}) between $q_1$ and $q_2$ may be
re-expressed in terms of cross ratios of points $p_i$, $i=1,\cdots,6$ and
$q_1$ and $q_2$,
\be
{(q_1-p_1)(q_1-p_2)(q_1-p_3)(q_2-p_4)(q_2-p_5)(q_2-p_6)
\over
(q_2-p_1)(q_2-p_2)(q_2-p_3)(q_1-p_4)(q_1-p_5)(q_1-p_6)}=1
\ee
so that the expression for $\R$, the ratios of various $\M^4$ and
the relation between $q_1$ and $q_2$ may all be expressed in terms of
cross ratios only and are thus M\" obius invariant.

\medskip

Using M\" obius invariance, we may set $p_1=\infty$, $p_2=0$ and $p_3=1$.
The relation between $q_1$ and $q_2$ simplifies and may be
expressed in terms of the symmetric polynomials
\bea
B_1 &=& p_4 + p_5 +p_6 \nonumber \\
B_2 &=& p_4 p_5 + p_5 p_6 + p_6 p_4 \nonumber \\
B_3 &=& p_4 p_5 p_6
\eea
via the following equation
\be
q_1 ^2 q_2 ^2 - q_1 q_2 (q_1 + q_2) + (B_1-B_2) q_1 q_2 + B_3 (q_1 +q_2)
-B_3 =0\, .
\ee
Similarly, the ratios of $\M$'s may be expressed in these terms
\be
{\M _{\nu _3 \nu _1} ^4 \over \M _{\nu _2 \nu _3} ^4} = \<\nu _1 |\nu
_2\> B_3
\qquad \qquad
{\M _{\nu _1 \nu _2} ^4 \over \M _{\nu _2 \nu _3} ^4} = \< \nu _1 |\nu
_3\> (1-B_1 +B_2 - B_3)
\, .
\ee
With the help of these expressions for the $B_i$, the relation between
$q_1$ and $q_2$ may be expressed solely in terms of the $\M$'s,
\be
\label{mrelation}
\<\nu _2 |\nu _3\> \M _{\nu _2 \nu _3} ^4 +{\<\nu _1 |\nu _2\> \over
(1-q_1) (1-q_2) } \M _{\nu _1 \nu _2} ^4 +{\<\nu _3 |\nu _1\> \over q_1
q_2} \M _{\nu _3 \nu _1} ^4 =0\, .
\ee
 Finally, the expression for $\R$ also simplifies considerably and we have
\bea
\R
&=& + \<\nu_2 |\nu _3\> \M _{\nu _2 \nu _3}^4 
 \biggl \{ q_1 (1-q_2) + q_2 (1-q_1) \biggr \}
           \nonumber \\
&& + \<\nu_3 |\nu _1\> \M _{\nu _3 \nu _1}^4 
 \biggl \{ {q_1-1 \over q_1 q_2} + {q_1-1 \over q_1 q_2} \biggr \}
           \nonumber \\
&& + \<\nu_1 |\nu _2\> \M _{\nu _1 \nu _2}^4 
 \biggl \{-{q_2\over (1-q_1) (1-q_2)} - {q_1 \over (1-q_1)(1-q_2)} \biggr \}
 \eea 
To see how the cancellation of $q$-dependence comes about, multiply 
(\ref{mrelation}) by a factor $(q_1 +q_2 -2 -2q_1 q_2)$ and regroup terms
as in the expression for ${\cal R}$. We find
\bea
0
&=& + \<\nu_2 |\nu _3\> \M _{\nu _2 \nu _3}^4 
 \biggl \{-2 + q_1 (1-q_2) + q_2 (1-q_1) \biggr \}
           \nonumber \\
&& +  \<\nu_3 |\nu _1\> \M _{\nu _3 \nu _1}^4 
 \biggl \{-2 + {q_1-1 \over q_1 q_2} + {q_1-1 \over q_1 q_2} \biggr \}
           \nonumber \\
&& +  \<\nu_1 |\nu _2\> \M _{\nu _1 \nu _2}^4 
 \biggl \{ -2 -{q_2\over (1-q_1) (1-q_2)} - {q_1 \over (1-q_1)(1-q_2)}
\biggr \}
 \eea 
This implies
\be
\R = 2 \<\nu_1 |\nu _2\> \M _{\nu _1 \nu _2}^4 + 2 \<\nu_2 |\nu _3\>
\M _{\nu _2 \nu _3}^4 + 2 \<\nu_3 |\nu _1\>  \M _{\nu _3 \nu _1}^4
\ee
so that our final answer is 
\be
\label{btc}
{\cal Z} \X_5 
= {\zeta ^1 \zeta ^2 \over 16 \pi ^2} \cdot  \tet [\delta ](0) ^4 \cdot 
{
\<\nu_1 |\nu _2\> \M _{\nu _1 \nu _2}^4 +
\<\nu_2 |\nu _3\> \M _{\nu _2 \nu _3}^4 + 
\<\nu_3 |\nu _1\> \M _{\nu _3 \nu _1}^4 
\over 
\M _{\nu _1 \nu _2}^2 \M _{\nu _2 \nu _3}^2
\M _{\nu _3 \nu _1}^2} 
\ee
This expression is entirely in terms of $\tet$-functions alone.
Given that $\M_{\nu _i \nu _j}$ and $\tet [\delta](0)^4$ both have
modular weight 2, the whole combination has modular weight -2, as
expected.

\vfill\eject

\section{The Chiral Measure in terms of $\tet$-Constants}
\setcounter{equation}{0}

In order to identify the measure factor derived above as a modular 
form  we begin by searching for formulas expressing $\M_{\nu_i\nu_j}$ in 
terms of $\tet$-constants. Modular forms of any weight are then related
to $\tet$-constants by standard expressions. 

\medskip

\subsection{A Key $\tet$-Constants and Bilinear $\tet$-Constants
Identity}

It turns out that there is a remarkable identity giving
{\sl the bilinear $\tet$-constant} ${\cal M}_{\nu_i\nu_j}$
in terms of $\tet$-constants. Here we shall show how
the existence of such an identity can be conjectured from
all the other identities which we have obtained so far.
Our derivation here is only up to signs, but the sign can
be determined later by degeneration arguments.

\medskip

We continue to use the standard notation of the previous section where
the branch points $p_i$ are associated with odd characteristics $\nu _i$.
Consider the Thomae-type formula (\ref{thomaeM}) for the bilinear
$\tet$-constant $\M_{\nu_i\nu_j}$ 
\be
{\M_{\nu_1 \nu_2} \M_{\nu_3 \nu_4} \over \M_{\nu_1 \nu_3}\M_{\nu_2 \nu_4}}
=
{(p_1 - p_2 ) (p_3-p_4) \over (p_1-p_3) (p_2-p_4)}
\ee
and compare this with the standard form for the Thomae formula 
given earlier in (\ref{thomae}) in terms of $\tet$ constants $\tet
[\delta] \equiv \tet [\delta ](0,\Omega)$,
$$
\tet[\delta] ^8 = c \prod _{a<b}  (p_{i_a} -p_{i_b})^2 
(p_{j_a} - p_{j_b})^2
$$
where $c$ depends on moduli but is independent of the spin structure.
The even spin structure  $\delta$ may be identified with the partition of
the branch points into two  groups 
\be
\delta \sim \{p_{i_1}, p_{i_2}, p_{i_3} \} \cup \{p_{j_1}, p_{j_2},
p_{j_3} \} \, .
\ee
Taking the ratio of two such expressions for different $\delta$ allows one to 
cancel out the spin structure independent factor $c$.
For example,
\bea
\delta_1 & \sim & \{p_1, p_2, p_3 \} \cup \{p_4, p_5, p_6 \}
\nonumber \\
\delta_2 & \sim & \{p_1, p_2, p_4 \} \cup \{p_3, p_5, p_6 \}
\eea
yields a fully determined expression of cross-ratios, which may be recast
in terms of $\M_{\nu_i\nu_j}$'s,
\bea
{\tet [\nu _1 + \nu _2 + \nu _3]^8 \over \tet [\nu_1 + \nu _2 + \nu _4]
^8} 
&=&
{ (p_1-p_3)^2 (p_2-p_3)^2 (p_4-p_5)^2 (p_4-p_6)^2 \over
(p_1-p_4)^2 (p_2-p_4)^2 (p_3-p_5)^2 (p_3-p_6)^2}
\nonumber \\
&=&
{\M_{\nu_1 \nu_3} ^2 \M_{\nu_4 \nu_6}^2 \M_{\nu_2 \nu_3} ^2  
\M_{\nu_4 \nu_5}^2
\over
\M_{\nu_1 \nu_4} ^2 \M_{\nu_3 \nu_6}^2 \M_{\nu_2 \nu_4} ^2 
\M_{\nu_3 \nu_5}^2}
\eea
Interchanging the roles of $p_2$ and $p_5$ and taking the ratio of
both, we find
\be
{\tet [\nu _1 + \nu _2 + \nu _3]^8 \tet [\nu_1 + \nu _4 + \nu_5]^8
\over 
\tet [\nu_1 + \nu _2 + \nu _4]^8 \tet [\nu_1 + \nu _3 + \nu _5] ^8}
=
{\M_{\nu_2 \nu_3} ^4 \M_{\nu_4 \nu_5} ^4 
\over \M_{\nu_2 \nu_4} ^4 \M_{\nu_3 \nu_5}^4 }
\ee
Making use of the  {\sl $\M_{\nu_i\nu_j}$ product formula},
(\ref{Mproductformula}) we get
\bea
{\M _{\nu _1 \nu _2}^{16} \over \M _{\nu _2 \nu _3}^{16}}
&=&
\prod _{i=4,5,6} 
{\M_{\nu_1 \nu_2} ^4  \M_{\nu_3 \nu_i} ^4  
\over \M_{\nu_2 \nu_3} ^4  \M_{\nu_1 \nu_i} ^4}
\nonumber \\
&=&
{\tet [\nu _1 + \nu _2 + \nu _4]^{16} \tet [\nu _1 + \nu _2 + \nu _5]^{16}
\tet [\nu _1 + \nu _2 + \nu _6]^{16}
\over 
\tet [\nu _2 + \nu _3 + \nu _4]^{16} \tet [\nu _2 + \nu _3 + \nu _5]^{16}
\tet [\nu _2 + \nu _3 + \nu _6]^{16}}
\eea
To arrange this result in a more symmetrical form, we
multiply numerator and denominator by 
the  same factor $\tet [\nu _1 + \nu _2 + \nu _3]^{16}$, so that
\be
{\M _{\nu _1 \nu _2}^{16} \over \M _{\nu _2 \nu _3}^{16} }
= {
\tet [\nu _1 + \nu _2 + \nu _3]^{16}
\tet [\nu _1 + \nu _2 + \nu _4]^{16} 
\tet [\nu _1 + \nu _2 + \nu _5]^{16}
\tet [\nu _1 + \nu _2 + \nu _6]^{16}
\over 
\tet [\nu _2 + \nu _3 + \nu _1]^{16}
\tet [\nu _2 + \nu _3 + \nu _4]^{16} 
\tet [\nu _2 + \nu _3 + \nu _5]^{16}
\tet [\nu _2 + \nu _3 + \nu _6]^{16} }\, .
\ee
This equation is solved by the following simple guess for
the identity between bilinear $\tet$-constants and standard
$\tet$-constants that we were looking for, (up to a 16-th root of unity)
\be
\label{identity}
\M _{\nu _1 \nu _2} \sim \prod _{i =3,4,5,6} 
\tet [\nu _1 + \nu _2 + \nu _i]
\ee
Since both sides transform covariantly under the modular group
and have modular weight 2, the factor of proportionality must be a
modular function. In subsection \S 5.2, we shall prove that the factor is
constant and equal to $\pm \pi ^2$. To prove this, we shall use
the fact that the rhs vanishes only at the boundary of moduli space, and
that the behavior of both sides at the boundary of moduli space
coincides, so that the factor of proportionality must be a modular
function holomorphic on (compactified) moduli space and thus constant.
Before we give this lengthy proof in section \S 5, we show in the next
subsection that this formula gives us the final result for the chiral
superstring measure on moduli.

\subsection{The Chiral Measure in terms of $\tet$-Constants}

We now make use of the crucial formula (\ref{identity}) to recast the
expression for the chiral superstring measure on moduli, derived in
(\ref{btc}), solely in terms of standard $\tet$-constants. First, we note
that each
$\M^2_{\nu_i\nu_j}$ has an overall common factor of $\tet [\delta]^2=
\tet [\nu_1 + \nu_2 +\nu_3]^2$,
\be
\label{factortetdelta}
\M _{\nu _1 \nu _2} ^2 =  \pi ^4 \tet [\delta ]^2 
\prod _{k=3,4,5}
\tet [ \nu_1 + \nu _2 +\nu _k ] ^2 \, .
\ee
It follows that 
\bea
&& {
\<\nu_1 |\nu _2\> \M _{\nu _1 \nu _2}^4 +
\<\nu_2 |\nu _3\> \M _{\nu _2 \nu _3}^4 + 
\<\nu_3 |\nu _1\> \M _{\nu _3 \nu _1}^4 
\over 
\M _{\nu _1 \nu _2}^2 \M _{\nu _2 \nu _3}^2
\M _{\nu _3 \nu _1}^2} 
\\ &&
\hskip 2in  =
{1 \over \pi ^4 \Psi _{10}} \sum _{1\leq i < j \leq 3}
\<\nu_i |\nu _j\> \prod _{k=4,5,6} \tet [\nu_i + \nu_j +\nu_k]^4
\nonumber 
\eea
Here, $\Psi _{10}$ is the weight 10 modular form for genus 2, 
\begin{eqnarray*}
\Psi _{10} (\Omega ) \equiv \prod _\delta \tet [\delta ]^2 (0,\Omega)
\end{eqnarray*}
and the product is taken over all even spin structures $\delta$.

\subsection{Mirror property}

The quantity 
\be
\label{q}
\sum _{1\leq i < j \leq 3}
\<\nu_i |\nu _j\> \prod _{k=4,5,6} \tet [\nu_i + \nu_j +\nu_k]^4
\ee
appears to depend not only upon the spin structure $\delta = \{ \nu _1,
\nu _2, \nu _3\} \cup \{ \nu _4, \nu _5, \nu _6\}$ but on the actual
triplet chosen to represent $\delta$. However, the odd spin structures
$\nu _1, \nu _2, \nu _3$ resulted from the choice of the points
$p_a$, and the
complete independence of the points $p_a$ established
in \cite{dpIII} guarantees that
(\ref{q}) is independent of the actual triplet chosen. Thus, (\ref{q})
must be invariant under the mirror transformation $\{ \nu _1, \nu _2, \nu
_3\}  \leftrightarrow \{ \nu _4, \nu _5, \nu _6\}$.  

\medskip

Here, we shall provide a direct proof of this mirror property. Given the
transitive action of modular transformations on spin structures, it
suffices to show its validity for any single spin structure. Consider the
case $\nu _1, \nu _2, \nu _3$ with the basis of (\ref{evenodd}), i.e.
$\delta _7$, so then the following quantity should vanish
\be
Q = 
\sum _{1\leq i < j \leq 3}
\<\nu_i |\nu _j\> \prod _{k=4,5,6} \tet [\nu_i + \nu_j +\nu_k]^4
-
\sum _{4\leq i < j \leq 6}
\<\nu_i |\nu _j\> \prod _{k=1,2,3} \tet [\nu_i + \nu_j +\nu_k]^4\, .
\ee
Using again the standard abbreviation $(i) = \tet [\delta _i] ^4$, $Q$
takes the form,
\be
Q = (8)(9)(0) - (1)(4)(6) + (2)(3)(5) - (2)(4)(9) - (5)(6)(8) +
(1)(3)(0)
\ee
To show that this quantity vanishes using the Riemann relations, given at
the end of subsection \S 2.4, we produce the following linear
superpositions of the Riemann relations first~: $\nu_1 \pm \nu_2$, $\nu_3
\pm \nu_4$ and $\nu_5 \pm \nu_6$. Retaining the equation $\nu _1 - \nu _2$
as defining $(7) = \tet [\delta _7]^4$ and eliminating
$(7)$ from all other equations and omitting the linearly dependent
equation, we are left with 4 Riemann relations that do not involve $(7)$.
We may cast these in the form where they express $(1),
\ (2), \ (5), \ (0)$ in terms of $(3), \ (4), \ (6), \ (8), \ (9)$,
\bea
(1)  & = & +(3) + (8) + (9) \nonumber \\
(2)  & = & +(3) - (6) + (8) \nonumber \\
(5)  & = & +(3) - (4) + (9) \nonumber \\
(0) & = & -(3) + (4) + (6) 
\eea
The expression for $Q$ now becomes
\bea
Q &=&
 (8)(9)\bigl [ -(3) + (4) + (6) \bigr ]
- \bigl [ (3) + (8) + (9) \bigr ] (4)(6)
\\ && 
+ \bigl [ (3) - (6) + (8) \bigr ] (3) \bigl [ +(3) - (4) + (9) \bigr ] 
- \bigl [ (3) - (6) + (8) \bigr ] (4)(9) 
\nonumber \\ &&
- \bigl [ (3) - (4) + (9) \bigr ] (6)(8) 
+ \bigl [ (3) + (8) + (9) \bigr ] (3) \bigl [ -(3) + (4) +(6) \bigr ]
\nonumber
\eea
This sum may be seen to cancel term by term and $Q=0$. 

\medskip

\subsection{The final formula}

In summary, we have then established the main formulas
of the present paper,
\bea
\A [\delta]
& = & \Z + {\zeta_1\zeta_2\over 16\pi^6} \ {\tet^4[\delta] (0,\Omega) \Xi
_6 [\delta ](\Omega) \over\Psi_{10} (\Omega)} 
\\ &&
\nonumber \\
\Xi _6 [\delta] (\Omega) & \equiv & \sum _{1\leq i < j \leq 3}
\<\nu_i |\nu _j\> \prod _{k=4,5,6} \tet [\nu_i + \nu_j +\nu_k]^4
(0,\Omega)
\eea
These formulas imply the ones for the chiral measure $d\mu[\delta](\Omega)$ 
stated in the Introduction. 
Due to the mirror property established in the preceding subsection,
$\Xi_6[\delta](\Omega)$ depends only upon $\delta$. {\sl It has modular
weight 6, but it is not a modular form}. Its precise modular
transformation properties will be derived in section \S 6. The above
formula thus gives the contribution of each even spin structure $\delta =
\nu_1+\nu_2+\nu_3$ to the chiral superstring measure entirely in terms of
$\tet$-constants. Determining the modular covariant assignments of the
GSO projection phase factors will be carried out in section \S 6.

\vfill\eject

\section{Degenerations -- Proofs of $\tet$-constant Relations}
\setcounter{equation}{0}

In this section, we shall derive three key results. The first result, in
subsection \S 5.1 below, is to obtain degeneration limits of
$\tet$-constants (and their derivatives) in the separating and
non-separating cases. The second result, in subsection \S 5.2 below, is to
prove {\sl the bilinear $\tet$-constant relation} between the bilinear
$\tet$ constants $\M_{\nu _i \nu _j}$ and ordinary $\tet$ constants,
\be
\label{tetbilinear}
\M _{\nu _i \nu _j} 
\equiv
\p_1 \tet [\nu_i] \p_2 \tet [\nu_j] - \p_1 \tet [\nu_j]
\p_2 \tet [\nu_i] 
= 
\pm \pi ^2 \prod _{k\not= i,j} \tet [\nu_i + \nu_j +\nu _k] \, .
\ee
The $\pm$ sign on the rhs is not intrinsic, since it will change
under interchange of $i \leftrightarrow j$ as well as under the addition
of certain complete periods to the $\nu _k$. However, in a given
basis of $\nu_k$, the sign is uniquely fixed and will be determined
below.  Finally, the third result, in subsection \S 5.3 below, is to prove
{\sl the $\M$ product Formula},
\bea
\label{Mproduct}
\< \nu _i | \nu _j \> \prod _{k \not= i,j} { \M_{\nu _j \nu _k} \over \M
_{\nu _i \nu _k}} =1\, .
\eea
Relation (\ref{tetbilinear}) is proven using the fact that both sides
transform covariantly under the modular group, so that the factor of
proportionality between the left and right sides, \` a priori, must be a
modular function, independent of the spin structures. Since genus 2
$\tet$-constants (for even spin structures) can vanish only at the
boundary of moduli space, the ratio of the lhs by the rhs must be a
modular function that is holomorphic on the inside of moduli space. Using
the degeneration limits of subsection \S 5.1, we shall show that this
modular function tends to 1 at both separating and non-separating boundary
components of moduli space, and must therefore be a constant. The
asymptotics then uniquely determines this constant. Relation
(\ref{Mproduct}) is an immediate consequence of relation
(\ref{tetbilinear}), with the precise sign assignments.

\subsection{Degenerations}

We begin by fixing a canonical homology basis of 1-cycles $A_I$, $B_I$,
such that $\# (A_I, B_J) = \delta _{IJ}$ for $I,J=1,2$. In this homology
basis, we parametrize the period  matrix $\Omega$  by the complex
variables $\tau_1$, $\tau_2$ and $\tau$,
\be
\Omega = \left ( \matrix{\tau _1 & \tau \cr \tau & \tau _2\cr} \right )
\ee
and decompose the (even or odd) spin structures $\kappa$ accordingly,
\be
\label{degenspin}
\kappa = \left [ \matrix{\kappa_1 \cr \kappa_2 \cr} \right ]
\hskip 1in
\left \{ 
\matrix{
\kappa_1   
= (\kappa_1 ' | \kappa_1 '') & \ \kappa_1 '\, , \ \kappa_1 '' =0, 1/2
\cr
\kappa_2 
= (\kappa_2 ' | \kappa_2 '') & \ \kappa_2 '\, , \ \kappa_2 '' =0, 1/2 \cr}
\right .
\ee
We may picture the genus 2 surface as two tori joined by a cylinder.
Torus 1 has homology basis $A_1,B_1$, spin structure $\kappa_1$ and
modulus $\tau_1$, while torus 2 has homology basis $A_2,B_2$, spin
structure $\kappa_2$ and modulus $\tau_2$. The parameter $\tau$
characterizes the (trivial from the genus 2 homological point of view)
cylinder that joins both tori.

\medskip

We shall need to work out degenerations of $\tet$-functions and it will
turn out to be convenient to have available the $\tet$-function in the
above variables, and $z=(z_1,z_2)$
\bea
\label{thetafunction}
\tet [\kappa] (z, \Omega)
& = &  \sum _{m,n \in {\bf Z}} \exp \biggl \{ 
i \pi (m+ \kappa_1 ')^2 \tau _1 
+ 2 \pi i (m+\kappa_1 ') (z_1 + \kappa_1 '')
\nonumber \\
&& \hskip .7in
+ i \pi (n+ \kappa_2 ')^2 \tau _2 
+ 2 \pi i (n+\kappa_2 ') (z_2 + \kappa_2 '')
\nonumber \\
&& \hskip .7in
+ 2 \pi i \tau (m+ \kappa_1 ') (n+ \kappa_2 ') \biggr \}
\eea
There are two possible degenerations (up to the action of the modular
group)~: {\sl the separating degeneration}, $\tau \to 0$,
keeping $\tau_1$ and $\tau_2$ fixed; and {\sl the non-separating
degeneration},  $\tau _2 \to + i \infty$, keeping $\tau _1$ and $\tau$
fixed. We discuss each case in turn below. 

\medskip

The resulting asymptotics may be expressed in terms of genus~1
$\tet$-functions, which will be denoted by $\tet _1$. We begin by
recalling the definition, mainly in order to fix our conventions 
\bea
\tet _1 [\kappa _i ] (z _i ,\tau _i)  \equiv  \sum _{m\in {\bf Z}} 
\exp \biggl \{  i \pi (m+ \kappa _i ')^2 \tau  _i
+ 2 \pi i (m+\kappa _i ') (z_i + \kappa _i '') \biggr \}
\qquad i=1,2
\eea
The $\tet_1$ function satisfies the heat equation
\bea
\label{heateq}
\p _{z_i} ^2 \tet _1 [\kappa _i] (z_i ,\tau _i) 
= 4 \pi i \p _{\tau_i}  \tet _1 [\kappa _i ](z_i , \tau_i)
\qquad i=1,2
\eea
a product relation,
\bea
\label{tetproduct}
\tet '_1 [\nu _0](0,\tau_1) &=& - \pi \ \tet _1 [\mu _1] \tet _1 [\mu _3]
\tet _1 [\mu _5] (\tau _1) = -2 \pi \eta (\tau_1)^3
\nonumber \\
\tet '_1 [\nu _0](0,\tau_2) &=& - \pi \ \tet _1 [\mu _2] \tet _1 [\mu _4]
\tet _1 [\mu _5] (\tau _2) = -2 \pi \eta (\tau_2)^3
\eea
and a doubling equation,
\bea
\label{doubling}
\tet _1[\nu_0] (2z,\tau) \eta (\tau)^3 =
\tet _1 [\nu_0] (z,\tau) \prod _{i=1,3,5} \tet _1 [\mu_i] (z,\tau)
\eea 
where $\kappa_i$, $i=1,2$, stand for any genus 1 spin structures while the
spin structures $\mu_1$, $\mu_3$, $\mu_5$ are the three distict even spin
structures (same for $\mu_2$, $\mu_4$ and $\mu_6$).

\subsubsection{Separating degeneration, $\tau \to 0$ keeping $\tau
_{1,2}$ fixed}

The $\tet$-function has the following Taylor expansion around $\tau=0$,
given in terms of genus 1 $\tet$-functions, which we denote $\tet _1$ for
clarity,
\be
\tet [\kappa ](z,\Omega)
=
\sum _{p=0}^\infty {1 \over p!} \left ( {\tau \over 2 \pi i} \right )^p
\p _{z_1} ^p \tet _1 [\kappa _1](z_1,\tau_1) \p _{z_2}^p \tet
_1[\kappa_2](z_2, \tau_2)
\, .
\ee
As general functions of $z_1$ and $z_2$, all $p$ contribute. 
Specializing to $\tet$-constants, or derivatives thereof (such as in
$\M_{\nu _i \nu _j}$), only even or odd $p$ contribute depending on the
parity of the genus 1 spin structures $\kappa_1$ and $\kappa_2$. We denote
the unique genus 1 odd spin structure by $\nu _0 \equiv [\half \half]$,
and any even  spin structure by $\mu$. Furthermore, we use the heat
equation of (\ref{heateq}) satisfied by $\tet_1$ to express double
$z_i$-derivatives in terms of single  $\tau_i$ derivatives, 
We then have the following cases : the $\tet$-constants for even spin 
structure  are given by
\bea
\tet \left [ \matrix{\mu _1 \cr \mu _2 \cr} \right ](0,\Omega )
&=&
\sum _{p=0} ^\infty {(2 \tau) ^{2p} \over (2p)!} 
\p _{\tau _1}^p \tet _1[\mu _1] (0,\tau _1) 
\p _{\tau _2}^p \tet _1[\mu _2] (0,\tau _2) 
\nonumber \\
\tet \left [ \matrix{\nu_0 \cr \nu_0 \cr} \right ](0,\Omega )
&=&
{1 \over 4 \pi i} \sum _{p=0} ^\infty {(2 \tau )^{2p+1} \over (2p+1)!} 
\p _{\tau _1}^p \tet _1 '[\nu_0] (0,\tau _1) 
\p _{\tau _2}^p \tet _1 '[\nu_0] (0,\tau _2) 
\eea
while the first derivatives $\tet$-constants for odd spin structure are 
given  by
\bea
\p _1 \tet \left [ \matrix{\nu _0 \cr \mu  \cr} \right ](0,\Omega )
&=&
\sum _{p=0} ^\infty {(2 \tau )^{2p} \over (2p)!} 
\p _{\tau _1}^p \tet _1 '[\nu _0] (0,\tau _1) 
\p _{\tau _2}^p \tet _1[\mu ] (0,\tau _2) 
\nonumber \\
\p _2 \tet \left [ \matrix{\nu_0 \cr \mu \cr} \right ](0,\Omega )
&=&
\sum _{p=0} ^\infty {(2 \tau )^{2p+1} \over (2p+1)!} 
\p _{\tau _1}^p \tet _1 '[\nu_0 ] (0,\tau _1) 
\p _{\tau _2}^{p+1} \tet _1 [\mu] (0,\tau _2) 
\nonumber \\
\p _1 \tet \left [ \matrix{\mu \cr \nu_0 \cr} \right ](0,\Omega )
&=&
\sum _{p=0} ^\infty {(2 \tau )^{2p+1} \over (2p+1)!} 
\p _{\tau _1}^{p+1} \tet _1 [\mu] (0,\tau _1) 
\p _{\tau _2}^p \tet _1 '[\nu_0] (0,\tau _2) 
\nonumber \\
\p _2 \tet \left [ \matrix{\mu \cr \nu_0  \cr} \right ](0,\Omega )
&=&
\sum _{p=0} ^\infty {(2 \tau )^{2p} \over (2p)!} 
\p _{\tau _1}^p \tet _1 [\mu] (0,\tau _1) 
\p _{\tau _2}^p \tet _1 '[\nu _0 ] (0,\tau _2) 
\eea
The leading asymptotics of the $\tet$-constants may now be read off,
\bea
\label{tetasym}
\tet \left [ \matrix{\mu _1 \cr \mu _2 \cr} \right ] (0,\Omega) & = & 
\tet _1 [\mu _1](0,\tau _1) \tet _1 [\mu_2] (0,\tau_2) +\O(\tau ^2)
\nonumber \\
\tet \left [ \matrix{\nu _0 \cr \nu _0  \cr} \right ] (0,\Omega) & = &
- 2 \pi i \tau \eta (\tau _1)^3 \eta (\tau _2)^3 +\O(\tau ^3)
\eea 
whence follows the leading asymptotics of the  modular form
$\Psi _{10} (\Omega)=\prod _\delta \tet [\delta ]^2$,
\bea
\label{psitenasym}
\Psi _{10} (\Omega ) & = & 
- (2 \pi \tau )^2 \cdot 2^{12} \cdot \eta (\tau _1)^{24} \eta (\tau
_2)^{24}  +\O(\tau ^4)
\eea
and the limits of the objects $\Xi_6 [\delta ](\Omega )$,
\bea
\label{Xiasym}
\Xi _6 \left [ \matrix{\mu _1 \cr \mu _2 \cr} \right ] (\Omega)
& = &
- 2^8 \cdot \<\mu_1 |\nu_0\> \<\mu _2 |\nu_0\> \eta (\tau _1)^{12} \eta
(\tau _2)^{12}  +\O(\tau ^2)
\nonumber \\
\Xi _6 \left [ \matrix{\nu _0 \cr \nu _0  \cr} \right ] (\Omega)
& = & 
-3 \cdot 2^8 \cdot \eta (\tau _1)^{12} \eta (\tau _2)^{12}  +\O(\tau ^2)
\eea

\subsubsection{Non-separating degeneration, $\tau_2 \to +i\infty$ keeping
$\tau _1$ and $\tau$ fixed}

The genus 2 $\tet$-function (\ref{thetafunction}) manifestly has a Taylor
expansion in powers of $q^\14$, where we introduce the standard
abbreviation
\be
q\equiv \exp \{ i \pi \tau _2 \}
\ee
We shall be interested only in the leading asymptotics as $q\to 0$. 
The behavior of the $\tet$-function in this limit depends upon the value
of $\kappa_2 '$, following the same notation as in (\ref{degenspin}). This
component of the spin structure is singled out because it is associated
with the torus with modulus $\tau_2$ that is degenerating to a wire in
the limit $q \to 0$. 

\medskip

We begin by examining the behavior of the ordinary $\tet$-constants and
even characteristics. If $\kappa_2 '=0$, only the $n=0$ terms in
(\ref{thetafunction}) will contribute to the leading asymptotics, while
for $\kappa_2 '=1/2$, both $n=0,-1$ will contribute. The leading
asymptotics is now easily retained from (\ref{thetafunction}),
\bea
\label{nonseptetlimits}
\tet \left [ \matrix{\mu \cr \mu_j   \cr} \right ] (0,\Omega)
& = &
\tet _1 [\mu ] (0,\tau _1 ) + \O (q)
\hskip 1.34in 
{\rm all \ even} \ \mu\, , \ {\rm and} \ j=2,4
\nonumber \\
\tet \left [ \matrix{\mu \cr \mu_6   \cr} \right ] (0,\Omega)
& = &
2 q^\14 \tet _1 [\mu] \biggl ( {\tau \over 2} , \tau_1 \biggr ) + \O
(q^{{5/4}})
\hskip .8in 
{\rm all \ even} \ \mu
\nonumber \\
\tet \left [ \matrix{\nu_0 \cr \nu_0   \cr} \right ] (0,\Omega)
& = &
2 i q^\14 \tet _1 [\nu_0 ] \biggl ( {\tau \over 2} , \tau_1 \biggr ) + \O
(q^{{5/4}})
\eea
Using this asymptotic behavior, we readily calculate that of the
modular form $\Psi _{10}(\Omega)$,
\bea
\Psi _{10} (\Omega ) =
- 2^{12}\,  q^2 \, \eta (\tau_1) ^{18} \, \tet _1 [\nu _0]^2
(\tau,\tau_1) 
+\O(q^3) 
\eea
Notice that $q$ enters here via $q^2$ only, as expected from 
the modular covariance of $\Psi_{10}$.

\medskip

The asymptotics of the objects $\Xi_6 [\delta]$ necessary to determine
the asymptotics of the measure up to and including order $\O(q)$ are given
by (we abbreviate $\tet _1 [\mu ] \equiv \tet _1 [\mu ] (0,\tau_1)$),
\bea
\Xi_6 \left [ \matrix{\mu_1 \cr \mu_2  \cr} \right ]  = 
- \Xi_6 \left [ \matrix{\mu_1 \cr \mu_4  \cr} \right ]
& = &
16 q \tet _1 [\mu_3]^4 \tet _1 [\mu_5]^4  \biggl (
-\tet _1 [\mu_1] ({\tau \over 2}, \tau_1) ^4 - \tet _1 [\nu_0] ({\tau
\over 2}, \tau_1) ^4 \biggr ) + \O(q^2)
\nonumber \\
\Xi_6 \left [ \matrix{\mu_3 \cr \mu_2  \cr} \right ]  = 
- \Xi_6 \left [ \matrix{\mu_3 \cr \mu_4  \cr} \right ]
& = &
16 q \tet _1 [\mu_1]^4 \tet _1 [\mu_5]^4  \biggl (
\tet _1 [\mu_3] ({\tau \over 2}, \tau_1) ^4 - \tet _1 [\nu_0] ({\tau
\over 2}, \tau_1) ^4 \biggr ) + \O(q^2)
\nonumber \\
\Xi_6 \left [ \matrix{\mu_5 \cr \mu_2  \cr} \right ]  = 
- \Xi_6 \left [ \matrix{\mu_5 \cr \mu_4  \cr} \right ]
& = &
16 q \tet _1 [\mu_1]^4 \tet _1 [\mu_3]^4  \biggl (
\tet _1 [\mu_5] ({\tau \over 2}, \tau_1) ^4 - \tet _1 [\nu_0] ({\tau
\over 2}, \tau_1) ^4 \biggr )  + \O(q^2)
\nonumber \\
\Xi_6 \left [ \matrix{\mu_i \cr \mu_6  \cr} \right ]
& = & \O(q) \qquad i=1,3,5
\\
\Xi_6 \left [ \matrix{\nu_0 \cr \nu_0  \cr} \right ]
& = &
- 16 q \sum _{i=1,3,5} \<\mu_i | \nu_0\> \tet _1 [\mu_i]^8 \tet _1
[\mu_i]^4 ({\tau \over 2}, \tau _1) + \O(q^2)
\nonumber 
\eea

\medskip

Next, we derive the asymptotics for the first derivatives of $\tet$
evaluated on odd spin structures $\nu$ which enter into $\M_{\nu _i \nu
_j}$, for example. This is easily done by inspecting
(\ref{thetafunction}). Differentiating by $\p_1$ brings down in the sum
of (\ref{thetafunction}) a factor of $2\pi i (m+ \kappa_1 ')$ while
differentiating by $\p_2$ brings down a factor of $2 \pi i (n+\kappa_2
')$. The asymptotic behavior depends upon the value of $\kappa_2 '$. When
$\kappa_2 '=0$, the $n=0$ term is leading in the $\p_1$ derivative, but
cancels out in the $\p_2$ derivative where the leading contribution comes
from the $n=\pm 1$ terms instead. We have the following limits
\bea
\label{nonsepder}
\p _1 \tet \left [ \matrix{\nu _0 \cr \mu_i  \cr} \right ] (0,\Omega)
& = & - 2 \pi \eta (\tau_1)^3 + \O(q)  \hskip 1.8in i=2,4
\nonumber \\
\p _2 \tet \left [ \matrix{\nu _0 \cr \mu_i  \cr} \right ] (0,\Omega)
& = & 4 \pi i q (-) ^{2 \mu _i ''} \tet _1 [\nu _0] (\tau, \tau_1)
+ \O(q^2)  \hskip .9in i=2,4
\nonumber \\
\p _1 \tet \left [ \matrix{\nu _0 \cr \mu_6  \cr} \right ] (0,\Omega)
& = &
4 q^\14 {\p \over \p \tau} \tet _1 [\nu_0] ({\tau \over 2}, \tau _1) + \O
(q^{{5/4}})
\nonumber \\
\p _2 \tet \left [ \matrix{\nu _0 \cr \mu_6  \cr} \right ] (0,\Omega)
& = &
2 \pi i q^\14 \tet _1 [\nu _0 ] ({\tau \over 2}, \tau _1) + \O
(q^{{5/4}})
\nonumber \\
\p _1 \tet \left [ \matrix{\mu \cr \nu_0  \cr} \right ] (0,\Omega)
& = &
4 i q^\14 {\p \over \p \tau} \tet _1 [\mu] ({\tau \over 2}, \tau _1) + \O
(q^{{5/4}}) \hskip 1in {\rm all \ even} \ \mu
\nonumber \\
\p _2 \tet \left [ \matrix{\mu \cr \nu_0  \cr} \right ] (0,\Omega)
& = &
-2 \pi q^\14 \tet _1 [\mu ] ({\tau \over 2}, \tau _1) + \O
(q^{{5/4}}) \hskip 1.05in {\rm all \ even} \ \mu
\eea

\subsection{Proof of the bilinear $\tet$-constant relation}

We shall now prove {\sl the bilinear $\tet$-constant relation} of
(\ref{tetbilinear}) and determine the multiplicative $\pm$ sign in the
formula for later use in subsection \S 5.3. Since the sign is {\sl not
intrinsic}, it is necessary to fix a definite basis of characteristics.
In the separating degeneration limit, every odd spin structure descends
to a spin structure assignment on the two resulting tori which is odd on
one torus while even on the other. Two distinct cases emerge,
depending on whether the genus 2 spin structures $\nu_i$ and $\nu_j$ in
$\M_{\nu _i \nu _j}$ descend to the odd spin structure on opposite tori
(first case below) or on the same torus (second case below).

\subsubsection{Separating Degenerations : First Case}

Let us now check the proposed formula for the first case,
\be
\label{sepfirst}
\nu _1 = \left [ \matrix{ \mu_1  \cr \nu _0 \cr} \right ]
\qquad \qquad 
\nu _2 = \left [ \matrix{ \nu_0 \cr \mu_2 \cr} \right ]
\ee 
where $\mu _1$ and $\mu _2$ are any two even genus 1 spin structures (not
necessarily taking the expressions of the Table (\ref{listodd})). We work
to the two lowest orders $\tau^0$ and $\tau^2$, (lowest order is
sufficient, but it is interesting to also have the next to leading order
available),
\bea
\p _1 \tet [\nu_1] (0,\Omega )
&=&
2 \tau 
\p _{\tau _1} \tet _1 [\mu_1] (\tau _1) \ \tet _1 '[\nu_0] (\tau _2) 
\nonumber \\
\p _2 \tet [\nu_1](0,\Omega )
&=&
\tet _1 [\mu_1] (\tau _1) \ \tet _1 '[\nu _0 ] (\tau _2) 
+ 2 \tau ^2
\p _{\tau _1} \tet _1 [\mu_1] (\tau _1) \
\p _{\tau _2} \tet _1 '[\nu _0 ] (\tau _2) 
\nonumber \\
\p _1 \tet [\nu_2] (0,\Omega )
&=&
\tet _1 '[\nu _0] (\tau _1) \ \tet _1[\mu_2 ] (\tau _2)  
+ 2 \tau ^2
\p _{\tau _1} \tet _1 '[\nu _0] (\tau _1) \ 
\p _{\tau _2} \tet _1[\mu_2 ] (\tau _2) 
\nonumber \\
\p _2 \tet [\nu_2] (0,\Omega )
&=&
2 \tau
\tet _1 '[\nu_0 ] (\tau _1) \ \p _{\tau _2} \tet _1 [\mu_2] (\tau _2) 
\eea
The remaining 4 odd spin structures are 
\be
\nu _3 = \left [ \matrix{ \mu_3  \cr \nu _0 \cr} \right ]
\qquad \qquad 
\nu _4 = \left [ \matrix{ \nu_0 \cr \mu_4 \cr} \right ]
\qquad \qquad 
\nu _5 = \left [ \matrix{ \mu_5  \cr \nu _0 \cr} \right ]
\qquad \qquad 
\nu _6 = \left [ \matrix{\nu_0 \cr \mu_6 \cr } \right ]
\ee 
where $\mu _1$, $\mu_3$ and $\mu_5$ are three distinct even genus 1 spin 
structures and $\mu _2$, $\mu_4$ and $\mu_6$ are also three distinct even 
genus  1 spin structures. They satisfy
\bea
\mu _1 + \mu_3 + \mu _5 &=& \nu _0 \nonumber \\
\mu _2 + \mu _4 + \mu _6 &=& \nu _0\, .
\eea
The quantity $\M_{\nu_i\nu_j}$ may now be easily expressed to this 
order
\bea
\M _{\nu _1 \nu _2} &=&
- \tet _1 [\mu_1] (\tau _1) \ \tet _1 '[\nu _0 ] (\tau _2) \
\tet _1 '[\nu _0] (\tau _1) \ \tet _1[\mu_2 ] (\tau _2) 
\nonumber \\ && 
+ 4\tau^2 \p _{\tau _1} \tet _1 [\mu_1](\tau _1) \ \tet _1
'[\nu_0](\tau_2) \
\tet _1 '[\nu_0 ] (\tau _1) \ \p _{\tau _2} \tet _1 [\mu_2] (\tau _2)
\nonumber \\ && 
-2 \tau^2 \tet _1 [\mu_1] (\tau _1) \ \tet _1 '[\nu _0 ] (\tau _2) \
\p _{\tau _1} \tet _1 '[\nu _0] (\tau _1) \
\p _{\tau _2} \tet _1[\mu_2 ] (\tau _2) 
\nonumber \\ &&
-2 \tau^2 
\tet _1 '[\nu _0] (\tau _1) \ \tet _1[\mu_2 ] (\tau _2) \
\p _{\tau _1} \tet _1 [\mu_1] (\tau _1) \
\p _{\tau _2} \tet _1 '[\nu _0 ] (\tau _2) 
\eea
We make use of the well-known identity for genus one $\tet$-functions,
(\ref{tetproduct}). Substituting these expressions, we see that the $\tau
_1$ derivative terms on $\tet [\mu _1]$ cancel, as well as the $\tau_2$
derivatives on
$\tet [\mu _2]$. One is left with
\bea
\M _{\nu _1 \nu _2} &=&
- \pi ^2 \tet _1 [\mu_1] ^2  \tet _1 [\mu_3 ]  \tet _1 [\mu_5](\tau_1)
\cdot
\tet _1 [\mu_2] ^2 \tet _1[\mu_4 ] \tet _1 [\mu _6] (\tau _2) 
\nonumber \\ && 
- 2 \pi ^2 \tau^2 
\tet _1 [\mu_1] \p _{\tau _1} \tet _1 [\mu_1](\tau _1) 
\tet _1 [\mu _3] \tet _1 [\mu_5] (\tau_1) \cdot
\tet _1 [\mu _2 ] ^2  \p _{\tau _2} \bigl ( \tet _1 [\mu_4] \tet_1
[\mu_6] \bigr ) (\tau_2)
\nonumber \\ && 
- 2 \pi ^2 \tau^2 
\tet _1 [\mu _1 ] ^2  \p _{\tau _1} \bigl ( \tet _1 [\mu_3] \tet_1
[\mu_5] \bigr ) (\tau_1) \cdot 
\tet _1 [\mu_2] \p _{\tau _2} \tet _1 [\mu_2](\tau _2) \
\tet _1 [\mu _4] \tet _1 [\mu_6] (\tau_2) 
\nonumber
\eea
We now wish to compare these asymptotics with the product of
$\tet$-constants for even characteristics, appearing on the rhs of
(\ref{tetbilinear}).  Worked out to the same order, the $\tet$-constants
behave as,
\be
\tet \left [ \matrix{ \kappa _1  \cr \kappa _2 \cr} \right ](0,\Omega)
=
\tet _1[\kappa _1](\tau_1) \tet _1 [\kappa _2](\tau_2)
+ 2 \tau ^2 
\p_{\tau _1} \tet _1[\kappa _1](\tau_1) \p _{\tau_2} \tet _1
[\kappa _2](\tau_2)
\ee
Up to the addition of complete periods, this assignment of even
characteristics is precisely what corresponds to {\sl the
bilinear $\tet$ relation} that we seek to prove.
\bea
\label{evensep}
\nu _1 + \nu _2 +\nu _3 &=&
\left [ \matrix{ \nu_0 + \mu_1 +\mu _3 \cr 2 \nu _0 + \mu_2  \cr} \right]
\equiv 
\left [ \matrix{  \mu_5 \cr \mu_2  \cr} \right ]
\nonumber \\ && \nonumber \\
\nu _1 + \nu _2 +\nu _4  &=&
\left [ \matrix{2 \nu_0 + \mu_1 \cr \nu _0 + \mu_2 +\mu_4  \cr} \right]
\equiv 
\left [ \matrix{  \mu_1 \cr \mu_6  \cr} \right ]
\nonumber \\ && \nonumber \\
\nu _1 + \nu _2 +\nu _5 &=&
\left [ \matrix{ \nu_0 + \mu_1 +\mu _5 \cr 2 \nu _0 + \mu_2  \cr} \right]
\equiv 
\left [ \matrix{  \mu_3 \cr \mu_2  \cr} \right ]
\nonumber \\ && \nonumber \\
\nu _1 + \nu _2 +\nu _6 &=&
\left [ \matrix{ 2\nu_0 + \mu_1 \cr \nu _0 + \mu_2 +\mu_6  \cr} \right]
\equiv 
\left [ \matrix{  \mu_1 \cr \mu_4  \cr} \right ]
\eea
The effects of the full periods $2 \nu_0$ cancel out and we are left with
our final formula, 
\be
\label{sepproduct}
\M _{\nu _1 \nu _2} = - \pi ^2
\tet \left [ \matrix{  \mu_3 \cr \mu_2  \cr} \right ]
\tet \left [ \matrix{  \mu_5 \cr \mu_2  \cr} \right ]
\tet \left [ \matrix{  \mu_1 \cr \mu_4  \cr} \right ]
\tet \left [ \matrix{  \mu_1 \cr \mu_6  \cr} \right ](0,\Omega)
\ee
It must be stressed here that this formula is exact, including a proper
assignment of the overall (non-intrinsic but important) sign. In the
expressions for the superstring chiral measure, only 
$\M_{\nu_i\nu_j}^2$ will enter
and the non-intrinsic sign will be unimportant.

\subsubsection{Separating Degenerations : Second Case}

Let us now check the proposed formula for the second type 
\be
\label{sepsecond}
\nu _1 = \left [ \matrix{ \mu_1  \cr \nu _0 \cr} \right ]
\qquad \qquad 
\nu _3 = \left [ \matrix{ \mu_3 \cr \nu_0 \cr} \right ]
\ee 
where $\mu _1$ and $\mu _3$ are any two distinct even genus 1 spin 
structures.  We work to lowest orders in $\tau$,
\bea
\p _1 \tet [\nu_1] (0, \Omega )
&=&
2 \tau 
\p _{\tau _1} \tet _1 [\mu_1] (\tau _1) \ \tet _1 '[\nu_0] (\tau _2) 
\nonumber \\
\p _2 \tet [\nu_1](0, \Omega )
&=&
\tet _1 [\mu_1] (\tau _1) \ \tet _1 '[\nu _0 ] (\tau _2)  
\nonumber \\
\p _1 \tet [\nu_3] (0, \Omega )
&=&
2 \tau 
\p _{\tau _1} \tet _1 [\mu_3] (\tau _1) \ \tet _1 '[\nu_0] (\tau _2) 
\nonumber \\
\p _2 \tet [\nu_3] (0, \Omega )
&=&
\tet _1 [\mu_3] (\tau _1) \ \tet _1 '[\nu _0 ] (\tau _2) 
\eea
The remaining 4 odd spin structures are 
\be
\nu _2 = \left [ \matrix{ \nu_0  \cr \mu_2 \cr} \right ]
\qquad \qquad 
\nu _4 = \left [ \matrix{ \nu_0 \cr \mu_4 \cr} \right ]
\qquad \qquad 
\nu _5 = \left [ \matrix{ \mu_5  \cr \nu _0 \cr} \right ]
\qquad \qquad 
\nu _6 = \left [ \matrix{\nu_0 \cr \mu_6 \cr } \right ]
\ee 
where, as before, $\mu _1$, $\mu_3$ and $\mu_5$ are three distinct even 
genus 1  spin structures and $\mu _2$, $\mu_4$ and $\mu_6$ are also three
distinct even  genus  1 spin structures. They satisfy
\bea
\mu _1 + \mu_3 + \mu _5 &=& \nu _0 \nonumber \\
\mu _2 + \mu _4 + \mu _6 &=& \nu _0\, .
\eea
The expression for $\M_{\nu _1 \nu _3}$ is then given by
\bea
\M_{\nu _1 \nu _3} &=&
2 \tau \biggl (\p _{\tau _1} \tet _1 [\mu_1] \tet _1 [\mu_3] -\p _{\tau _1} 
\tet  _1 [\mu_3]\tet _1 [\mu_1] \biggr ) (\tau _1) \cdot \tet _1 '[\nu_0]
(\tau _2) ^2
\\
&=&
2 \pi ^2 \tau \biggl ( \p _{\tau _1} \tet _1 [\mu_1] \tet _1 [\mu_3] - \p
_{\tau  _1} \tet _1 [\mu_3]\tet _1 [\mu_1] \biggr ) (\tau _1) \cdot \tet _1
[\mu _2]^2 
\tet _1 [\mu_4]^2 \tet _1 [\mu_6]^2 (\tau_2)
\nonumber
\eea
Furthermore, the expansion to lowest order for the even spin structure 
$\tet$-constants is
\bea
\tet \left [ \matrix{\sigma _1 \cr \sigma _2 \cr} \right ](0,\Omega )
&=&
 \tet _1[\mu _1] (0,\tau _1) 
 \tet _1[\mu _2] (0,\tau _2) 
\nonumber \\
\tet \left [ \matrix{\nu_0 \cr \nu_0 \cr} \right ](0,\Omega )
&=&
{2 \tau \over 4 \pi i} 
\tet _1 '[\nu_0] (0,\tau _1)  \tet _1 '[\nu_0] (0,\tau _2) 
\eea

\medskip

To make contact between these two formulas, we need the following equation 
between derivatives of ratios of genus one theta constants \cite{igusa3}
\be
\p _{\tau _1} \ln {\tet _1 [\mu_1] \over \tet _1 [\mu _3] }
= {i \pi \over 4} \tet _1 [\mu _5]^4 \cdot \sigma (\mu_1,\mu_3)
\ee
where $\sigma$ obeys $\sigma (\mu_1, \mu_3) = - \sigma (\mu _3, \mu _1)$ 
and is  given by
\be 
\label{signaturesigma}
\sigma ([10],[00]) = \sigma ([00],[01]) = \sigma ([10],[01]) =+1\, .
\ee
Thus, we have the expression
\bea
\p _{\tau _1} \tet _1 [\mu_1] \tet _1 [\mu_3] -
\p _{\tau _1} \tet _1 [\mu_3] \tet _1 [\mu_1] 
&=&
{i \pi \over 4} \tet _1 [\mu_1] \tet _1 [\mu_3] \tet _1 [\mu_5]^4 (\tau _1)
\sigma (\mu _1, \mu_3)
\nonumber \\
&=&
-{i \over 4} \tet _1 '[\nu_0] \tet _1 [\mu _5]^3 (\tau_1) \sigma 
(\mu_1,\mu_3)
\eea

\medskip

Using this formula to eliminate the derivative terms in the expression for 
$\M_{\nu_i\nu_j}$  and using the product formula for $\tet _1 '[\nu_0]$,
we get
\be
\M _{\nu _1 \nu _3} =
- \pi ^2 \ \sigma (\mu _1, \mu _3) \ 
\tet \left [ \matrix{  \nu_0 \cr \nu_0  \cr} \right ]
\tet \left [ \matrix{  \mu_5 \cr \mu_2  \cr} \right ]
\tet \left [ \matrix{  \mu_5 \cr \mu_4  \cr} \right ]
\tet \left [ \matrix{  \mu_5 \cr \mu_6  \cr} \right ](0,\Omega)
\ee
which is our final formula for the second case. 
Up to the addition of complete periods, (whose effects cancel out
completely) this assignment of even characteristics is precisely what
corresponds to the product formula that we proposed to prove.
\bea
\nu _1 + \nu _3 +\nu _2 &=&
\left [ \matrix{ \nu_0 + \mu_1 +\mu _3 \cr 2 \nu _0 + \mu_2  \cr} \right]
\equiv 
\left [ \matrix{  \mu_5 \cr \mu_2  \cr} \right ]
\nonumber \\ && \nonumber \\
\nu _1 + \nu _3 +\nu _4  &=&
\left [ \matrix{ \nu_0 + \mu_1 + \mu_3 \cr 2\nu _0 +\mu_4  \cr} \right]
\equiv 
\left [ \matrix{  \mu_5 \cr \mu_4  \cr} \right ]
\nonumber \\ && \nonumber \\
\nu _1 + \nu _3 +\nu _5 &=&
\left [ \matrix{ \mu_1 + \mu_3 + \mu _5 \cr 3\nu _0   \cr} \right]
\equiv 
\left [ \matrix{  \nu_0 \cr \nu_0  \cr} \right ]
\nonumber \\ && \nonumber \\
\nu _1 + \nu _3 +\nu _6 &=&
\left [ \matrix{ \nu_0 + \mu_1 +\mu_3 \cr 2\nu _0 +\mu_6  \cr} \right]
\equiv 
\left [ \matrix{  \mu_5 \cr \mu_6  \cr} \right ]
\eea

\medskip

For both cases a completely intrinsic formula may be obtained by squaring
the above
\be
\M _{\nu _1 \nu _3} ^2 =  \pi ^4
\prod _{i\not=1,3}
\tet [ \nu_1 + \nu _3 +\nu _i ] ^2 (0,\Omega)
\ee
and so this formula is universall valid.

\subsubsection{Non-Separating Degenerations : First Case}

Next, we show that the same formulas with the same signs are reproduced
in the non-separating degeneration limit. Recall that the first case
corresponded to the spin structure assignments
\be
\nu _i = \left [ \matrix{ \mu_i  \cr \nu _0 \cr} \right ]
\qquad \qquad 
\nu _j = \left [ \matrix{ \nu_0 \cr \mu_j \cr} \right ]
\ee 
where $\mu _i$ and $\mu_j$ are any even genus 1 characteristics.
Actually, when analyzing non-separating degenerations, this case itself
falls into two subcases; $\mu_i$ can be any even spin structure, but the
cases $j=2,4$ and $j=6$ generate different asymptotic behaviors and must
be treated differently. For concreteness, and without loss of generality,
we make definite assignments~: $\mu_i =\mu _1, \ \mu_j=\mu_ 2$ for the
first subcase, while $\mu_i=\mu_1, \ \mu_j= \mu _6$ for the second
subcase. All other 5 cases in this class are analogous to one of these
two. 

\medskip

The asymptotics of $\M_{\nu_i \nu_j}$ is easily computed in both cases,
using (\ref{nonsepder}), and we find,
\bea
\M _{\nu _1 \nu _2} & = & - 4 \pi ^2 q^\14 \eta (\tau_1) ^3 \tet _1
[\mu_1] ({\tau \over 2}, \tau_1)
 \\
\M _{\nu _1 \nu_6} & = & - 8 \pi q^\half \biggl \{
{\p \over \p \tau} \tet _1 [\mu_1] ({\tau \over 2}, \tau_1)
\tet _1 [\nu_0] ({\tau \over 2}, \tau_1)
-
{\p \over \p \tau} 
\tet _1 [\nu_0] ({\tau \over 2}, \tau_1)
\tet _1 [\mu_1] ({\tau \over 2}, \tau_1) \biggr \}
\nonumber
\eea
This result needs to be compared with the limit of the product of
$\tet$-constants for these spin structure assignments. The corresponding
even spin structures occurring in the product were already determined in
(\ref{evensep}), but we now need to adapt the notation to that of the
current situation.  The products of the corresponding genus two
$\tet$-functions as well as their asymptotic behaviors are given by the
following limits,
\bea
\tet \left [ \matrix{ \mu_1  \cr \mu _4 \cr} \right ] 
\tet \left [ \matrix{ \mu_1  \cr \mu _6 \cr} \right ] 
\tet \left [ \matrix{ \mu_3  \cr \mu _2 \cr} \right ] 
\tet \left [ \matrix{ \mu_5  \cr \mu _2 \cr} \right ] (0,\Omega)
& = &
2 q^\14 \tet _1 [\mu_1] ({\tau \over 2}, \tau_1)
 \prod _{i=1,3,5} \tet _1 [\mu _i] (0,\tau_1) 
\\
\tet \left [ \matrix{ \mu_1  \cr \mu _2 \cr} \right ] 
\tet \left [ \matrix{ \mu_1  \cr \mu _4 \cr} \right ] 
\tet \left [ \matrix{ \mu_3  \cr \mu _6 \cr} \right ] 
\tet \left [ \matrix{ \mu_5  \cr \mu _6 \cr} \right ] (0,\Omega)
& = & 
4 q^\half \tet _1 [\mu_1] (0,\tau_1)^2 
\tet _1 [\mu_3 ] ({\tau \over 2}, \tau_1) 
\tet _1 [\mu_5 ] ({\tau \over 2}, \tau_1)
\nonumber 
\eea 
Using the product formula (\ref{tetproduct}), it is manifest that the
subcase $\nu_1 \nu_2$ is consistent with the following equality,
\bea
\M _{\nu _1 \nu _2} = - \pi ^2
\tet \left [ \matrix{ \mu_1  \cr \mu _4 \cr} \right ] 
\tet \left [ \matrix{ \mu_1  \cr \mu _6 \cr} \right ] 
\tet \left [ \matrix{ \mu_3  \cr \mu _2 \cr} \right ] 
\tet \left [ \matrix{ \mu_5  \cr \mu _2 \cr} \right ] (0,\Omega)
\eea
which is indeed the precise same form as we had obtained for the
separating degenerations in (\ref{sepproduct}), including the
non-intrinsic sign.

\medskip

Comparison for the subcase $\nu_1 \nu_6$ is more complicated because the
limits of $\M_{\nu_1 \nu_6}$ and the product are rather different looking.
The required genus 1 identity does not appear to be familiar when
expressed in terms of $\tet$-functions, but it is well known when
translated into the
Jacobian elliptic functions $sn(u)$, $cn(u)$ and $dn(u)$.
The correspondence between the
functions $sn(u)$, $cn(u)$,
$dn(u)$ and $\tet$-functions 
is given by\footnote{A convenient reference is
\cite{bateman}. We notice, however, that the precise correspondence
between our notations for $\tet$-functions and those of \cite{bateman}
involves subtle signs,  given by $\tet _1 [\mu_1] (v) = + \tet _3 (v)$,
$\tet _1 [\mu_3] (v) = + \tet _4 (v)$, $\tet _1 [\mu_5] (v) = + \tet _2
(v)$,
$\tet _1 [\nu_0] (v) = - \tet _3 (v)$.}
\bea
\label{jacobitet}
sn (u) & = & -{\tet _1 [\mu_1] (0) \tet _1 [\nu_0] (v) \over 
         \tet _1 [\mu_5] (0) \tet _1 [\mu_3] (v)}
\qquad \qquad
dn(u) = +{\tet _1 [\mu_3] (0) \tet _1 [\mu_1] (v) \over 
         \tet _1 [\mu_1] (0) \tet _1 [\mu_3] (v)}
\nonumber \\
cn(u) & = & +{\tet _1 [\mu_3] (0) \tet _1 [\mu_5] (v) \over 
         \tet _1 [\mu_5] (0) \tet _1 [\mu_3] (v)}
\hskip .95in 
u \equiv v \pi \tet _1 [\mu _1](0)^2
\eea
Here, the modular parameter has been suppressed. 
The standard derivative formulas for Jacobian elliptic functions ($sn'(u)
= cn(u) dn(u)$, $cn'(u) = -sn(u) dn(u)$ and $dn'(u) = - k^2 sn(u) cn(u)$
together with the standard quadratic relations between these functions
produce the following relations,
\bea
{\p \over \p u} \ln {sn(u) \over dn(u) } & = & { cn(u) \over sn(u) dn(u)}
\hskip 1in 
{\p \over \p u} \ln {sn(u) \over cn(u) } = { dn(u) \over sn(u) cn(u)}
\nonumber \\
{\p \over \p u} \ln sn(u)  & = & { cn(u) dn(u) \over sn(u) }
\eea
Translating these formulas into $\tet$-functions using
(\ref{jacobitet}), and changing variables from $u$ to $v$ gives
\bea
{\p \over \p v} \ln {\tet _1[\nu_0] (v) \over \tet _1 [\mu_1](v)}
= - \pi  \tet _1 [\mu_1] ^2 (0)
{\tet _1 [\mu_3 ] (v) \tet _1 [\mu_5] (v) \over 
\tet _1 [\nu_0](v) \tet _1 [\mu_1](v)}
\eea
valid for any even spin structure $\mu_1$ and its two distinct partmers
$\mu_3$ and $\mu_5$. Multiplying both sides by $\tet _1[\nu_0] (v) \tet _1
[\mu_1](v)$ yields,
\bea
{\p \over \p v} \tet _1 [\nu_0] (v) \tet _1 [\mu_1](v) -
{\p \over \p v} \tet _1 [\mu_1] (v) \tet _1[\nu_0] (v) 
= - \pi  \tet _1 [\mu_1] ^2 (0)
\tet _1 [\mu_3 ] (v) \tet _1 [\mu_5] (v) 
\eea
Replacing $v=\tau /2$, and restoring the genus 1 modulus dependence, we
finally get the desired formula,
\bea
&& {\p \over \p \tau} \tet _1 [\nu_0] ({\tau \over 2}, \tau_1) 
                   \tet _1 [\mu_1] ({\tau \over 2}, \tau_1) - 
{\p \over \p \tau} \tet _1 [\mu_1] ({\tau \over 2}, \tau_1)
                   \tet _1 [\nu_0] ({\tau \over 2}, \tau_1)
\nonumber \\ && \hskip 1.35in 
= - {\pi \over 2}  
\tet _1 [\mu_1] ^2 (0) \tet _1 [\mu_3 ] ({\tau \over 2}, \tau_1) 
\tet _1 [\mu_5] ({\tau \over 2}, \tau_1) 
\eea
This reproduces the following formula
\bea
\M_{\nu_1 \nu_6} = - \pi ^2 
\tet \left [ \matrix{ \mu_1  \cr \mu _2 \cr} \right ] 
\tet \left [ \matrix{ \mu_1  \cr \mu _4 \cr} \right ] 
\tet \left [ \matrix{ \mu_3  \cr \mu _6 \cr} \right ] 
\tet \left [ \matrix{ \mu_5  \cr \mu _6 \cr} \right ]
\eea
in agreement with the general formula proposed and with the sign of the
separating case.

\subsubsection{Non-Separating Degenerations : Second Case}

Finally, we show that the same formulas with the same signs are also
reproduced in the non-separating degeneration limit for the second case.
Here, {\sl we must further distinguish between two subcases}. 

\medskip

The {\sl first subcase} has both genus 1 even spin structures on the
degenerating torus, 
\be
\nu _i = \left [ \matrix{ \nu_0  \cr \mu_i \cr} \right ]
\qquad \qquad 
\nu _j = \left [ \matrix{ \nu_0 \cr \mu_j \cr} \right ]
\qquad i,j=2,4,6
\ee 
and we have (the case $\M_{\nu_4\nu_6}$is analogous
to $\M_{\nu_2 \nu_6}$)
\bea
\M _{\nu _2 \nu_4 } & = & 16 i \pi ^2 q \eta (\tau_1)^3 \tet _1 [\nu_0]
(\tau, \tau_1) 
\nonumber \\
\M _{\nu _2 \nu_6 } & = & - 4 i \pi^2 q^\14 \eta (\tau_1)^3 \tet _1
[\nu_0] ({\tau\over 2},\tau_1)
\eea
For these two assignments, the product of $\tet$-constants is given by
\bea
\tet \left [ \matrix{  \nu_0 \cr \nu_0  \cr} \right ]
\tet \left [ \matrix{  \mu_1 \cr \mu_6  \cr} \right ]
\tet \left [ \matrix{  \mu_3 \cr \mu_6  \cr} \right ]
\tet \left [ \matrix{  \mu_5 \cr \mu_6  \cr} \right ](0,\Omega)
& = & 16 i q \eta (\tau_1)^3 \tet _1 [\nu_0] (\tau , \tau_1)
\nonumber \\
\tet \left [ \matrix{  \nu_0 \cr \nu_0  \cr} \right ]
\tet \left [ \matrix{  \mu_1 \cr \mu_4  \cr} \right ]
\tet \left [ \matrix{  \mu_3 \cr \mu_4  \cr} \right ]
\tet \left [ \matrix{  \mu_5 \cr \mu_4  \cr} \right ](0,\Omega)
& = &
4 i q^\14 \eta (\tau_1)^3 \tet _1 [\nu_0] ({\tau\over 2}, \tau_1)
\eea
Both are manifestly in agreement with the relations obtained for
separating degenerations.

\medskip

The {\sl second subcase} has both genus 1 odd spin structures on the
degenerating torus,
\be
\nu _i = \left [ \matrix{ \mu_i \cr \nu_0 \cr} \right ] 
\qquad \qquad 
\nu _j = \left [ \matrix{ \mu_j \cr \nu_0 \cr} \right ] 
\qquad i,j=1,3,5
\ee 
For this subcase, we have
\bea
\M_{\nu _i \nu_j} =
- 8 i \pi q^\half \biggl (
{\p \over \p \tau} \tet _1 [\mu_i] ({\tau \over 2}, \tau_1) 
\tet _1 [\mu_j] ({\tau \over 2}, \tau_1) -
{\p \over \p \tau} \tet _1 [\mu_j] ({\tau \over 2}, \tau_1) 
\tet _1 [\mu_i] ({\tau \over 2}, \tau_1) \biggr )
\eea
and the product formula (here, $k\not=i,j$)
\bea
\tet \left [ \matrix{  \nu_0 \cr \nu_0  \cr} \right ]
\tet \left [ \matrix{  \mu_k \cr \mu_2  \cr} \right ]
\tet \left [ \matrix{  \mu_k \cr \mu_4  \cr} \right ]
\tet \left [ \matrix{  \mu_k \cr \mu_6  \cr} \right ](0,\Omega)
=
4 i q^\half \tet _1 [\mu_k](0,\tau_1) ^2 \tet _1 [\mu _k] ({\tau \over
2}, \tau_1) \tet _1 [\nu_0] ({\tau \over 2}, \tau_1)
\eea
The genus 1 identities needed here are obtained by working out the
following combinations of derivatives 
\bea
{\p \over \p u} \ln dn(u)  & = & - {\tet _1 [\mu_5]^4 \over \tet_1
[\mu_1]^4} { sn(u) cn(u) \over dn(u)}
\hskip 1in 
{\p \over \p u} \ln cn(u) =  - { sn(u) dn(u) \over cn(u) }
\nonumber \\
{\p \over \p u} \ln {dn(u) \over cn(u) } &=& + {\tet _1 [\mu_3]^4 \over
\tet_1 [\mu_1]^4}  { sn(u) \over cn(u) dn(u)}
\eea
and translating these equations into $\tet _1$ functions using
(\ref{jacobitet}), 
\bea
&&
{\p \over \p \tau} \tet _1 [\mu_i] ({\tau \over 2}, \tau_1) 
\tet _1 [\mu_j] ({\tau \over 2}, \tau_1) -
{\p \over \p \tau} \tet _1 [\mu_j] ({\tau \over 2}, \tau_1) 
\tet _1 [\mu_i] ({\tau \over 2}, \tau_1)
\\
&& \hskip 1in =
+ \sigma (\mu_i,\mu_j) {\pi \over 2} \tet _1 [\mu_k] (0,\tau_1)^2 \tet _1
[\mu _k] ({\tau
\over 2}, \tau_1) \tet _1 [\nu_0] ({\tau \over 2}, \tau_1)
\nonumber 
\eea
where the signature $\sigma (\mu_1, \mu_2)$ was introduced in
(\ref{signaturesigma}). Using these relations, we again recover the formula
established in the case of separating degenerations.

\subsection{Proof of the $\M$ Product Formula}

As we had stressed, the ${\cal M}_{\nu_i\nu_j}$ product formula follows
from the independence of the chiral measure from the points $q_{\alpha}$.
Here, we give a direct and independent proof of the identity using only
$\tet$-function results.
Again, we must split this treatment into two cases according to the form
of the spin structures, since we  need the detailed sign assignments.

\medskip

We begin with the first case (\ref{sepfirst}), and need to prove the
formula
\be
\<\nu_1 |\nu _2\> \prod _{i=3,4,5,6} 
{\M _{\nu _2 \nu _i} \over \M _{\nu _1 \nu _i} } =1\, .
\ee 
It is convenient to use the abbreviations,
\be
(00) = \tet \left [ \matrix{  \nu_0 \cr \nu_0  \cr} \right ](\Omega)
\qquad \quad
(ij) = \tet \left [ \matrix{  \mu_i \cr \mu_j  \cr} \right ](\Omega)\, .
\ee
Then, we have the following expressions for the $\M_{\nu_i\nu_j}$'s 
entering the product
\bea
\M_{\nu _1 \nu _3} &=& - \pi ^2 (00)(52)(54)(56) \cdot \sigma 
(\mu_1,\mu_3)
\nonumber \\
\M_{\nu _1 \nu _4} &=& - \pi ^2 (34)(54)(12)(16) 
\nonumber \\
\M_{\nu _1 \nu _5} &=& - \pi ^2 (00)(32)(34)(36) \cdot \sigma (\mu_1,\mu_5)
\nonumber \\
\M_{\nu _1 \nu _6} &=& - \pi ^2 (36)(56)(12)(14) 
\eea
as well as
\bea
\M_{\nu _2 \nu _3} &=& + \pi ^2 (12)(52)(34)(36) 
\nonumber \\
\M_{\nu _2 \nu _4} &=& - \pi ^2 (00)(16)(36)(56) \cdot \sigma (\mu_2,\mu_4)
\nonumber \\
\M_{\nu _2 \nu _5} &=& + \pi ^2 (12)(32)(54)(56) 
\nonumber \\
\M_{\nu _2 \nu _6} &=& - \pi ^2 (00)(14)(34)(54) \cdot \sigma (\mu_2,\mu_6)
\eea
Putting all factors together, we find that all the factors of
$\pi$ and $(ij)$ precisely cancel one  another, so that we are left with
\be
 \prod _{i=3,4,5,6} 
{\M _{\nu _2 \nu _i} \over \M _{\nu _1 \nu _i} } =
{\sigma (\mu_2, \mu_4) \cdot \sigma (\mu_2, \mu_6) 
\over \sigma (\mu_1,\mu_3) \cdot \sigma (\mu_1, \mu_5)}
\ee 
It remains to evaluate the product of $\sigma$'s. By inspection of all
cases, it  is easy to see that one has
\bea
\sigma (\mu_1,\mu_3) \cdot \sigma (\mu_1, \mu_5) &=& - \< \mu_1 | \nu_0 \>
\nonumber \\
\sigma (\mu_2, \mu_4) \cdot \sigma (\mu_2, \mu_6) &=& - \< \mu_2 |\nu_0\>
\eea
and that 
\be
\< \nu_1 | \nu_2 \> = \< \mu_1 | \nu_0\> \<\mu_2 | \nu_0 \>
\ee
which proves the formula for the first case.

\medskip

For the second case (\ref{sepsecond}), 
we need to prove the formula
\be
\<\nu_1 |\nu _3\> \prod _{i=2,4,5,6} 
{\M _{\nu _3 \nu _i} \over \M _{\nu _1 \nu _i} } =1\, .
\ee 
We use the same abbreviation as for the first case. Then, we have the
following expressions for the $\M_{\nu_i\nu_j}$'s entering the product
\bea
\M_{\nu _1 \nu _2} &=& - \pi ^2 (32)(52)(14)(16) 
\nonumber \\
\M_{\nu _1 \nu _4} &=& - \pi ^2 (34)(54)(12)(16) 
\nonumber \\
\M_{\nu _1 \nu _5} &=& - \pi ^2 (00)(32)(34)(36) \cdot \sigma (\mu_1,\mu_5)
\nonumber \\
\M_{\nu _1 \nu _6} &=& - \pi ^2 (36)(56)(12)(14) 
\eea
as well as
\bea
\M_{\nu _3 \nu _2} &=& - \pi ^2 (12)(52)(34)(36) 
\nonumber \\
\M_{\nu _3 \nu _4} &=& - \pi ^2 (14)(54)(32)(36) 
\nonumber \\
\M_{\nu _3 \nu _5} &=& - \pi ^2 (00)(12)(14)(16) \cdot \sigma (\mu_3,\mu_5)
\nonumber \\
\M_{\nu _3 \nu _6} &=& - \pi ^2 (16)(56)(32)(34) 
\eea
Putting all factors together, we find that all the factors of $\pi$ and
of $(ij)$ precisely cancel  one  another, and we are left with
\be
 \prod _{i=2,4,5,6} 
{\M _{\nu _3 \nu _i} \over \M _{\nu _1 \nu _i} } =
{\sigma (\mu_3, \mu_5)  \over \sigma (\mu_1,\mu_5) } =
\sigma (\mu_3, \mu_5)  \sigma (\mu_1,\mu_5) 
\ee 
 The last product of $\sigma$'s is easily re-expressed, by inspection of 
all possible cases
\be
\sigma (\mu_3, \mu_5)  \sigma (\mu_1,\mu_5) =\<\nu_1 |\nu_3\>
\ee
and this proves the formula for the second case.

\vfill\eject

\section{The GSO Projection and Cosmological Constant}
\setcounter{equation}{0}

The correct degrees of freedom of the superstring are obtained only after
enforcing the GSO projection. In the string path integral formulation,
this corresponds to summing the contributions to
the chiral measure of each spin structure.
Now, due to chiral splitting, the contribution of each
spin structure $\delta$ is only determined up to a global phase
factor, which is independent of the moduli, but may depend
on $\delta$. The issue in enforcing the GSO projection is then
to determine the relative phases between the contributions of
various spin structures. The key criterion that the
relative phases must satisfy is modular invariance
for the full chiral measure.   

\subsection{The GSO Projection}

It was shown in subsection \S 4.3 that $\Xi_6 [\delta](\Omega)$, defined
earlier by
\be
\Xi _6 [\delta] (\Omega) \equiv \sum _{1\leq i < j \leq 3}
\<\nu_i |\nu _j\> \prod _{k=4,5,6} \tet [\nu_i + \nu_j +\nu_k]^4
(0,\Omega)
\ee
depends only on $\delta$ and not on the particular triplet of
points in the partition used to represent $\delta$. The modular
transformation properties of $\Xi _6 [\delta](\Omega) $ can be now read
off from Table 2. It turns out that they are closely related to those of
$\tet [\delta ]^4(0,\Omega)$ and given by
\bea
\tet [\tilde \delta ]^4 (0,\tilde \Omega) 
& = & 
\epsilon (\delta , M) ^4 \ \det (C\Omega + D)^2 \ \tet [\delta ] ^4
(0,\Omega)
\nonumber \\
\Xi _6 [\tilde \delta ] (\tilde \Omega) 
& = & 
\epsilon (\delta, M)^4 \ \det (C\Omega + D)^6 \ \Xi _6 [\delta ] (\Omega )
\eea
so that the product of the two transforms as
\be
\label{phase1}
\tet [\tilde \delta ]^4 (0, \tilde \Omega) 
\Xi _6 [\tilde \delta ] (\tilde \Omega)
=
\det (C\Omega + D)^8 
\tet [\delta ]^4 (0, \Omega) \Xi _6 [\delta ] (\Omega )
\ee
where there are NO SIGNS in front of the right hand side.

\medskip

To enforce the GSO projection, we have to sum over spin structures
the chiral measures $d\mu[\delta](\Omega)$ with suitable relative phases 
$\eta_{\delta}$. The criterion is that  the total chiral measure
$d\mu(\Omega) =\sum_{\delta}\eta_{\delta}d\mu[\delta](\Omega)$ has to
lead to  a full measure (\ref{meas}) invariant under $Sp(4,{\bf Z})$.
Now recall that a modular form $\Psi_k(\Omega)$
of weight $k$ is a holomorphic
function of $\Omega_{IJ}$ which transforms as follows under
$Sp(4,{\bf Z})$
\be
\label{modularform}
\Psi_k(\tilde\Omega)
=
\det(C\Omega+D)^{k}\Psi_k(\Omega)
\ee
Since $\det\,\Im\Omega$
and the measure $\prod_{I\leq J}d\Omega_{IJ}$ transform 
under $Sp(4,{\bf Z})$ as
\bea
\det\,\Im\tilde\Omega
&=&
|\det(C\Omega+D)|^{-2}\det\,\Im\Omega\nonumber\\
\prod_{I\leq J}d\tilde\Omega_{IJ}
&=&
\det(C\Omega+D)^{-3}\prod_{I\leq J}d\Omega_{IJ}
\eea 
we find that the holomorphic coefficient $d\mu(\Omega)/\prod_{I\leq 
J}d\Omega_{IJ}$ must be a modular form of weight $-2$
\be
{d\mu(\tilde\Omega)\over\prod_{I\leq J}d\tilde\Omega_{IJ}}
=
\det(C\Omega+D)^{-2}
{d\mu(\Omega)\over\prod_{I\leq J}d\Omega_{IJ}}
\ee

\medskip

Returning now to the expression for $d\mu[\delta](\Omega)$ in terms of
$\Xi_6[\delta](\Omega)$, we observe that the expressions
$\tet^4[\delta](\Omega)\Xi_6[\delta](\Omega)$ for each spin structure
$\delta$ are {\sl not} modular forms, because they transform 
into each other. However, in view of  the preceding phases (\ref{phase1})
for modular transformations, there is a {\sl unique choice of relative
phases} which will make the sum of spin structures into a modular form.
It is obtained by taking all the relative phase factors to be the same,
say $+1$. We obtain in this way the following modular form of weight 8
\be
\Upsilon _8 (\Omega ) = \sum _\delta \tet [\delta ]^4 (\Omega) \Xi _6
[\delta ] (\Omega )\, .
\ee
The resulting chiral superstring measure is then 
$$
d\mu(\Omega)
={1\over 16\pi^6}\prod_{I\leq J}d\Omega_{IJ}\Upsilon_8(\Omega)
\Psi_{10}^{-1}(\Omega)
$$ 
with the modular form  $\Upsilon_8(\Omega)\Psi_{10}^{-1}(\Omega)$
of weight -2 as coefficient, as desired.

\medskip

{\sl Notice that there is a unique way of constructing this modular
form, as opposed to, say, a mere superposition
of $\tet [\delta]^4$, where
there would be 6 independent choices.} For example, had 
$\Xi_6[\delta](\Omega)$ been independent of $\delta$ (and hence been
already a modular form $\Xi_6(\Omega)$ of weight 6, instead of merely
transforming covariantly into $\Xi_6[\tilde\delta]$), we would have had
the following six independent choices
\be
\label{badddd}
\Xi_6(\Omega)\sum_{\delta}\<\nu|\delta\>
\tet^4[\delta](\Omega)
\ee
for each odd spin structure $\nu$. They would 
all lead to a modular form of weight $8$, which is
identically $0$, by the Riemann identity.

\subsection{The Ring of Genus 2 Modular Forms}

It may be helpful to summarize here some basic facts
about modular forms in genus $h=2$. The ring of modular forms
in genus $2$ has been identified by Igusa \cite{igusa1,igusa2,igusa3}
as a polynomial ring with generators $\Psi_4$, $\Psi_6$,
$\Psi_{10}$ and $\Psi_{12}$. Of particular interest to us
are $\Psi_4$, $\Psi_6$, and $\Psi_{10}$.
We have encountered $\Psi_{10}=\prod_{\delta}\tet^2 [\delta](\Omega)$
before, while $\Psi_4(\Omega)$ and $\Psi_6(\Omega)$ are defined
respectively by
\be
\label{psi4}
\Psi _4 (\Omega) \equiv \sum _\delta \tet^8 [\delta ] (0,\Omega)
\ee
and by
\be
\label{psi6}
\Psi_6(\Omega)
=
{1 \over 4} \sum_{syz(\delta_1, \delta _2,\delta_3)} \pm \tet [\delta _1]^4 
\tet[\delta_2]^4 \tet[\delta_3]^4
\ee
Here, the sum is taken over {\sl syzygous triples} of even characteristics 
$\delta _1$, $\delta _2$ and $\delta_3$ which are defined to satsify 
\bea
\< \delta _1 | \delta _2\> \< \delta _1 | \delta _2\> \< \delta _1 | \delta _2\>
=1
\eea
while {\sl antisyzygous triples} have $-1$ instead. Of the 120 triples of 
pairwise distinct even spin structures, 60 are syzygous and 60 anti-syzygous, 
and these characterizations are invariant under the action of the modular group.
The expression (\ref{psi6}) looks deceptively similar to
$\Xi_6[\delta](\Omega)$, but it is of course different.
It does not depend on a spin structure $\delta$ and is a genuine
modular form. Had it taken the place of $\Xi_6[\delta](\Omega)$
in the expression (\ref{finamppq})
for $d\mu[\delta](\Omega)$, there would have been
no unique way of enforcing the GSO projection.

\subsection{Vanishing of the Cosmological Constant}

Now that the full chiral measure for the superstring has been determined, we can 
turn to the evaluation of the string cosmological constant for the Type II 
superstrings, \cite{gs82} which is given by
\be
\Lambda _{II}
=
\int \det^{-5}\Im\,\Omega\ 
\bigg|{\prod_{I\leq J}d\Omega_{IJ}
\over 16\pi^6}\bigg|^2
\bigg |{\Upsilon _8(\Omega)\over\Psi_{10}(\Omega)} \bigg |^2
\ee
We shall show that the cosmological constant $\Lambda _{II}$ vanishes by
showing that the modular form $\Psi_8(\Omega)$ vanishes identically in
$\Omega$. It is very instructive to do this in two different ways. 

\medskip

The first proof of the vanishing of $\Upsilon _8 (\Omega)$ is obtained by
exploiting the fact that $\Upsilon _8$ is a modular form of weight 8. 
Recall that there is in genus $2$
a unique modular form $\Psi _4$ of weight 4 given by (\ref{psi4}).
The sum in (\ref{psi4})
is over all even spin structures $\delta$. By the classification result of Igusa 
quoted above, any modular form of weight 8 must be proportional to
$\Psi _4 ^2$, so that $\Upsilon _8 (\Omega) = r \Psi _4 (\Omega)^2$,
with $r$ a constant. Since $r$ is independent of $\Omega$, it may be
evaluated in the separating degeneration limit of
\S 5.1.1. It is well-known that the limit $\tau \to 0$ of $\Psi_4$ is
non-vanishing; from (\ref{tetasym}), it is given by
\bea
\Psi _4 (\Omega ) = 
\biggl (\sum _{\mu _1} \tet _1 [\mu_1] (0,\tau_1)^8 \biggr ) 
\biggl (\sum _{\mu _2} \tet _1 [\mu_2] (0,\tau_2)^8 \biggr ) + \O(\tau ^2)
\eea
where the summations are over all genus 1 even spin structures $\mu_1$
and $\mu_2$. The limit of $\Upsilon _8$ may be obtained from the limits
of $\tet ^4 [\delta] \Xi _6 [\delta]$, derived in turn
from (\ref{tetasym}) and (\ref{Xiasym}),
\bea
\tet \left [ \matrix{ \mu_1  \cr \mu _2 \cr} \right ]^4 
\Xi _6 \left [ \matrix{ \mu_1  \cr \mu _2 \cr} \right ] 
& = &
- 2^8  \<\mu_1 |\nu_0\> \<\mu_2|\nu_0\> \tet _1 [\mu_1] ^4 (0,\tau_1)
 \tet _1 [\mu_2] ^4 (0,\tau_2) \eta (\tau_1) ^{12} \eta (\tau_2)^{12}
+\O(\tau^2)
\nonumber \\
\tet \left [ \matrix{ \nu_0  \cr \nu_0 \cr} \right ]^4 
\Xi _6 \left [ \matrix{ \nu_0  \cr \nu_0 \cr} \right ] 
& = &
- 3 \cdot 2^{12} \pi ^4 \tau^4   \eta (\tau_1) ^{24} \eta (\tau_2)^{24}
+ \O(\tau^6)
\eea
Clearly the last case above tends to 0 as $\tau \to 0$, and will not
contribute in this limit. The summation over all even spin structures 
thus reduces to a summation over all genus 1 even spin structures $\mu_1$
and $\mu_2$, and the limit is given by
\bea
\Upsilon _8 (\Omega ) = - 2^8 \eta (\tau_1) ^{12} \eta (\tau_2)^{12}
\prod _{i=1,2} \biggl (\sum _{\mu _i} \<\mu_i |\nu_0\> \tet _1 [\mu_i]
(0,\tau_i)^4 \biggr )   + \O(\tau ^2)
\eea
but this vanishes by the genus 1 Riemann identity
\bea
\sum _{\mu _i} \<\mu_i |\nu_0\> \tet _1 [\mu_i] (0,\tau_i)^4 =0
\eea
Thus, the constant $r$ vanishes and therefore we have $\Upsilon _8(\Omega
)=0$ identically. As a result, the cosmological constant vanishes for both
the Type II and the heterotic string theories to two loop order.

\medskip

The second proof of the vanishing of $\Upsilon _8 (\Omega)$ is
constructed by using the Riemann relations to recast $\Upsilon _8
(\Omega)$ in terms of the modular form $\Psi _4^2$ and the modular form 
\bea
\Psi _8 (\Omega) \equiv \sum _\delta \tet [\delta ]^{16} (0,\Omega)
\eea
and exploiting the well-known relation $4\,\Psi _8 = \Psi _4 ^2$.
To compute $\Upsilon _8$ in terms of $\Psi _4 ^2 $ and $\Psi _8$, we
operate as follows. We write out the sum over $\delta$ as a sum over
triples $\delta = \nu_l + \nu_m + \nu_n$, taking into account the mirror
property,
\be
\Upsilon _8 (\Omega )
=
\half \sum _{1 \leq i<j<k \leq 6} \ \sum _{l<m\in \{i,j,k\}} 
\< \nu _l |\nu _m\>
\prod _{n\not= l,m } \tet [\nu_l + \nu_m +\nu_n]^4
\ee
Given that a pair $(l,m)$ belongs to 4 different triples, we may rewrite
this sum simply as a sum over all pairs $(m,n)$, as follows
\be
\Upsilon _8 (\Omega) =
2\sum _{l<m}  \< \nu _l |\nu _m\>
\prod _{n\not= l,m } \tet [\nu_l + \nu_m +\nu_n]^4
\ee
There are 15 such pairs, giving rise to 15 different contributions to
$\Upsilon_8$,
\bea
\half \Upsilon _8 &=&
-(1)(4)(5)(8) - (1)(2)(6)(9) +(2)(3)(5)(7)
\nonumber \\ &&
+(2)(3)(6)(8) + (3)(4)(5)(9) -(1)(4)(6)(7)
\nonumber \\ &&
+(2)(4)(7)(9) + (5)(6)(7)(8) -(1)(3)(8)(9)
\nonumber \\ &&
-(1)(3)(7)(0) + (2)(4)(8)(0) +(5)(6)(9)(0)
\nonumber \\ &&
+(7)(8)(9)(0) - (1)(2)(5)(0) +(3)(4)(6)(0)\, .
\eea
Next, we use the Riemann bilinear relations $R[\nu]= \sum _\delta
\<\nu |\delta \> \tet [\delta ]^4=0$, each of which is associated with an
odd spin structure $\nu$. Each of the 15 terms of $\Upsilon_8$ is in one
to one correspondence with one of the 15 linear combinations $R[\nu_i]
- \<\nu_i |\nu_j \> R[\nu_j]$ of pairs $\nu_i$ and $\nu_j$, $i \not= j$.
Indeed, for any given pair, there are precisely 4 non-zero terms in each
the Riemann relation $R[\nu_i] - \<\nu_i |\nu_j \> R[\nu_j]$, determined
by the non-vanishing of the quantity
\bea
\< \delta |\nu_i \> - \<\delta |\nu_j \> \<\nu_j |\nu_i\>
=\left \{ \matrix{
2 \<\nu_j | \nu _k\> & {\rm if} \quad \delta  = \nu_i + \nu_j +\nu_k \cr
0 & {\rm if} \quad \delta   = \nu_i + \nu_k +\nu_l } \right .
\eea
where $\nu_i$, $\nu_j$, $\nu_k$ and $\nu_l$ are all distinct. The case
where $\delta$ is the sum of three $\nu$'s, all distinct from $\nu_i$
and $\nu_j$ is equivalent to the first case above using the mirror
property.  

\medskip

There are two cases, depending upon whether $(1)$ enters the product or
not. The corresponding Riemann identities are then of the form
\bea
0 & = & (1) - (i) - (j) - (k)
\nonumber \\
0 & = & (h) + (i) - (j) - (k)
\eea
for $h,i,j,k$ pairwise distinct and different from 1, and such that the
corresponding products $(1) (i) (j) (k)$ and $(h)(i) (j) (k)$ occur in
the sum $\Upsilon _8$. By placing two terms on the lhs and the other two
terms on the rhs and squaring both sides, one gets
\bea
+ 2 (1) (i) + 2 (j) (k)  & = & (1)^2 + (i)^2 - (j)^2 - (k)^2
\nonumber \\
- 2 (h) (i) + 2 (j) (k) & = & (h)^2 + (i)^2 - (j)^2 - (k)^2
\eea 
Taking again the square, we obtain
\bea
\pm 8 (g) (i) (j) (k) & = &
(g)^4 + (i)^4 + (j)^4 + (k)^4 - 2 (g)^2 (i)^2 - 2 (g)^2 (j)^2
\nonumber  \\
&&   - 2 (g)^2
(k)^2 - 2 (i)^2 (j)^2 - 2 (i)^2 (k)^2 - 2 (j)^2 (k)^2
\eea
where the sign on the left is $+$ for $g=1$ and $-$ for $g=h\not=1$.
using this formula and summing all terms in $\Upsilon_8$, we readily get
\bea
\Upsilon _8 (\Omega) = 2 \Psi _8 (\Omega) - \half \Psi _4 (\Omega )^2
\eea
and this vanishes by the well-known identity $4\Psi _8 (\Omega ) = \Psi
_4 (\Omega )^2$. However, this identity does not follow from the Riemann
relations alone.

\medskip

In conclusion, we observe that the preceding mechanism for enforcing
the GSO projection and producing a vanishing cosmological constant
provides yet another distinction with the many earlier efforts to
treat supermoduli \cite{fms,superm,superg,adp90,mand,mhns,neveu,opf}
and resolve ambiguities \cite{ars,ln,v87}
in superstring multiloop amplitudes. In particular, the earlier
proposals based on the picture-changing operator Ansatz \cite{fms}
inserted at various special points had all relied only on
Riemann identities to insure both modular invariance and the
vanishing of the cosmological constant \cite{ams,lp,fact}. 

\vfill\eject

\section{The Bosonic and Heterotic Strings}
\setcounter{equation}{0}

To complete our treatment of closed orientable string theories, we shall
discuss the chiral measures of the purely bosonic string in 26 dimensions
and of the heterotic strings in 10 dimensions.

\subsection{The chiral bosonic string measure}

A closed expression in terms of modular forms for the bosonic measure was
obtained in \cite{moore} by exploiting the constraints of modular
invariance and the behavior at the boundary of moduli space known on
physical grounds. Here, as another illustration of the methods
in this paper, we shall {\sl present a first principles
derivation} following the calculational scheme used for the superstring.
The starting point is the chiral measure of \cite{vv87, dp88}, when
bosonic ghosts are inserted at points $p_a$, $a=1,2,3$ and the period
matrix is used as moduli coordinates on the surface,
\bea
d \mu _B (\Omega) = \Z_B (\Omega) d\Omega _{11} d\Omega _{12} d\Omega
_{22}
\eea
where the chiral matter-ghost partition function is given by
\bea
\label{chiralbosonic}
\Z_B =
{\tet (\sum _a p_a -3 \Delta) \prod _{a<b} E(p_a,p_b) \prod _a \sigma
(p_a)^3
\over Z^{27} \det \omega _I \omega _J (p_a)}
\eea
The chiral partition function is completely independent of the ghost
insertion points $p_a$, as may be checked by matching the zeros of the
numerator and denominator in $\Z_B$. We recall the expression for the
chiral partition function $Z$ of a single chiral boson,
\bea
\label{littleZ}
Z = {\tet (z_1 + z_2 -w-\Delta) E(z_1 , z_2)  \sigma (z_1) \sigma (z_2)
\over
E(z_1,w) E(z_2,w) \sigma (w) \det \omega_I (z_J)}
\eea
where the points $z_1$, $z_2$ and $w$ are arbitrary generic points.

\medskip

We now wish to evaluate the quantity $\Z_B$ as a modular form. We proceed
as follows. First, we place the three ghost insertion points at three
distinct branch points. Each branch point uniquely corresponds to an odd
spin structure and the partition of three odd spin structures fixes a
unique even spin structure, which we shall denote by $\delta$,
\bea
\label{peeandnu}
\delta = \nu_1 + \nu _2 + \nu_3
\hskip 1in
p_a = \Delta + \nu _a  \qquad a=1,2,3
\eea
Second, we choose the points $z_1$, $z_2$ and $w$ in (\ref{littleZ}) to
coincide with the points $p_a$, which may be done in three different ways.
Multiplying these three different ways together, and expressing the
points $p_a$ in terms of the odd spin structures inside the
$\tet$-functions, we have
\bea
\label{zeenine}
Z^9 = {\tet (\nu _1+ \nu_2-\nu_3) \tet (\nu_2+\nu_3-\nu_1)
\tet (\nu_3+\nu_1-\nu_2) \sigma (p_1) \sigma (p_2) \sigma (p_3)
\over
E(p_1,p_2) E(p_2,p_3) E(p_3,p_1) \det \omega _I(p_1,p_2)  \det \omega
_I(p_2,p_3) \det \omega _I(p_3, p_1)}
\eea
Third, we use the following remarkable identity,\footnote{The sign
factor in this formula corresponds to ordering the pairwise index
$IJ$ on the lhs as follows 11, 12, 22; and the single index $I$
on the rhs as follows 1, 2.}  
\bea
\det \omega _I \omega _J (p_a) = - \det \omega _I(p_1,p_2)  \det \omega
_I(p_2,p_3) \det \omega _I(p_3, p_1)
\eea
which may be derived by noticing that the 3$\times$3 determinant on the
left hand side is of the Vandermonde form and the right
hand side is its product representation.  Using (\ref{zeenine}) for one
power $Z^9$ in the denominator of (\ref{chiralbosonic}), the
determinants cancel and we are left with
\bea
\label{ZBintermediate}
\Z _B = {\tet (\nu_1+\nu_2 +\nu_3) \prod _{a<b} E(p_a,p_b) \prod _a
\sigma (p_a)^2
\over 
Z^{18} \tet (\nu _1+ \nu_2-\nu_3) \tet (\nu_2+\nu_3-\nu_1) 
\tet (\nu_3+\nu_1-\nu_2)}
\eea
Next, we produce an alternative expression for $Z^3$. Using
(\ref{theMformula})  to eliminate the finite-dimensional determinant in
$Z^3$ in facvor of $\M$, we obtain
\bea
Z^3 =  
{\tet (\nu_a + \nu_b - \nu_c) E(p_a,p_b) \sigma (p_a) \sigma (p_b)
\over 
E(p_a,p_c) E(p_b,p_c) \sigma (p_c) \omega _{\nu_a} (p_b) \omega
_{\nu_b}(p_a)} \cdot \M_{\nu_a \nu_b}
\eea
where $c \not= a,b$.
Again, given the points $p_a$, $a=1,2,3$, there are 3 different ways of
expressing $Z^3$ this way. Multiplying together all 3 expressions, we get
\bea
Z^9 & = & { \tet (\nu _1+ \nu_2-\nu_3) \tet (\nu_2+\nu_3-\nu_1) 
\tet (\nu_3+\nu_1-\nu_2) \sigma (p_1) \sigma (p_2) \sigma (p_3)
\over 
E(p_1,p_2) E(p_2,p_3) E(p_3,p_1) \prod _{a\not= b} \omega _{\nu_a}(p_b)}
\nonumber \\
&& \hskip 1in \times
\M_{\nu_1 \nu_2} \M_{\nu_2 \nu_3}  \M_{\nu_3 \nu_1}
\eea
Using this expression to eliminate the remaining factor of $Z^{18}$ in
(\ref{ZBintermediate}), we get
\bea
\Z_B = {\tet (\nu_1+\nu_2 +\nu_3) E(p_1,p_2)^4 E(p_2,p_3)^4 E(p_3,p_1)^4
\prod _{a< b} \omega _{\nu_a} (p_b)^4
\over 
\tet (\nu _1+ \nu_2-\nu_3)^3 \tet (\nu_2+\nu_3-\nu_1) ^3
\tet (\nu_3+\nu_1-\nu_2)^3 
\M_{\nu_1 \nu_2} ^2 \M_{\nu_2 \nu_3} ^2  \M_{\nu_3 \nu_1}^2}
\eea
Evaluating the prime forms, we use their form in terms of
$\tet$-functions at odd spin structure and we choose to evaluate
$E(p_a,p_b)$ with the help of the third odd spin structure $\nu_c$,
associated with the third point $p_c $, $c\not=a,b$. This way, one obtains
\bea
E(p_a,p_b)^2 \omega _{\nu_c} (p_a) \omega_{\nu_c}(p_b)
= \tet [\nu_c](\nu_a-\nu_b)^2
\eea
Using this result for each of the prime form factors above, we obtain an
expression for $\Z_B$ in terms of
$\tet$-constants and bilinear $\tet$-constants $\M$ only,
\bea
\label{ZBalmostfinal}
\Z_B = {\tet (\nu_1+\nu_2 +\nu_3) \tet [\nu_3](\nu_1 - \nu_2)^4 \tet
[\nu_1](\nu_2 - \nu_3)^4 \tet [\nu_2](\nu_3 - \nu_1)^4
\over 
\tet (\nu _1+ \nu_2-\nu_3)^3 \tet (\nu_2+\nu_3-\nu_1) ^3
\tet (\nu_3+\nu_1-\nu_2)^3 
\M_{\nu_1 \nu_2} ^2 \M_{\nu_2 \nu_3} ^2  \M_{\nu_3 \nu_1}^2}
\eea
Now, all the $\tet$-constants in the above relation are proportional to
$\tet [\delta ](0,\Omega)$, with the following proportionality
exponential factors,
\bea
\tet (\nu_1+\nu_2 +\nu_3) & = & C[\delta] \tet [\delta ](0,\Omega)
\nonumber \\
\tet (\nu_a + \nu_b - \nu _c) & = & C_c [\delta] \tet [\delta ](0,\Omega)
\nonumber \\
\tet [\nu_a ] (\nu_b - \nu _c) & = & \tilde C_a [\delta] \tet [\delta
](0,\Omega)
\eea
The exponential factors $C[\delta]$, $C_a[\delta]$ and $\tilde
C_a[\delta]$ are easily determined; suffice it to know here that the
combination that enters into (\ref{ZBalmostfinal}) reduces to 1 since
$C[\delta ] \tilde C_1 [\delta]^4 \tilde C_2 [\delta]^4 \tilde C_3
[\delta]^4 =  C_1 [\delta ]^3 C_2 [\delta ]^3 C_3 [\delta ]^3$. It is
then manifest that 9 powers of $\tet [\delta] (0,\Omega)$ cancel between
numerator and denominator in  (\ref{ZBalmostfinal}), so that we are left
with
\bea
\label{bosonmeasure}
d\mu_B (\Omega) = {\cal Z}_B \prod _{I\leq J} d\Omega _{IJ} 
\hskip .8in
\Z_B = {\tet [\delta ]^4 (0,\Omega) \over 
\M_{\nu_1 \nu_2} ^2 \M_{\nu_2 \nu_3} ^2  \M_{\nu_3 \nu_1}^2}
= {1 \over \pi ^{12} \Psi _{10}(\Omega)}
\eea
in view of the property (\ref{factortetdelta}) of $\M$ and the definition
(\ref{Psi10}) of the modular form $\Psi _{10} (\Omega)$. Notice that the
measure  $d\mu_B (\Omega)$ is a modular form of weight $-13$, as indeed
expected for the bosonic string in 26 dimensions.

\subsection{The Heterotic String Measure and Cosmological Constant}

The heterotic string amplitudes and measure are constructed by assembling,
at fixed internal momenta $p^\mu_I$, the chiral amplitude and measure 
for the left moving (holomorphic) part of the superstring with the chiral
amplitude for the right moving (anti-holomorphic) part of the bosonic
string, of which 16 dimensions have been compactified on the Cartan tori
of $Spin (32)/Z_2$ or of $E_8 \times E_8$, \cite{heter}. For the two-loop 
measure, the
contribution of the compactified bosons produces a winding contribution
which is given by an extra factor of the $Spin (32)/Z_2$ or $E_8
\times E_8$ root lattice $\tet$-functions. These are modular forms of
weight 8 and must thus be proportional to $\overline{\Psi _8 (\Omega)}$.
Thus, the two-loop right chiral heterotic string measure is given by
\bea
d\mu _H (\bar \Omega) = \overline{\Psi _8 (\Omega)} \ 
{ 1 \over \pi ^{12} \overline{\Psi _{10} (\Omega)} } \ d\bar \Omega
_{11} d\bar
\Omega _{12} d\bar \Omega _{22}
\eea
In the fermionic realization, the same factor arises by removing 16
chiral bosons (i.e. multiplying by a factor of $\bar Z^{16}$) and adding
in 32 Majorana-Weyl fermions (i.e. multiplying by the associated chiral
partition function $\overline{\Psi _8 (\Omega )}/ \bar Z^{16}$. The
heterotic string cosmological constant is given by integration of the
product of the measures,
\bea
\Lambda _H = \int d\mu _H (\bar \Omega) d \mu (\Omega)
= 
\int \det^{-5}\Im\,\Omega\ 
\bigg|{\prod_{I\leq J}d\Omega_{IJ}
\over 16\pi^6 \Psi_{10}(\Omega) }\bigg|^2
\Upsilon _8(\Omega) \overline{\Psi _{8} (\Omega)}
\eea
Which also vanishes in view of the fact that $\Upsilon _8=0$.

\vfill\eject

\section{Asymptotic Behavior of the Measure}
\setcounter{equation}{0}

In this last section, we derive the asymptotic behaviors of the chiral
superstring measure at fixed even spin structure as the genus 2 surface
degenerates, and we interpret the leading behaviors in terms of physical
processes. Two types of degenerations may be distinguished. In a given
canonical homology basis, we may write the period matrix as
\bea
\Omega = \left (\matrix{\tau_1 & \tau \cr \tau & \tau _2}\right )
\eea
The two degenerations are then : separating degeneration as $\tau \to 0$,
with $\tau_1$ and $\tau_2$ kept fixed; and non-separating degeneration as 
$\tau _2 \to + i \infty$, with $\tau_1$ and $\tau$ kept fixed. We 
investigate these two types of degenerations of the measure in turn. (The
degenerations for the bosonic string are well-known \cite{div} and may be
recovered from (\ref{bosonmeasure}).)

\subsection{Separating Degenerations}

We shall be interested in the asymptotics of the measure $d\mu
[\delta](\Omega)$ to leading order. Therefore, we shall need
the leading asymptotics given in (\ref{tetasym}), (\ref{psitenasym}) and
(\ref{Xiasym}). Putting all together, we find the following asymptotic
behavior of the chiral superstring measure,
\bea
d\mu \left [ \matrix{  \mu_1 \cr \mu_2  \cr} \right ] (\Omega)
& = &
{1 \over 2^{10} \pi ^8 \tau ^2} \ \< \mu _1 |\nu _0\> \<\mu _2 |\nu_0 \> 
{\tet _1 [\mu_1 ]^4 (\tau_1) \tet [\mu _2 ]^4 (\tau_2) \over 
\eta (\tau _1)^{12} \eta (\tau _2)^{12}} \ d\tau _1 \ d\tau_2 \ d\tau
+\O(\tau^0)
\nonumber \\
d\mu \left [ \matrix{ \nu_0 \cr \nu_0  \cr} \right ](\Omega)
& = &
{ 3 \tau ^2 \over 2^6 \pi ^4} \ d\tau _1 \ d\tau_2 \ d\tau
+\O(\tau^0)
\eea
The physical interpretation of the divergences is as follows.

\begin{itemize}

\item
The measure for the single spin structure $\delta _0$ with R sectors
on each genus 1 component behaves in a completely regular way as $\tau \to
0$. Neither intermediate tachyon nor massless particles arise, and this
is as expected. 

\item 
Viewed as a measure on the genus 1 components, the limiting measure $d\mu
[\delta _0]$ is also completely regular if we further let $\tau_2
\to +i\infty$ in the R sector, again as expected.

\item
The measure for the remaining 9 spin structures $\delta _i$, $i=1,\cdots
,9$, with NS sectors on each genus 1 component exhibits an intermediate
tachyon pole in the form $d\tau /\tau^2$, as expected. It also exhibits a
massless pole, which may be seen by combining the left and right measures
and expanding the measure factor $(\det \Im \Omega )^{-5}$ for small
$\tau$, as expected.

\item
Viewed as a measure on the genus 1 components, the limiting measure
$d\mu[\delta _i]$ further exhibits a tachyon divergence as $\tau_2 \to
+i\infty$, as expected. 

\item
Performing a partial sum over all even spin structures on a single genus
1 component cancels the tachyon divergence as expected as well,
$$
\sum _{i=1,3,5} d\mu \left [ \matrix{ \mu_i \cr \mu_k  \cr} \right ]
(\Omega) = \sum _{j=2,4,6} d\mu \left [ \matrix{ \mu_l \cr \mu_j  \cr} 
\right ] (\Omega) =0
\qquad k=2,4,6;\ l=1,3,5
$$

\end{itemize}

\subsection{Non-Separating Degenerations}

We shall be interested in the singular behavior as $q\to 0$ of the
measure $d\mu[\delta ](\Omega)$. Therefore, it suffices to use the
asymptotics of the $\tet$-functions, of $\Psi _{10}(\Omega)$ and of the
objects $\Xi _6 [\delta ](\Omega)$ obtained in Section \S5,
\bea
\label{nonseplimits}
d\mu \left [ \matrix{\mu_1 \cr \mu_2  \cr} \right ]  
= - 
d\mu \left [ \matrix{\mu_1 \cr \mu_4  \cr} \right ] 
& = &
{\tet _1 [\mu_1] ({\tau \over 2}, \tau_1) ^4 + \tet _1 [\nu_0] ({\tau
\over 2}, \tau_1) ^4 \over 2^8 \pi ^6 \cdot q \cdot \eta (\tau_1)^6 \tet
_1 [\nu_0] (\tau, \tau_1) ^2 } \ d\tau _1 \ d\tau_2 \ d\tau  + \O(q^0)
\nonumber \\
d\mu \left [ \matrix{\mu_3 \cr \mu_2  \cr} \right ]  
=  - 
d\mu \left [ \matrix{\mu_3 \cr \mu_4  \cr} \right ] 
& = &
{-\tet _1 [\mu_3] ({\tau \over 2}, \tau_1) ^4 + \tet _1 [\nu_0] ({\tau
\over 2}, \tau_1) ^4  \over 2^8 \pi ^6 \cdot q \cdot \eta (\tau_1)^6 \tet
_1 [\nu_0] (\tau, \tau_1) ^2}  \ d\tau _1 \ d\tau_2 \ d\tau  + \O(q^0)
\nonumber \\
d\mu \left [ \matrix{\mu_5 \cr \mu_2  \cr} \right ]  
=  - 
d\mu \left [ \matrix{\mu_5 \cr \mu_4  \cr} \right ] 
& = &
{-\tet _1 [\mu_5] ({\tau \over 2}, \tau_1) ^4 + \tet _1 [\nu_0] ({\tau
\over 2}, \tau_1) ^4 \over 2^8 \pi ^6 \cdot q \cdot \eta (\tau_1)^6 \tet
_1 [\nu_0] (\tau, \tau_1) ^2} \ d\tau _1 \ d\tau_2 \ d\tau + \O(q^0)
\nonumber \\
d\mu \left [ \matrix{\nu_0 \cr \nu_0  \cr} \right ] = \ \
d\mu \left [ \matrix{\mu_i \cr \mu_6  \cr} \right ]  & = & \O(q^0) \qquad
i=1,3,5
\eea
The physical interpretation of the divergences is as follows.

\begin{itemize}

\item
The $1/q$ divergence signals the tachyon passing along the $B_2$ cycle
(i.e. traversing the $A_2$ cycle) of the second genus 1 component of the
degeneration.

\item
Summing the measures for spin structures $\mu_2=[00]$ and $\mu_4
=[0\half]$ on the genus 1 component with canonical homology cycles $A_2$
and $B_2$ corresponds to a partial GSO projection in the NS sector of the
long $B_2$ cylinder. This partial projection eliminates the tachyon in
the $B_2$ loop and as the cycle $B_2$ becomes infinitely long the
tachyon cancels out, as expected. 

\item The remaining even spin structures (last line in
(\ref{nonseplimits})) correspond to Ramond states moving along the $B_2$ cycle 
and therefore have no poles, as expected.

\end{itemize}

\bigskip

\noindent
{\large \bf  Acknowledgements}

\medskip

We are happy to acknowledge conversations with David Gieseker. 
We would especially like to thank Edward Witten for his continued 
interest in this project, and for helpful correspondence and 
conversations. In particular, it was he who stressed to us the 
importance of degenerations as consistency checks for the chiral 
measure, which influenced a substantial part of our work.
We also wish to thank the Aspen Center for Physics, the Santa Barbara Institute
for Theoretical Physics (ITP), the Ecole Normale Sup\' erieure, the
Institut Henri Poincar\' e (IHP), and the National Center for Theoretical
Science in Hsin-Chu, Taiwan for their hospitality while part of this work
was being carried out.

\vfill\eject

\end{document}